\documentclass[amsmath,amssymb,aps,twocolumn,superscriptaddress]{revtex4-2}

\usepackage{graphicx}
\usepackage{hyperref}
\usepackage{MnSymbol}
\usepackage[normalem]{ulem}
\usepackage{xcolor}

\DeclareMathOperator{\Tr}{Tr}
\DeclareMathOperator{\tr}{tr}
\DeclareMathOperator{\Prob}{Prob}
\DeclareMathOperator{\const}{const}
\DeclareMathOperator{\sign}{sign}
\DeclareMathOperator{\diag}{diag}

\begin{document}

\title{Theory of free fermions under random projective measurements}

\author{Igor Poboiko}
\author{Paul P{\"o}pperl}
\author{Igor V. Gornyi}
\author{Alexander D. Mirlin}
\affiliation{\mbox{Institute for Quantum Materials and Technologies, Karlsruhe Institute of Technology, 76021 Karlsruhe, Germany}}
\affiliation{\mbox{Institut f\"ur Theorie der Kondensierten Materie, Karlsruhe Institute of Technology, 76128 Karlsruhe, Germany}}

\date{\today}

\begin{abstract}
We develop an analytical approach to the study of one-dimensional free fermions subject to random projective measurements of local site occupation numbers, based on the Keldysh path-integral formalism and replica trick.
In the limit of rare measurements, $\gamma / J \ll 1$ (where $\gamma$ is measurement rate per site and $J$ is hopping constant in the tight-binding model), we derive a non-linear sigma model (NLSM) as an effective field theory of the problem. Its replica-symmetric sector is described by a $U(2) / U(1) \times U(1) \simeq  \text{S}^2$ sigma model with diffusive behavior, and the replica-asymmetric sector is a two-dimensional NLSM defined on $SU(R)$ manifold with the replica limit $R \to 1$. 
On the Gaussian level, valid in the limit $\gamma / J \to 0$, this model predicts a logarithmic behavior for the second cumulant of number of particles in a subsystem and for the entanglement entropy. 
However, the one-loop renormalization group analysis allows us to demonstrate that this logarithmic growth saturates at a finite value $\sim(J / \gamma)^2$ even for rare measurements, which corresponds to the area-law phase. This implies the absence of a measurement-induced entanglement phase transition for free fermions. The crossover between logarithmic growth and saturation, however, happens at exponentially large scale, $\ln l_\text{corr} \sim J / \gamma$. This makes this crossover very sharp as a function of the measurement frequency $\gamma/J$, which can be easily confused with a transition from the logarithmic to area law in finite-size numerical calculations. We have performed a careful numerical analysis, which supports our analytical predictions. 
\end{abstract}

\maketitle

\section{Introduction}
\label{sec:intro}
The problem of measurement-induced entanglement phase transitions has recently attracted much interest. It is closely related to the general problem of the dynamics of open systems in contact with environment, with the measurement apparatus being a specific realization of such environment. A lot of interest in this field has been motivated by ongoing developments in quantum information processing, with environment-induced noise being one of the main obstacles irrespective of specific architectures \cite{Aharonov2000a,Preskill2018a,Bharti2022}. Interestingly, measurements can be used as a source of a controllable noise that governs the properties of a quantum  system, in particular, entanglement.  

Quite generally, measurement-induced transitions are driven by a competition between unitary dynamics, which favors the spreading of entanglement through the system, and stochastic non-unitary evolution induced by the interaction with the measurement apparatus, which tends to reduce entanglement. Originally explored in quantum circuits \cite{Li2018a, Skinner2019a, Chan2019a, Szyniszewski2019a, Li2019a,  Bao2020a, Choi2020a, Gullans2020a, Gullans2020b, Jian2020a, Zabalo2020a, Iaconis2020a, Turkeshi2020a, Zhang2020c, Nahum2021a, Ippoliti2021a,    Ippoliti2021b,Lavasani2021a,Lavasani2021b,Sang2021a,Fisher2022,Block2022a,Sharma2022,Jian2023,Kelly2023}, measurement-induced entanglement transitions have also been studied in other systems, such as free fermionic systems \cite{Cao2019a,Alberton2021a,Chen2020a,Tang2021a, Agrawal2022, Barratt2022, Coppola2022,Ladewig2022,Carollo2022,Buchhold2022,Yang2022,Szyniszewski2022,Buchhold2021a,Buchhold2021a,VanRegemortel2021a,Youenn2023,Loio2023,Turkeshi2022b,Kells2023}, Majorana fermions \cite{Fava2023,Swann2023}, spin systems with Ising-type interaction \cite{Lang2020a,Rossini2020a,Biella2021a,Turkeshi2021,Tirrito2022,Yang2023,Weinstein2023,Murciano2023,Sierant2022a,Turkeshi2022a}, Bose-Hubbard type models \cite{Tang2020a,Goto2020a,Fuji2020a,Doggen2022a,Doggen2023}, disordered systems in the context of Anderson \cite{Szyniszewski2022} or many-body localization \cite{Lunt2020a,Yamamoto2023}, and extensions of Sachdev-Ye-Kitaev model \cite{Jian2021a,Altland2022}. While most of the efforts were either computational or analytical, signatures of measurement-induced phase transitions have also been reported in experimental studies of systems based on trapped ions \cite{Noel2022a} and superconducting quantum processors \cite{Koh2022, Hoke2023}.

An important quantitative measure that is commonly used to distinguish phases of the system subject to measurements in the context of measurement-induced phase transitions is the entanglement entropy that characterizes entanglement between a large subsystem and the rest of the system. Depending on the scaling of the entanglement entropy with the subsystem size, the possible phases include:
\begin{itemize}
    \item Volume-law phase, with the entanglement entropy proportional to the volume of the subsystem. Such behavior is characteristic for a typical highly-entangled pure many-body state.
    \item Area-law phase, where entanglement entropy scales linearly with the area of the boundary of the subsystem (thus independent of system size for one-dimensional systems). This behavior is characteristic of weakly entangled states with finite correlation length.
    \item Intermediate (``critical'') phases with the sublinear (e.g., power-law or logarithmic, etc.) growth of the entanglement entropy with the subsystem's volume.
\end{itemize}

A major part of the activity in the field was dealing with random quantum circuits
\cite{Li2018a, Skinner2019a,Szyniszewski2019a,Chan2019a,Choi2020a,Bao2020a,Zhang2020c,Turkeshi2020a,Iaconis2020a,Zabalo2020a,Jian2020a,Ippoliti2021a,Lavasani2021a,Sang2021a}. For this class of systems, a transition between the area-law and volume-law phases (with a logarithmic behavior of the entropy at criticality) was found numerically in most of the works. This result was also obtained analytically in certain limiting cases by a mapping onto known statistical mechanics models \cite{Skinner2019a,Jian2020a,Iaconis2020a,Lavasani2021a}. A similar behavior was also found for interacting many-body Hamiltonian models \cite{Tang2020a,Goto2020a,Fuji2020a,Doggen2022a,Doggen2023}.

On the contrary, the behaviour of non-interacting fermionic systems (and related Ising  models) remains a subject of debates. Several works 
reported a transition between the critical and area-law phases \cite{Buchhold2021a,Alberton2021a,Turkeshi2021,Szyniszewski2022,Kells2023}. At the same time, it was argued in Ref.~\cite{Cao2019a} that the area-law always holds in the presence of measurements. Numerical simulations in Ref.~\cite{Coppola2022} also favor the area law but with an intermediate logarithmic behavior for a small rate of measurements. 
For a model where measurements are replaced by random non-unitary dynamics, an emergent conformal field theory has been reported \cite{Chen2020a} with a single critical (logarithmic) phase.

In several papers, field-theoretical approaches to the problem of free fermions subjected to continuous monitoring have been proposed. In Ref. \cite{Buchhold2021a}, a replicated Keldysh bosonic theory was derived,
resulting in an effective Luttinger-liquid description, which yields a Berezinsky-Kosterlitz-Thouless-type transition between the area-law and logarithmic phases.  
However, the prediction of Ref. \cite{Buchhold2021a} that the ``central charge'' (a prefactor in front of the logarithm in the scaling of the entanglement entropy) is less than unity appears to be inconsistent with the numerical evidence \cite{Alberton2021a}. 

A step towards a derivation of a Keldysh non-linear sigma model (NLSM) for monitored free fermions was done in Ref.~\cite{Yang2022}. This approach yields a field theory that is similar to the NLSM describing the replica-symmetric sector, as derived in the present paper.
However, the description in Ref.~\cite{Yang2022} lacks the replica structure of soft modes that are relevant for the entanglement entropy. Further, the role of measurement-induced ``heating'', which inevitably happens in the monitored systems, was not addressed in that work. 

In this work, we derive and analyze a replicated Keldysh NLSM for one-dimensional free fermions under random local projective measurements. Its replica-symmetric sector is described by a $U(2) / U(1) \times U(1)$ sigma model, and the replica-asymmetric sector (which is of main interest for the behavior of entanglement) is a two-dimensional NLSM with the $SU(R)$ manifold subject to the replica limit $R \to 1$. On the Gaussian level, this field theory yields a logarithmic behavior for the second cumulant of number of particles in a subsystem and for the entanglement entropy. 
However, the one-loop renormalization group (RG) analysis shows that this logarithmic growth is affected by ``weak-localization corrections'' and saturates even for arbitrarily rare measurements. 
This corresponds to the area-law phase and thus implies the absence of a measurement-induced entanglement phase transition for free fermions. For a small measurement rate, the true thermodynamic limit revealing the area law requires exponentially large system sizes. We also perform numerical simulations that confirm these analytical predictions.

While our work neared completion, a related replicated NLSM was proposed for continuously monitored Majorana fermions in Refs.~\cite{Jian2023,Fava2023}. 
The replica limit $R \to 1$ was established there as crucial for taking into account the Born rule for the probabilities of measurement outcomes, as opposed to the case of ``forced measurements'' \cite{Jian2023}, where the $R\to 0$ limit should be taken. 
The sigma-model manifold for the case of monitored Majorana fermions was found to be the orthogonal group $SO(R)$, which differs from the special unitary group derived in the present work. As a consequence, the RG flow for the NLSM of Refs.~\cite{Jian2023,Fava2023} has the opposite sign of $\beta$-function compared to our case. This behavior is reminiscent of the weak anti-localization RG in two-dimensional disordered systems with spin-orbit interaction. It yields, for a weak monitoring of Majorana fermions, a critical phase with the $\ln^2 l$ scaling of the entanglement entropy. 

Thus, the complex fermions considered here and Majorana fermions addressed in Refs.~\cite{Jian2023,Fava2023} demonstrate essentially different types of behavior. This is a result of different symmetries of the models and, as a consequence, of associated NLSMs.   
More specifically, the system studied in the present work obeys particle number conservation, which does not hold for Majorana fermion random circuits. 

This paper is organized as follows. 
The model is defined in Sec.~\ref{sec:Model}. In  Sec.~\ref{sec:Keldysh}, we develop a field-theoretical approach based on the replica trick and fermionic Keldysh path integral. 
In Sec.~\ref{sec:Gaussian}, we analyze the model at the Gaussian level and obtain results for the density correlation function. As discussed in the following sections, these ``mean-field'' results are valid at intermediate sizes of the subsystem, $l \ll l_{\text{corr}}$, where $l_{\text{corr}}$ is the scale at which the quantum correction equals (up to a sign) the leading term.
Section \ref{sec:NLSM} is devoted to the derivation of $U(2R) / U(R) \times U(R)$ NLSM. Its replica-symmetric analysis is relevant for the dynamics of the density matrix averaged over the measurement trajectories. We further focus on the $SU(R)$ replica-asymmetric sector of the theory (describing particle-number fluctuations and entanglement) and analyze it by the RG means. Based on the results for the particle-number cumulant, we discuss the scaling of the entanglement entropy in Sec.~\ref{sec:Entropy}.
Our analytical findings are supported by direct numerical simulations in Sec.~\ref{sec:Numerics}.  Finally, we summarize the results of this work and discuss its possible implications and generalizations in Sec.~\ref{sec:Conclusions}. Some technical aspects of our calculations are presented in Appendices \ref{sec:appendix:Keldysh}, \ref{sec:appendix:HubbardStratonovich}, \ref{App:NLSM}, \ref{sec:appendix:BallisticCrossover}, \ref{sec:appendix:RepliconAction} and \ref{sec:appendix:RG}.

\section{Model and observables}
\label{sec:Model}

\subsection{Measurement protocol}

We study the one-dimensional tight-binding free fermion model described by the following Hamiltonian:
\begin{equation}
\label{eq:HoppingHamiltonian}
\hat{H}_{0}=-J\sum_{x=1}^{L}\left[\hat{\psi}^{\dagger}(x)\hat{\psi}(x+1)+h.c.\right].
\end{equation}
During the time interval $[t_i, t_f]$ of duration $T = |t_f - t_i|$,
we randomly pick $M$ uniformly distributed time moments $t_m$, $m=1,\ldots, M$.
At each of these times $t_m$, we randomly choose a site $x_m \in \{1,\ldots, L\}$ (also from a uniform distribution) and perform a projective measurement of the site occupation number 
$$\hat{n}(x_m)\equiv\hat{\psi}^{\dagger}(x_{m})\hat{\psi}(x_{m}).$$
The outcome of this measurement $n_m$ can be either zero or unity. We are interested in the thermodynamic limit $M, L, T \to \infty$, keeping the measurement rate per site $\gamma \equiv M / L T$ finite. The protocol is similar to that in Ref.~\cite{Paul2023}, where random local projective measurements were considered for a single-particle (in contrast to the many-body here) problem in a disordered chain. 

We describe the system in terms of a non-normalized time-dependent density matrix $\hat{D}(t)$, which is defined as follows. Initially ($t=t_i$), it coincides with the system's density matrix, $\hat{D}(t_i) \equiv \hat{\rho}_0$. Between two consecutive measurements at times $t_{m}$, $t_{m+1}$ it undergoes the standard unitary evolution with the evolution operator 
$$\hat{U}_{0}(t_{m+1},t_{m})=\exp\left[-i\hat{H}_{0}(t_{m+1}-t_{m})\right],$$  
according to
\begin{equation}
\hat{D}(t_{m+1}) = \hat{U}_0(t_{m+1},t_{m}) \hat{D}(t_{m}) \hat{U}_0(t_{m}, t_{m+1}).
\end{equation}
A measurement of the site occupation with a given outcome $n_m = 0,1$ changes this matrix discontinuously:
\begin{equation}
\hat{D}(t_{m}+0)=\hat{\mathbb{P}}_{n_{m}}(x_{m})\hat{D}(t_{m}-0)\hat{\mathbb{P}}_{n_{m}}(x_{m}),
\end{equation}
where $\hat{\mathbb{P}}_{n_{m}}$ is a projection operator onto the corresponding eigensubspace of $\hat{n}(x_m)$. These projectors are explicitly given by
\begin{equation}
\hat{\mathbb{P}}_{0}(x)=1-\hat{n}(x),\quad\hat{\mathbb{P}}_{1}(x)=\hat{n}(x).
\end{equation}

The (normalized) density matrix for a given measurement trajectory $\{x_m, t_m, n_m\}$ can be expressed as:
\begin{equation}
\label{eq:DensityMatrixD}
\hat{\rho}(t)=\hat{D}(t)/\Tr \hat{D}(t).
\end{equation}
The normalization factor $\Tr \hat{D}(t)$ has its own physical meaning. Specifically, it provides a generalization of Born's rule for a set of consecutive projective measurements, i.e., it gives the probability for the sequence of measurement outcomes $\{n_m\}$ for a given set of points and time moments $\{x_m, t_m\}$:
\begin{equation}
\label{eq:BornRule}
\Prob(\{n_{m}\}|\{x_{m},t_{m}\})=\Tr\hat{D}(\{x_{m},t_{m},n_{m}\}).
\end{equation}
For the purposes of this work, we will focus on pure initial states,
$$\hat{\rho}_0 = \left|\Psi_{0}\right\rangle \left\langle \Psi_{0}\right|.$$
The purity of the quantum state is maintained both by the unitary evolution and by measurements for any given quantum trajectory $\{x_m, t_m, n_m\}$.

\subsection{Quantities of interest}
 
The key quantity of interest in the present context is the entanglement entropy, which is defined as follows. 
Consider a subsystem $A$ and the rest of the system $\bar{A}$, and introduce a reduced density matrix in the standard way via the partial trace $\hat{\rho}_A = \Tr_{\bar{A}} \hat\rho$. 
The entanglement entropy $\mathcal{S}_{E}$ is given by the usual von Neumann entropy of the reduced density matrix:
\begin{equation}
    \mathcal{S}_{E} = -\Tr(\hat{\rho}_{A} \ln \hat{\rho}_{A}).
\end{equation}

Let us focus for simplicity on the case when the initial state $\left|\Psi_{0}\right\rangle$ is a pure Gaussian state (a Slater determinant). The projective measurements do not change the Gaussian property of the state $\left|\Psi(t)\right>$, which allows us to relate the entanglement entropy to the full counting statistics of the number of particles in the subsystem via the formula by Klich and Levitov \cite{KlichLevitov} 
(see also Refs.~\cite{Song2011,Song2012,Thomas2015,Burmistrov2017}):
\begin{equation}
\label{eq:KlichLevitov}
\mathcal{S}_{E}=\sum_{q=1}^{\infty}2\zeta(2q){\cal C}_{A}^{(2q)}
=
\frac{\pi^{2}}{3}{\cal C}_{A}^{(2)}+\frac{\pi^{4}}{45}{\cal C}_{A}^{(4)}+\dots,
\end{equation}
where 
\begin{equation}
\label{eq:ParticleCumulants}
{\cal C}_{A}^{(N)}=\left\llangle \left(\sum_{x\in A}\hat{n}(x)\right)^{N} \right\rrangle 
\end{equation}
is the $N$-th cumulant (as denoted by double angular brackets) of the number of particles in the subsystem $A$. 
This relation holds for an arbitrary measurement trajectory, and thus it holds for quantities averaged over trajectories as well. What makes such an averaging highly nontrivial is that $N$-th cumulant is a nonlinear functional of the density matrix $\hat{\rho}_A$: it contains terms up to the $N$-th order with respect to the density matrix. This means that one should be able to average an arbitrary power of the density matrix over the measurement trajectories. We are now going to discuss how to deal with this problem analytically.

\section{Replicated Keldysh field theory}
\label{sec:Keldysh}

\subsection{Replica trick and Keldysh action}
\label{sec:replica-trick}

The problem of averaging the $N$-th cumulant of number of particles reduces to the problem of simultaneous averaging of $N$ copies of the density matrix:
\begin{equation}
\hat{\rho}_N \equiv \overline{\otimes_{r=1}^{N} \hat{\rho}_r},
\end{equation}
where the overbar denotes the averaging over quantum trajectories
${(x_m, t_m, n_m)}$. 
The crucial step is then to rewrite the averaged replicated density matrix in terms of matrices $\hat{D}$ using 
Eq.~\eqref{eq:DensityMatrixD}. Performing averaging over the measurement outcomes with the Born rule probabilities given by Eq.~\eqref{eq:BornRule}, we get
\begin{equation}
\hat{\rho}_{N}= \overline{\sum_{\{n_{m}=0,1\}} 
\left(\otimes_{r=1}^{N}\hat{D}_{r}
\right)/(\Tr \hat{D})^{N-1}
}^{(x_{m},t_{m})}.
\label{eq11}
\end{equation}
Here, the overbar with the label $(x_m,t_m)$ stands for averaging over positions and times of measurements for fixed outcomes.
In order to perform the averaging within the field theory, we get rid of denominators by utilizing the replica trick:
\begin{equation}
\hat{\rho}_{N}=\lim_{R\to1}\overline{\sum_{\{n_{m}=0,1\}}\Tr_{r=N+1,\dots,R}\otimes_{r=1}^{R}\hat{D}_{r}}^{(x_{m},t_{m})}.
\label{rhoQ-replica-limit}
\end{equation}
Here, the product of first $N$ out of $R$ replicas produces the numerator in Eq.~(\ref{eq11}), while the trace over the rest of replicas, $N+1,\ldots,R$, yields $(\Tr \hat{D})^{R-N}$, which in the limit $R\to 1$ gives the denominator of Eq.~(\ref{eq11}), $(\Tr \hat{D})^{N-1}$.
Note that the total number of replicas of $\hat{D}$-matrices, denoted as $R$, is independent of $N$: the replica limit relevant to this problem is
\begin{equation}
    R\to 1.
    \label{replica-lim-R-1}
\end{equation}

The expression on the right-hand side of Eq.~(\ref{rhoQ-replica-limit}) is explicitly defined only for integers $R \geq N$, so that for $N > 1$ an analytic continuation $R \to 1$ is needed. To calculate the observables in $N$ replicas and to implement the trace over replicas $N+1,\ldots,R$ in Eq.~(\ref{rhoQ-replica-limit}), we will introduce the corresponding sources in $N$ replicas, see Sec.~\ref{Sec:VD} below for details. For $N = 1$, which incorporates only the properties of the average density matrix, no analytic continuation is required, as seen from Eq.~(\ref{eq11}) which does not contain a denominator in this case. It is worth emphasising that this replica trick differs from the usual replica trick used to perform averaging over quenched disorder, where the replica limit $R \to 0$ should be taken. The limit \eqref{replica-lim-R-1} in our calculation is a direct consequence of the Born's rule, which gives an extra $\Tr \hat{D}$ factor in the numerator.

Remarkably, the average of $R$ copies of matrix $\hat{D}$ over random Poissonian statistics of measurement times and uniform distribution of their location, together with summation over outcomes, can be performed exactly within the Keldysh formalism. As detailed in Appendix \ref{sec:appendix:Keldysh}, the averaging yields the following local action:
\begin{equation}
\label{eq:Action:Fermionic}
iS[\bar{\psi},\psi]=i\sum_{r=1}^{R}\bar{\psi}_{r}\hat{G}_{0}^{-1}\psi_{r}+i\gamma\int d^2\textbf{x}{\cal L}_{M}[\bar{\psi},\psi],
\end{equation}
where we have introduced the short-hand notation $\textbf{x} = (x, t)$ and $\int d^2 \textbf{x} = \sum_x \int_{t_i}^{t_f} dt$.
The quadratic part in Eq.~\eqref{eq:Action:Fermionic} describes free fermions:
\begin{equation}
\label{eq:BareFermionicGreenFunction}
\hat{G}_{0}^{-1}=i\partial_{t}-\hat{H}_{0}+i\delta\hat{\Lambda}_{0},\quad \hat{\Lambda}_{0}(\epsilon)=\begin{pmatrix}1 & 2F_{0}(\epsilon)\\
0 & -1
\end{pmatrix}_{K},
\end{equation}
where $\hat{H}_0$ is the Hamiltonian \eqref{eq:HoppingHamiltonian}.
The term with infinitesimal $\delta \to +0$ fixes the correct causality properties of the retarded (advanced) Green's functions and  contains the information about the initial Keldysh distribution function $F_0(\epsilon) = 1 - 2 f_0(\epsilon)$. 
The additional term in Eq.~(\ref{eq:Action:Fermionic})
involving
\begin{equation}
\label{eq:Action:Measurement:Grassman}
i{\cal L}_{M}[\bar{\psi},\psi]=\sum_{n=0,1}\prod_{r}V_{n}[\bar{\psi}_{r},\psi_{r}]-1,
\end{equation}
where
\begin{equation}
V_{0,1}[\bar{\psi},\psi]=\frac{1}{4}\mp\frac{1}{2}\left(\bar{\psi}_{2}\psi_{1}+\bar{\psi}_{1}\psi_{2}\right)-\bar{\psi}_{1}\psi_{1}\bar{\psi}_{2}\psi_{2},
\label{eq17}
\end{equation}
results from the measurements.
We note that this term is local in space and time: $(x, t)$-arguments of all fields in Eqs.~(\ref{eq:Action:Measurement:Grassman}) and (\ref{eq17}) are the same.

It is convenient to rewrite the interaction vertices in the exponential form (here $\hat{\tau}_x$ is Pauli matrix acting in the Keldysh space):
\begin{equation} \label{V01}
V_{0,1}[\bar{\psi}_r,\psi_r]=\frac{1}{4}\exp\left(\mp2\bar{\psi}_{r}\hat{\tau}_{x}\psi_{r}\right). 
\end{equation}               
This can be done because of the Grassmanian nature of fields $\psi$, which terminates the series expansion of the exponential at the second term. Substituting Eq.~(\ref{V01}) in Eq.~(\ref{eq:Action:Measurement:Grassman}), we arrive at  
\begin{equation}
\label{eq:Action:Measurement:Exponential}
i{\cal L}_{M}[\bar{\psi}, \psi]=\frac{2}{4^{R}}\cosh\left(2\bar{\psi}\hat{\tau}_{x}\psi\right)-1.
\end{equation}

It is worth emphasizing that the time integration in the Keldysh action is performed up to time $t = t_f$, which is the time at which one calculates the observables with a non-trivial replica structure (such as the entanglement entropy or the particle-number cumulant). The introduction of this upper limit in the time integral is an important feature of the Keldysh formalism for the measurement problem. In the conventional Keldysh technique, one can continue the time-integration contour from $t = t_f$ to $t = +\infty$ and this part of the contour exactly cancels with the backward part because of the unitarity of quantum-mechanical evolution. However, in the presence of measurements for $R \neq 1$ (when the evolution is manifestly non-unitary owing to the insertion of projectors associated with the measurement-induced collapse of the wave function), this cancellation does not generically occur. A convenient way to take this into account is to ``switch off'' the measurements directly after the observation time $t_f$, i.e., to put $\gamma(t > t_f) = 0$. Clearly, this does not influence observation results. At the same time, this allows us to extend the Keldysh contour up to $t=+\infty$, since the evolution after $t=t_f$ is now unitary, so that the corresponding forward and backward contributions cancel out, as usual.

In what follows, when considering correlation functions with a nontrivial replica structure, we will be interested in the case of equal times, $t=t' = t_f$. Indeed, the particle-number cumulants and the entanglement entropy belong to this class of observables. Observables with a non-trivial replica structure and $t \ne t'$, such as $\overline{\left\langle \hat{n}(x,t)\right\rangle \left\langle \hat{n}(x^{\prime},t^{\prime})\right\rangle }$, require simultaneous averaging of two density matrices taken at different times. Our replica approach described above would require some modification in order to calculate such quantities. This is beyond the scope of this paper.

\subsection{Generalized Hubbard-Stratonovich transformation}
\label{sec:HS-transformation}

We switch from the Grassmanian integration to the integration over bosonic modes incorporated in two auxiliary $2R \times 2R$ matrices, ${\cal G}$ and $\Sigma$, utilizing the generalized Hubbard-Stratonovich transformation (see Appendix \ref{sec:appendix:HubbardStratonovich} for the detailed derivation). The matrix ${\cal G}_{ij}(x,t)\sim -i\psi_{i}(x,t)\bar{\psi}_{j}(x,t)$ (with indices $i,j$ incorporating both Keldysh and replica structure) is related to the local fermionic Green's function, and $\Sigma(x,t)$ to the fermionic self-energy. These matrices are originally introduced as Hermitian ones with a flat integration measure; however, adjustment of the integration contour over the eigenvalues of ${\cal G}$ to the complex plane is required to ensure the convergence of the integral at infinity.

The resulting action has the form 
\begin{equation}
S[{\cal G},\Sigma]=S_{0}[{\cal G},\Sigma]+\gamma\int d^{2}\textbf{x}{\cal L}_{M}[{\cal G}]
\label{eq:S-S0-LM}
\end{equation}
with:
\begin{align}
\label{eq:Action:Matrix}
iS_0[{\cal G},\Sigma] &=\Tr\left[\ln(i\partial_{t}-\hat{H}_{0}+i\hat{\Sigma})-i\hat{\Sigma}\hat{{\cal G}}\right],\\
\label{eq:Action:Measurement:Matrix}
i{\cal L}_M[{\cal G}] &= \det\left(\frac{1}{2}-i\hat{{\cal G}}\hat{\tau}_{x}\right)+\det\left(\frac{1}{2}+i\hat{{\cal G}}\hat{\tau}_{x}\right)-1.
\end{align}
The trace $\Tr$ is calculated in replica and Keldysh spaces as well as real space and time, and the infinitesimal ${\delta\to+0}$ term from Eq. \eqref{eq:BareFermionicGreenFunction} is omitted for brevity. In this action, $\cal G$ and $\Sigma$ are assumed to be slow variables. This was used to derive Eq.~(\ref{eq:Action:Measurement:Matrix}), which is obtained by decoupling Eq.~(\ref{eq:Action:Measurement:Exponential}) in all slow channels, as detailed in Appendix \ref{sec:appendix:HubbardStratonovich}.

\section{Gaussian approximation}
\label{sec:Gaussian}

We start an analysis of the action given by Eqs.~\eqref{eq:Action:Matrix} and \eqref{eq:Action:Measurement:Matrix} by treating it in the Gaussian approximation. This approximation is controlled by the parameter $\gamma / J \ll 1$, which corresponds to rare measurements.

\subsection{Saddle-point analysis}
\label{sec:saddle-point-analysis}

We first consider the $R=1$ case, where the measurement action \eqref{eq:Action:Measurement:Matrix} reduces to a manifestly $U(2)$-invariant expression:
\begin{equation}
\label{eq:Action:Measurement:Matrix:invariant}
i{\cal L}_{M}^{(\text{inv})}[{\cal G}]=2\det\hat{{\cal G}}-\frac{1}{2}.
\end{equation}
With this action, we proceed by finding spatially homogeneous saddle points of Eq.~(\ref{eq:S-S0-LM}). The saddle point equations then read:
\begin{equation}
\hat{{\cal G}}=\int_{-\infty}^{\infty}\frac{d\epsilon}{2\pi}\int_{-\pi}^{\pi}\frac{dk}{2\pi}\frac{1}{\epsilon-\xi(k)+i\hat{\Sigma}}\equiv -\frac{i}{2}\sign\hat{\Sigma},
\end{equation}
\begin{equation}
\hat{\Sigma}(\textbf{x})=-2 i \gamma \det\hat{{\cal G}}(\textbf{x})\cdot \hat{{\cal G}}^{-1}(\textbf{x}),
\end{equation}
where $\xi(k)=-2J\cos k$ corresponds to the bare fermionic spectrum. For the energy integration, the principal value is taken, in agreement with the regularization procedure described in Appendix \ref{sec:appendix:Keldysh}.

These equations have a manifold of solutions parametrized by the $2 \times 2$ matrix $\hat{Q}$, which satisfies the non-linear constraint $\hat{Q}^2 = 1$ as well as $\Tr \hat{Q}=0$, as follows:
\begin{equation}
\label{eq:SCBA:SigmaG}
\hat{{\cal G}}=-i\hat{Q}/2,\quad\hat{\Sigma}=\gamma \hat{Q}.
\end{equation}
This will be basis for the derivation of the NLSM in Sec.~\ref{sec:NLSM}. For the purposes of Sec.~\ref{sec:Gaussian}, we need a particular 
solution, which has a form characteristic for Green's functions in Keldysh space [cf. Eq.~\eqref{eq:BareFermionicGreenFunction}]: it should satisfy the causality and be consistent with the initial conditions incorporated in the Keldysh distribution function $F_0(\epsilon)=1-2f_0(\epsilon)$.
As usual, such a saddle point corresponds to the solution of 
the self-consistent Born approximation (SCBA):
\begin{equation}
\label{eq:SCBA:Q}
\hat{Q}_{\text{SCBA}}=\hat{\Lambda}=\begin{pmatrix}1 & 2(1-2n)\\
0 & -1
\end{pmatrix}_{K},
\end{equation}
where $n=\int(dk / 2\pi)f_{0}(\xi_{k})  \in [0, 1]$ is the average fermionic density, i.e., the filling factor of the band. The number of particles is the only physical conserved quantity in the problem, and this filling factor is the only parameter that keeps any information about the initial state that was parametrized by the Keldysh distribution function.  

The one-particle Green's functions that correspond to the SCBA solution $\hat{\Lambda}$ are given by:
\begin{align}
\begin{split}
\label{eq:SCBA:G}
G_{R/A}(\boldsymbol{k}) &= \frac{1}{\epsilon-\xi(k)\pm i/2\tau_{0}},\quad\tau_{0}\equiv1/2\gamma,\\
G_{K}(\boldsymbol{k}) &= (1-2n)[G_R(\boldsymbol{k}) - G_A(\boldsymbol{k})],
\end{split}
\end{align}
where $\boldsymbol{k}=(k,\epsilon)$ and $\tau_0$ plays the role of the (inelastic) mean free time. Physically, this solution describes the steady-state of the fermions heated to infinite temperature. 
This should not be a surprise, given that projective measurements are inelastic processes that heat up the system (in the sense of fully randomizing its energy). In order to study the non-equilibrium transient regime (before the system achieves the steady state), one needs to introduce matrix fields that can depend on two time indices, $\hat{{\cal G}}(x,t,t^{\prime})$; this will be studied elsewhere.

The SCBA solution \eqref{eq:SCBA:G} can be shown to be exact for an arbitrary $\gamma / J$ ratio for $R = 1$, by using the fermionic diagram technique and rewriting the ``interaction'' \eqref{eq:Action:Measurement:Grassman} in a form 
\begin{equation}
   i{\cal L}_{M}[\bar{\psi},\psi]=-(\bar{\psi}\psi)\cdot(\bar{\psi}\psi) - 1/2. 
   \label{LMR1}
\end{equation} 
At variance with the general form of the vertex (\ref{V01}) involving the matrix $\tau_x$, this interaction vertex involves only the identity matrix in Keldysh space.
Since the interaction line is instantaneous, intersections of these lines are forbidden by causality, and only ``rainbow'' diagrams that are included in SCBA contribute to the Green functions. The exactness of the SCBA for the systems in the presence of random dynamical white noise, which are equivalent on the level of the Keldysh action to our case $R = 1$, was previously noted in Ref. \cite{Giamarchi2022}.

For arbitrary $R \neq 1$, one should instead consider the full form of the measurement-induced action \eqref{eq:Action:Measurement:Matrix}. 
One can check (see Appendix \ref{App:NLSM}) that $\hat{Q}_{\text{SCBA}}=\hat{\Lambda}$ remains a saddle point of the action for a half-filled band, $n=1/2$. For this case, $\hat{\Lambda}$ becomes the $\tau_z$ matrix in Keldysh space and the relation between $\hat{Q}$ and $\hat{\Sigma}$ in Eq.~(\ref{eq:SCBA:SigmaG}) is modified by the replacement $\gamma\to\gamma/2^{R-1}$. We expect that the half-filling case $n=1/2$ is representative for the problem that we consider, that is the physics should not qualitatively depend on $n$. 
For $n\neq 1/2$ and $R\neq 1$, the SCBA solution (\ref{eq:SCBA:Q}) ceases to be an exact saddle point, since the full action explicitly involves $\tau_x$ for $R\neq 1$. 
For an arbitrary filling factor, 
the terms that violate the saddle-point property of Eq.~\eqref{eq:SCBA:Q} have coefficients that vanish at $R\to 1$, so that the SCBA saddle point $\hat{\Lambda}$ is restored in this limit.
The saddle-point solution thus depends on the order of limits $R \to 1$ and $\delta \to 0$ [see Eq. \eqref{eq:BareFermionicGreenFunction}], and the correct order of limits should be the following: first take the limit $R \to 1$, and only then $\delta \to 0$. In this way, $\hat{Q}=\hat{\Lambda}$ is the correct saddle-point solution yielding the same physics for any $n$, as expected on physical grounds.

\subsection{Quadratic fluctuations around the saddle point}
\label{sec:quadratic-fluctuations}

We proceed with the Gaussian analysis by performing a second-order expansion of the full matrix action \eqref{eq:S-S0-LM}-\eqref{eq:Action:Measurement:Matrix}. We parametrize fluctuations around $\hat{\Lambda}$ from \eqref{eq:SCBA:Q} as $\hat{\Sigma}=\gamma(\hat{\Lambda}+\delta\hat{Q}_{\Sigma})/2^{R-1}$ and 
$\hat{{\cal G}}=-i\left(\hat{\Lambda}+\delta\hat{Q}_{{\cal G}}\right)/2$, and perform an expansion up to second order in $\delta\hat{Q}_{\Sigma}$ and $\delta\hat{Q}_{{\cal G}}$. In the $R \to 1$ limit (and also for arbitrary $R$ and $n=1/2$), the expansion starts from quadratic terms. The corresponding contribution to $S_0$ from Eq.~(\ref{eq:Action:Matrix}) reads
for $R\to 1$ as (see Appendix \ref{App:NLSM}):
\begin{multline}
i\delta S_{0}=\frac{1}{16 \tau_0^2}\int d^2\textbf{x}_{1}d^2\textbf{x}_{2}{\cal B}(\textbf{x}_{1}-\textbf{x}_{2}),\\
\times \Tr\left[\delta\hat{Q}_{\Sigma}(\textbf{x}_{1})(1+\hat{\Lambda})\delta\hat{Q}_{\Sigma}(\textbf{x}_{2})(1-\hat{\Lambda})\right]\\
-\frac{1}{4 \tau_0}\int d^2\textbf{x}
\Tr\left[\delta\hat{Q}_{\Sigma}(\textbf{x})\delta\hat{Q}_{{\cal G}}(\textbf{x})\right]
\label{eq:Action:MatrixGaussian}
\end{multline}
Here, trace $\Tr$ stands for replica and Keldysh spaces.
The expansion of the measurement part of the action (\ref{eq:Action:Measurement:Matrix}) contains two terms: 
$\delta{\cal L}_{M}=\delta{\cal L}_{M}^{(1)}+\delta{\cal L}_{M}^{(2)}$, where
\begin{multline}
i\gamma\delta{\cal L}_{M}^{(1)}=-\frac{1}{64n\tau_0}\Tr\left[\left((\hat{\Lambda}-\hat{\tau}_{x})\delta\hat{Q}_{{\cal G}}\right)^{2}\right]\\
-\frac{1}{64(1-n)\tau_0}\Tr\left[\left((\hat{\Lambda}+\hat{\tau}_{x})\delta\hat{Q}_{{\cal G}}\right)^{2}\right],
\label{eq:LM1}
\end{multline}
\begin{multline}
i\gamma \delta{\cal L}_{M}^{(2)}=\frac{1}{64n \tau_0}\Tr^{2}\left[(\hat{\Lambda}-\hat{\tau}_{x})\delta\hat{Q}_{{\cal G}}\right]\\
+\frac{1}{64(1-n)\tau_0}\Tr^{2}\left[(\hat{\Lambda}+\hat{\tau}_{x})\delta\hat{Q}_{{\cal G}}\right].
\label{eq:LM2}
\end{multline}

In Eq.~(\ref{eq:Action:MatrixGaussian}) we introduced the notation ${\cal B}(\textbf{x})$ for the elementary block of a diffuson ladder,
${\cal B}(\textbf{x})=G_{R}(\textbf{x})G_{A}(-\textbf{x})$, whose Fourier transform reads:
\begin{equation}
\label{eq:DiffusiveBlock}
{\cal B}^{-1}(\boldsymbol{q})=\sqrt{(1/\tau_0-i\omega)^{2}+\left(4J\sin\frac{q}{2}\right)^{2}},
\end{equation}
where $\boldsymbol{q}=(q,\omega)$.
For the smallest frequencies and momenta, $\omega \tau_0 \ll 1$ and $q l_0 \ll 1$, the block ${\cal B}$ acquires the following ``diffusive'' form:
\begin{equation}
{\cal B}^{-1}(\boldsymbol{q})\approx\tau_{0}^{-1}-i\omega+Dq^{2},\quad D\equiv v_{0}^{2}\tau_{0}=J^{2}/\gamma.
\end{equation} 
Here, we introduced the mean square velocity $v_0$ and the mean free path $l_0$:
\begin{equation}
\label{eq:MeanFreeParameters}
v_{0}^{2}\equiv\int_{-\pi}^{\pi}\frac{dk}{2\pi}\left(\frac{\partial\xi}{\partial k}\right)^{2}=2J^{2},\quad l_{0}=v_{0}\tau_{0}=\frac{J}{\gamma\sqrt{2}}.
\end{equation}

The structure of the measurement action, Eqs.~(\ref{eq:LM1}) and (\ref{eq:LM2}), suggests splitting $2R \times 2R$ matrices into two sectors in the replica space, ``longitudinal'' (``replica-symmetric'') and ``transversal'' (``replicon''):
\begin{equation}
\delta\hat{Q}^{(\parallel)}=\frac{1}{R}\tr_{R}\delta\hat{Q},\quad\delta\hat{Q}_{rr^{\prime}}^{(\perp)}=\delta\hat{Q}_{rr^{\prime}}-\delta\hat{Q}^{(\parallel)}\delta_{rr^{\prime}}.
\end{equation}
Such splitting is natural because these modes are orthogonal, and the transversal mode is traceless and does not contribute to $\delta{\cal L}_{M}^{(2)}$. On the Gaussian level, the theory then splits into two independent sectors that can be studied separately.
We proceed with this analysis in Sec.~\ref{Sec:IVC}, where we derive the density correlation functions at the Gaussian level.

\subsection{Density correlations}
\label{Sec:IVC}

Within our formalism, the density operator has a single replica index and two Keldysh components, ``classical'' (denoted without superscript) and ``quantum'' (with superscript $q$):
\begin{equation}
\delta\rho_{r}=-\frac{1}{4}\tr_{K}(\delta\hat{Q}_{{\cal G},rr}\hat{\tau}_{x}),\quad\delta\rho_{r}^{(\text{q})}=-\frac{1}{4}\tr_K(\delta\hat{Q}_{{\cal G},rr}).
\end{equation}
The correlation functions involving ``quantum'' component (i.e., retarded and advanced correlation functions) vanish, since the system is heated to the infinite temperature, and response functions $\propto 1/T$ at $T\gg J$. Thus, we focus on the Keldysh component of the density correlation function, which is a matrix in the replica space, with the following structure:
\begin{multline}
\label{eq:DensityCorrelator:ReplicaStructure}
C_{rr^{\prime}}(\textbf{x}) = \left\llangle \delta\rho_{r}(x, t)\delta\rho_{r^\prime}(x^\prime, t^\prime)\right\rrangle \\
=C_{0}(x-x^\prime, t-t^\prime)-C_{\text{repl}}(x-x^\prime, t, t^\prime)(1-\delta_{rr^{\prime}}).
\end{multline}

The ``replica-symmetric'' correlation function $C_0(\textbf{x})$ is determined by the evolution of the average density matrix, as typically described by the Lindblad equation (cf. a related problem of dissipative dynamics in number-conserving open systems \cite{Bernard2018,Bernard2022,nosov2023}), and depends only on the time difference in the steady-state regime:
\begin{equation}
C_0(\textbf{x}) = \overline{\left\langle \left\{ \hat{n}(\textbf{x}),\hat{n}(0)\right\} \right\rangle /2}-n^{2}.
\end{equation}

On the other hand, as was pointed out at the end of Sec.~\ref{sec:replica-trick}, when considering the off-diagonal density correlation function $C_{\text{repl}}(x-x^\prime, t, t^\prime)$, we will be interested in the case of equal times, 
$t = t^\prime = t_f$, as is relevant to the particle-number cumulant, Eq.~\eqref{eq:ParticleCumulants},
\begin{multline}
C_{\text{repl}}(x-x^{\prime},t=t^{\prime}=t_{f})
\equiv C(x-x^{\prime})\\
=\overline{\left\langle \left\{ \hat{n}(x,t_{f}),\hat{n}(x^{\prime},t_{f})\right\} \right\rangle/2 }-\overline{\left\langle \hat{n}(x,t_{f}\right\rangle \left\langle \hat{n}(x^{\prime},t_{f})\right\rangle }.
\label{eq:ReplicaOffdiagonalCorrelationFunction}
\end{multline}
This correlation function is of central interest in the present paper. As discussed above, for determining this correlation function, we will stop measurements at $t=t_f$ by setting $\gamma(t > t_f) = 0$. This will lead to an ``absorbing'' boundary condition at $t=t_f$ in the non-linear sigma-model formalism,  see Sec.~\ref{sec:NLSM}.

A key object that naturally arises when calculating the quadratic fluctuations is the diffuson ${\cal D}$, defined as a ladder series:
\begin{equation}
\label{eq:Diffuson}
{\cal D}^{-1}(\boldsymbol{q})\equiv{\cal B}^{-1}(\boldsymbol{q})-\tau_0^{-1}\approx Dq^{2}-i\omega.
\end{equation}
It is given by an average of the product of the retarded and advanced Green's functions,  ${\cal D}(\textbf{x}-\textbf{x}^{\prime})=\overline{G_{R}(\textbf{x},\textbf{x}^{\prime})G_{A}(\textbf{x}^{\prime},\textbf{x})}$ over measurement trajectories. For the same reason that rendered SCBA exact,  only ladder diagrams contribute to this average for arbitrary $\gamma / J$, since intersections of effective interaction lines are forbidden by causality. This implies the absence of corrections to the diffusion coefficient. Diffusive character of the associated Lindbladian dynamics, Eq.~\eqref{eq:Diffuson}, was obtained earlier in Ref. \cite{Esposito2005}.

Within the Gaussian approximation, Eqs.~(\ref{eq:Action:MatrixGaussian})-(\ref{eq:LM2}), the replica-symmetric density correlation function reads:
\begin{equation}
\label{eq:DensityCorrelator:GaussianResult:C0}
C_{0}(\boldsymbol{q}) =n(1-n)2 {\rm Re}{\cal D}(\boldsymbol{q}) \approx n(1-n)\frac{2Dq^{2}}{\omega^{2}+D^{2}q^{4}}.
\end{equation}

To determine the off-diagonal density correlation function, one should solve an integral equation which takes into account the presence of the boundary at $t = t_f$. The result for the equal-time density correlation function reads (see Appendix \ref{sec:appendix:BallisticCrossover} for details):
\begin{equation}
\label{eq:DensityCorrelator:GaussianResult:C}
C(q) \approx n(1-n)\times\begin{cases}
2ql_{0}, & ql_{0}\ll 1,
\\
1, & ql_{0}\gg 1.
\end{cases}
\end{equation}

The prefactor $n(1-n)$ ensures that correlations are completely absent for empty or filled bands, when no dynamics is happening.
The large-distance $x \gg l_0 \sim J/\gamma$ and long-time $t \gg \tau_0 \sim 1/\gamma$ behavior is dominated by the infrared behavior, yielding:
\begin{align}
\label{C0-diffusive}
C_{0}(x,t) &\approx n(1-n)\frac{\exp\left(-x^{2}/4D|t|\right)}{\sqrt{4\pi D|t|}},\\
\label{eq:DensityCorrelator:GaussianResult:Cr}
C(x) &\approx -\frac{2 n(1-n) l_0}{\pi x^2}
\end{align}
Equation (\ref{C0-diffusive}) describes the standard diffusive spreading of the averaged density. In contrast, Eq.~(\ref{eq:DensityCorrelator:GaussianResult:Cr}) makes manifest the nonlocal effect of measurements. 
Here and below, the time argument in the correlation functions refers to the difference of two times, $t=t^{\prime\prime}-t^\prime$,
in the long-time limit $t^\prime \to \infty$, when the measurements have already effectively ``thermalized'' the chain.

It is worth emphasizing that the diffusion coefficient and the replica-symmetric correlation function $C_0(x,t)$ obtained at the Gaussian level are in fact exact as a consequence of the structure of the effective interaction in the replica and time spaces. At the same time, loop corrections may arise (and do arise) for the off-diagonal density correlation functions. In fact, the quantum corrections to $C(x)$ are of crucial importance for our analysis, as discussed below.

\subsection{Fluctuations of number of particles}

The second cumulant of number of particles in a subsystem is directly related to the equal-time correlation function $C(x)$ via the following relation:
\begin{multline}
\label{eq:SecondCumulant}
{\cal C}_{l}^{(2)}=\int_{0}^{l}dx\int_{0}^{l}dy\, C(x-y)\\
=\frac{2}{\pi}\int_{0}^{\infty}\frac{dq}{q^{2}}C(q)(1-\cos ql).
\end{multline}
It follows from the structure of the Fourier representation \eqref{eq:DensityCorrelator:GaussianResult:C} that the correlation function $C(x)$ includes a delta-peak $n(1-n)\delta(x)$ and a negative tail $\propto 1/x^2$, which is described by Eq.~\eqref{eq:DensityCorrelator:GaussianResult:Cr}, see Appendix \ref{sec:appendix:BallisticCrossover}. The integral over this tail compensates for the contribution of the delta peak. Indeed, the integral $\int\! dx\, C(x)$ is exactly zero.

The delta-peak determines the behavior of the cumulant at distances $l \ll l_0$, yielding the ``volume law'' at such scales:
\begin{equation}
\label{eq:Cumulant:VolumeLaw}
{\cal C}_{l}^{(2)}\approx n(1-n)\,l,\quad l\ll l_{0}.
\end{equation}
On the other hand, at distances $l\gg l_0$, the contribution from this delta-peak is largely compensated by the integral over the ``tail'' of $C(x)$.
As a result, the linear growth of the cumulant crosses over to the logarithmic behavior originating from slow decay $\sim 1 / x^2$ of the tail:
\begin{equation}
\label{cumulant-log}
{\cal C}_{l}^{(2)}\approx \frac{4 n(1-n) l_0}{\pi}\ln\frac{l}{l_{0}},\quad l_{0}\ll l.
\end{equation}
We reiterate that this result holds only on the Gaussian level, i.e., in the leading order in $\gamma / J \ll 1$. Below, in Sec.~\ref{sec:RG}, we will demonstrate that the logarithmic growth saturates at an exponentially large length scale $l_\text{corr}$, satisfying $\ln(l_\text{corr} / l_0) \sim J/\gamma$. This gives rise to the area law in the thermodynamic limit.

\begin{figure}[ht]
    \centering
    \includegraphics[width=\columnwidth]{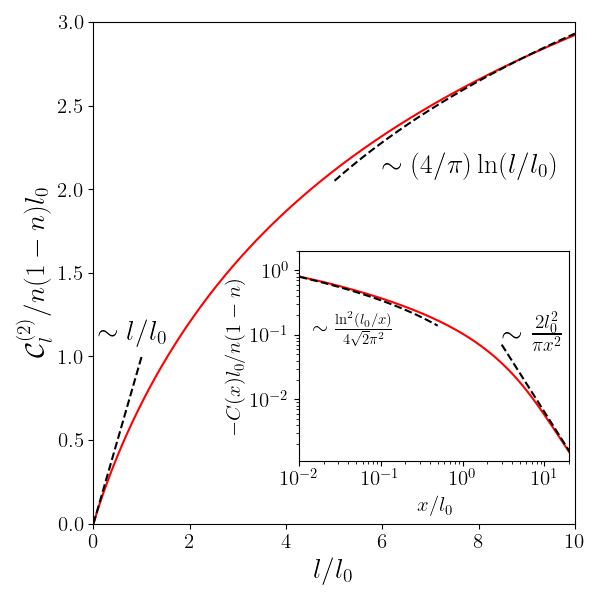}
    \caption{Second cumulant of the number of particles, Eq.~\eqref{eq:SecondCumulant}, in a subsystem of length $l$ in the Gaussian approximation. Inset: equal-time density correlation function, $C(x)$. The detailed calculation is performed in Appendix \ref{sec:appendix:BallisticCrossover}. Dashed curves are asymptotics for $l \ll l_0$ and $l \gg l_0$.}   
    \label{fig:DensityCorrelator:Analytics}
\end{figure}

The full analysis of the behavior of correlation function and cumulant in the Gaussian approximation, including the crossover between ballistic and diffusive regimes, is performed in Appendix \ref{sec:appendix:BallisticCrossover}. 
The results are summarized in Fig.~\ref{fig:DensityCorrelator:Analytics}.

\section{Non-linear sigma-model}
\label{sec:NLSM}

\subsection{Symmetries of the action and NLSM manifold}
\label{sec:symmetries}

To derive the effective field theory---the NLSM---it is instructive to inspect first symmetries of our problem with respect to rotations in replica and Keldysh spaces. The vector fields $\psi$ and $\bar\psi$ have $2R$ components, so that the group acting in this space is $U(2R)$, with $4R^2$ generators. 
The saddle point $\hat Q_{\rm SCBA} = \hat \Lambda$ is not rotated by $2R^2$ of these generators, which form a subgroup $U(R) \times U(R)$.  Thus, rotations of $\hat\Lambda$ yield a symmetric-space manifold
$U (2R)/U (R) \times U (R)$.

The fermionic action \eqref{eq:Action:Fermionic},
\eqref{eq:Action:Measurement:Exponential}
and its matrix counterpart \eqref{eq:Action:Matrix}, \eqref{eq:Action:Measurement:Matrix} have important symmetries which are responsible for soft modes studied on the Gaussian level in Sec. \ref{sec:Gaussian}. 
Out of $2R^2$ generators forming the above symmetric space, there is an exact symmetry of the full action with $R^2$ generators of rotations of the form $\hat{{\cal R}}_{\Phi}=e^{i\hat{\Phi}\hat{\tau}_{x}/2}$ since $\hat{{\cal R}}_{\Phi}$ commutes with $\hat\tau_x$ that enters the action for $R\ne 1$. Here
$\hat\Phi$ are $R\times R$ Hermitian matrices in replica space, so that matrices $\hat U = e^{i\hat{\Phi}}$ form a group $U(R)$. The remaining $R^2$ generators on the coset space are of the form
$\hat{{\cal R}}_{\Theta}=e^{i\hat{\Theta}\hat{\tau}_{y}/2}$, with $\hat{\Theta}$ being matrices in replica space. Out of these generators, there is a single---replica-symmetric---one, 
$\hat{{\cal R}}_{\theta}=e^{i\theta\hat{\tau}_{y}/2}$, which is an exact symmetry of the action for $R\to 1$ or, else, for any $R$ at $n=1/2$. As usual, symmetries of the action give rise to massless modes. 

The remaining $R^2-1$ replica-asymmetric generators of the form 
$\hat{{\cal R}}_{\Theta}=e^{i\hat{\Theta}\hat{\tau}_{y}/2}$ (those with traceless $\hat{\Theta}$) are not symmetries of the action and thus correspond
to massive modes. However, in the quadratic expansion they couple to $\hat \Phi$ modes. Hence, we take them into account and integrate them out in the Gaussian approximation, which yields a contribution to the action of massless ($\hat \Phi$) modes.

The $U(1)$ replica-symmetric (determinant) mode $\det \exp (i\hat\Phi)$ combines with $\hat{{\cal R}}_{\theta}$ into a replica-symmetric $U(2)/U(1)\times U(1)$ manifold that has a geometry of the sphere $ \text{S}^2$. We will denote the matrix field belonging to this manifold by $\hat Q_0$. The full manifold of matrices corresponding to the symmetry of the action is then obtained by rotating $\hat Q_0$ by matrices $\exp (i\hat\Phi) / \det \exp (i\hat\Phi)$ that form the group $SU(R)$. 

To derive the NLSM taking into account the above symmetries, it is convenient to proceed as follows. We will first consider the $U(2R)$-invariant part of the action $S^{(\text{inv})}=S_{0}+\gamma S_{M}^{(\text{inv})}$, which will produce the NLSM defined on the symmetric space $U(2R) / U(R) \times U(R)$. 
The difference $\gamma (S_{M}-S_{M}^{(\text{inv})})$ which vanishes at $R=1$ will then be restricted to this manifold and will provide an additional structure on it. As a result, we will obtain an $SU(R)$ NLSM for the replicon modes $\hat\Phi$ and an $ \text{S}^2$ theory for the replica-symmetric sector. The former will describe the correlation function $C(x)$ and the latter the correlation function $C_0(x,t)$. 

The presence of the ``boundary'' at time $t=t_f$  is taken into account by putting $\gamma(t > t_f) = 0$, 
see a discussion at the end of Sec.~\ref{sec:replica-trick} and in Sec.~\ref{Sec:IVC}. It is relevant for the effective field theory describing the massless $\hat{\Phi}$ modes. The condition $\gamma =0$ corresponds to $D \to \infty$ at $t>t_f$, thus leading to 
a boundary condition $\hat{\Phi}(t = t_f) = 0$. This is in full analogy with boundary conditions arising in a theory of diffusive disordered systems on boundary with an ideal metal ($D\to \infty$). This type of boundary condition was also discussed in a related context of random unitary circuits in Ref.~\cite{Jian2022}. 

Since we are interested in the replica limit $R \to 1$, we will set 
$R \to 1$ in numerical factors that arise in the derivation. At the same time, we will keep $R$ arbitrary in the dimensionality of corresponding symmetry groups.

\subsection{Field theory restricted to \texorpdfstring{$U(2R)/ U(R) \times U(R)$}{U(2R) / U(R) x U(R)} manifold}

The saddle point analysis of the matrix action was already performed in Sec. \ref{sec:Gaussian}, where it was identified that the solutions can be parametrized by a single $\hat{Q}$-matrix satisfying standard non-linear constraint $\hat{Q}^{2}=1$ according to Eq.~\eqref{eq:SCBA:SigmaG}.
As explained in Sec.~\ref{sec:symmetries}, we consider the manifold spanned by arbitrary $U(2R) / U(R) \times U(R)$ rotations of the saddle point $\Lambda$ as $\hat{Q}=\hat{{\cal R}}\hat{\Lambda}\hat{{\cal R}}^{-1}$, restricting ourselves to a smooth time and spatial dependence $\hat{{\cal R}}(\textbf{x})$.
Performing a gradient expansion of the $\Tr \ln$ term in the action \eqref{eq:Action:Matrix}
in a standard way (see, e.g., Ref. \cite{KamenevLevchenko}), we arrive at the following NLSM action:
\begin{align}
\label{eq:Action:NLSM}
i{\cal L}_{0}[\hat{Q}] &=\Tr\left(\frac{1}{2}\hat{\Lambda}\hat{{\cal R}}^{-1}\partial_{t}\hat{{\cal R}}-\frac{D}{8}(\partial_{x}\hat{Q})^{2}\right),\\
\label{eq:Action:Measurement:NLSM}
i{\cal L}_{M}[\hat{Q}] &= \det\left(\frac{1-\hat{Q}\hat{\tau}_{x}}{2}\right)+\det\left(\frac{1+\hat{Q}\hat{\tau}_{x}}{2}\right)-1,
\end{align}
where the trace is now taken over the Keldysh and replica spaces. The dynamic term [the first term in Eq.~\ref{eq:Action:NLSM}] has the form of Wess-Zumino term and cannot be written in terms of $\hat{Q}$-matrix itself. Equivalently, it has the meaning of the Berry phase of the $\hat{Q}(t)$ trajectory on the $U(2R) / U(R) \times U(R)$ manifold.
We also note a certain similarity of the measurement-induced part of the action, Eq.~(\ref{eq:Action:Measurement:NLSM}), to the disorder-induced action in the NLSM derived in Ref.~\cite{Ostrovsky2014} for a chiral metal with vacancies. In that work, Poissonian averaging over infinitely strong point-like scatterers (cf. averaging over local projective measurements in the present paper) also resulted in the appearance of determinants involving the $\hat{Q}$-matrix in the action. 

Note that, at variance with the case of the NLSM for quenched disorder, the measurements are inelastic and the system is heated to the infinite temperature. For this reason, the diffusion coefficient $D$ is expressed in terms of the root-mean-square velocity averaged over the whole Brillouin zone, in agreement with Eq. \eqref{eq:MeanFreeParameters}.

\subsection{Replica-symmetric sector}
\label{sec:replica-sym}

To explore the replica-symmetric sector, we can directly set $R=1$.
The term \eqref{eq:Action:Measurement:NLSM} then vanishes, leaving us with the $U(2)/U(1)\times U(1)$ NLSM with the action \eqref{eq:Action:NLSM}. 
This NLSM completely reproduces results for the diffuson ${\cal D}(\boldsymbol{q})$ and replica-symmetric density correlation function $C_{0}(\boldsymbol{q})$ obtained earlier in Sec. \ref{sec:Gaussian}. It is worth emphasizing  that the replica-symmetric sector does not contain any renormalization of diffuson degrees of freedom. 
All diagrams that come from the non-linear interaction between diffusons in arbitrary parametrization of the NLSM manifold vanish completely because of the retarded structure of the diffusons and instantaneous-in-time interaction vertices. 
Nevertheless, the theory is not Gaussian: non-linear vertices still can have non-trivial contribution to higher correlation functions of diffusive modes.

An interesting observation can be made by noting that for $R=1$, the sigma-model manifold is just a two-dimensional sphere $\text{S}^{2}$. One can then consider a parametrization of the manifold by conventional polar and azimuthal angles, $\theta$ and $\phi$, as follows:
\begin{equation}
\hat{Q}=\hat{F}e^{i\phi\hat{\tau}_{z}/2}e^{i\theta\hat{\tau}_{y}/2}\hat{\tau}_{z}e^{-i\theta\hat{\tau}_{y}/2}e^{-i\phi\hat{\tau}_{z}/2}\hat{F},
\end{equation}
with matrix 
\[
\hat{F}=\begin{pmatrix}1 & 1-2n\\
0 & -1
\end{pmatrix}
\]
The sigma-model action \eqref{eq:Action:NLSM} then reduces to:
\begin{equation}
i{\cal L}_{0}[\theta,\phi]=i s\,(1-\cos\theta)\partial_{t}\phi-D s^2\left[(\partial_{x}\theta)^{2}+\sin^{2}\theta(\partial_{x}\phi)^{2}\right].
\end{equation}
This action is formally equivalent to the imaginary-time action of the quantum spin-${s=1/2}$ chain with isotropic Heisenberg interaction. 
The spin components have the form
$s_x = \sin \theta \cos \phi / 2$, $s_y = \sin \theta \sin \phi / 2$, $s_z = \cos \theta / 2$.
The steady state of such a chain would then correspond to the ferromagnetic ground state $\left|\psi\right>=\left|\uparrow\uparrow\dots\uparrow\right>$. On the level of $\hat{Q}$-matrix, it corresponds to the north pole $\phi=0$, $\theta=0$, which is exactly the SCBA saddle point $\hat{Q} = \hat\Lambda$. 

The classical Keldysh component of fermionic density,
\begin{equation}
\rho_0= \frac{1}{4} \text{tr}_K (1- \hat Q_0 \hat\tau_x), 
\label{rho0Q0}
\end{equation}
can be expressed 
in this parametrization  in terms of spin variables as a projection of the spin onto a complex vector $\boldsymbol{h}$ of unit ``length'' $h_x^2 + h_y^2 + h_z^2 = 1$:
\begin{equation}
\rho = \frac{1}{2}-\boldsymbol{h}\hat{\boldsymbol{s}},\quad\boldsymbol{h}=\begin{pmatrix}1-(1-2n)^{2} / 2\\
i (1-2n)^{2}/2\\
1-2n
\end{pmatrix}.
\end{equation}
 Within the spin language, correlation functions of operators $\hat{\rho}$ are calculated on top of the ferromagnetic ground state. In the long-time and long-distance limit, the main contribution to such correlation functions will come from the lowest-energy excitations in the Heisenberg ferromagnet, which are known to be magnons with quadratic dispersion $\omega \sim D q^2$. This exactly reproduces the diffusive pole calculated earlier.

\subsection{Replicon modes: \textit{SU(R)} NLSM}
\label{Sec:VD}

In Sec.~\ref{sec:replica-sym}, we have discussed properties of the field theory on the replica-symmetric sub-manifold $\hat{Q}=\hat{Q}_{0}\otimes\hat{\mathbb{I}}_{R}$, with $\hat{Q}_0$ being a $2\times 2$ matrix in the Keldysh space. The whole $U(2R) / U(R) \times U(R)$ manifold is then obtained from the replica-symmetric configuration by arbitrary rotations
\begin{equation}
\label{eq:QMatrix:Replicon}
\hat{Q}=\hat{{\cal R}}_{\Phi}\hat{{\cal R}}_{\Theta}\hat{Q}_{0}\hat{{\cal R}}_{\Theta}^{-1}\hat{{\cal R}}_{\Phi}^{-1},
\quad
\hat{{\cal R}}_{\Phi}=e^{i\hat{\Phi}\hat{\tau}_{x}/2},~\hat{{\cal R}}_{\Theta}=e^{i\hat{\Theta}\hat{\tau}_{y}/2},
\end{equation}
with the replicon modes $\hat{\Theta}$ and $\hat{\Phi}$ being $R \times R$ traceless matrices in the replica space.  
As discussed in Sec.~\ref{sec:symmetries}, 
this parametrization is chosen in such a way that $\Phi$-rotations are an exact symmetry of measurement action \eqref{eq:Action:Measurement:NLSM} for arbitrary $R$, as they commute with $\hat{\tau}_x$. The modes $\Theta$, on the other hand, are massive and will be integrated out in the Gaussian approximation, contributing to the effective action for the massless mode $\Phi$. 

To calculate the replica-off-diagonal density correlation function, we introduce a generating functional for the fermion density, with different sources for different replicas incorporated into a replica-diagonal matrix $\hat{\xi} = \diag[\{\xi_r\}_{r=1}^R]$:
\begin{equation}
Z[\xi]=\left\langle \exp\left[i\int d^2 \boldsymbol{x}\sum_{r=1}^{R}\xi_{r}(\textbf{x})\rho_{r}(\textbf{x})\right]\right\rangle.
\end{equation}
This translates to an additional Lagrangian term:
\begin{equation}
\label{eq:Action:DensitySource}
i{\cal L}_{\text{source}}[\hat{Q},\xi]=\frac{i}{4}\Tr\left[\hat{\xi}\left(1-\hat{Q}\hat{\tau}_{x}\right)\right].
\end{equation}

Expansion of the effective action in $\Theta$ modes and subsequent Gaussian integration is performed in Appendix \ref{sec:appendix:RepliconAction}, bringing us to the following $SU(R)$ effective action for $\hat{U}=\exp\left(i\hat{\Phi}\right)$:
\begin{equation}
i{\cal L}_{\Phi}=-\frac{g[\hat{Q}_0]}{2}\tr\left[\frac{1}{v_{0}}(\partial_{t}^{\Xi}\hat{U})^{\dagger}\partial_{t}^{\Xi}\hat{U}+v_{0}\partial_{x}\hat{U}^{\dagger}\partial_{x}\hat{U}\right].
\label{LPhi}
\end{equation}
with $\tr \equiv \tr_R$ taken in the replica space only.
Here, $v_{0}$ is the root-mean-square velocity  defined in Eq.~\eqref{eq:MeanFreeParameters}, $\hat{\Xi} = \hat{\xi} - \xi_0$ is the replicon density source where $\xi_0 = \tr\hat{\xi}$, and the ``covariant derivative'' is defined as
\begin{equation}
\partial^\Xi_{t}\hat{U}=\partial_{t}\hat{U}+\frac{i}{2}\{\hat{U},\hat{\Xi}\}.
\end{equation}

The coupling constant in the NLSM action (\ref{LPhi})  connects replica-symmetric and replicon modes via its dependence on the replica-symmetric density (\ref{rho0Q0}):
\begin{equation}
g[\hat{Q}_{0}]=2l_{0}\rho_{0}(1-\rho_{0}).
\end{equation}
On the Gaussian level, 
\begin{equation}
g[\hat{Q}_{0}]\approx g_{0}=2l_{0}n(1-n),
\end{equation}
and the replicon sector decouples from the replica-symmetric modes. 

The resulting action contains a second derivative with respect to time, and, hence, it has to be supplied with the boundary conditions at $t = t_f$. As discussed above, the boundary is implemented by setting $\gamma(t > t_f) = 0$, which results in the boundary condition $\hat{\Phi}(t > t_f) = 0$ or, equivalently, $\hat{U}(t > t_f) = \hat{\mathbb{I}}$.

The $SU(R)$ symmetry of the action dictates the following form of the Green's function for the generators $\hat{\Phi}$:
\begin{multline}
\left\langle \Phi_{r_{1}r_{2}}(x,t)\Phi_{r_{1}^{\prime}r_{2}^{\prime}}(x^{\prime},t^{\prime})\right\rangle =\left[\delta_{r_{1}r_{2}^{\prime}}\delta_{r_{2}r_{1}^{\prime}}-\frac{1}{R}\delta_{r_{1}r_{2}}\delta_{r_{1}^{\prime}r_{2}^{\prime}}\right]\\
\times G_{\Phi}(x-x^{\prime},t,t^{\prime}).
\end{multline}
Within the Gaussian approximation, the Green's function reads:
\begin{multline}
G_{\Phi}(x,t,t^{\prime})=\int_{0}^{\infty}\frac{2d\omega}{\pi}\sin\omega(t-t_{f})\sin\omega(t^{\prime}-t_{f})\\
\times\int(dq)e^{iqx}\frac{v_{0}/g_{0}}{\omega^{2}+v_{0}^{2}q^{2}}    
\end{multline}
Finally, the off-diagonal density correlation function can be obtained by differentiating the generating functional with respect to sources $\hat{\Xi}$, yielding
\begin{multline}
C(x-x^\prime)=\lim_{t, t^\prime\to t_f}
\Big[\frac{\left\langle g[Q_{0}]\right\rangle}{v_{0}}\delta(\boldsymbol{r}-\boldsymbol{r}^\prime)\\
-\frac{1}{v_{0}^{2}}\left\langle g^{2}[\hat{Q}_{0}]\partial_{t}\hat{\Phi}(x,t)\partial_{t}\hat{\Phi}(x^{\prime},t^{\prime})\right\rangle \Big],
\end{multline}
which, within the Gaussian approximation, reproduces the results obtained in Sec.~\ref{sec:Gaussian}, see Eq.~\eqref{eq:DensityCorrelator:GaussianResult:C}:
\begin{equation}
C(q) \approx g_0 |q|.
\end{equation}

\subsection{Renormalization-group analysis}
\label{sec:RG}

The dependence of the coupling constant $g$ on $\hat{Q}_0$ provides interaction between replica-symmetric and replicon modes. 
This interaction does not renormalize the replica-symmetric correlation function $C_0$ since the dimensionality of the replicon subspace is $\dim SU(R)=R^{2}-1\to0$. 
At the same time, it can yield corrections to the effective action for $\Phi$-fields of the replicon sector. These corrections are however infrared-finite (and small for small $\gamma/J$), as can be seen from the structure of replica-diagonal correlation function $C_0(\boldsymbol{q})$, Eq. \eqref{eq:DensityCorrelator:GaussianResult:C0}.
We thus focus on the renormalization of the $SU(R)$ action itself.

It is known that the perturbative expansion for the $U(R)$ sigma-model in two dimensions exhibits logarithmic divergencies that can be resummed within the RG framework \cite{HikamiRG,WegnerRG}. In addition, we also should include a running coupling constant $Z_s$ describing renormalization of the source
 source terms in \eqref{LPhi} with $\Xi$ restricted to the boundary $t = t_f$:
\begin{equation}
\partial_{t}^{\Xi}\hat{U}\simeq\partial_{t}\hat{U}+\frac{iZ_{s}}{2}\left\{ \hat{U},\hat{\Xi}\right\} 
\end{equation}
The corresponding one-loop RG equations are derived in Appendix \ref{sec:appendix:RG}, and the result reads:
\begin{align}
\label{eq:RG}
\frac{\partial g}{\partial\ln \ell}&=-\frac{R}{4\pi} + O(1/g),\\[0.1cm]
\frac{\partial \ln Z_s}{\partial\ln \ell}&=0 + O(1/g^2).
\label{eq:RG-Z}
\end{align}
Here $\ell$ is the RG length scale and $g(\ell)$ and $Z(\ell)$ are the associated running couplings.  
Finally, we can take the limit $R \to 1$ in Eq.~\eqref{eq:RG}, which yields
\begin{equation}
\label{eq:RG1}
\frac{\partial g}{\partial\ln \ell}=-\frac{1}{4\pi} + O(1/g).
\end{equation}
The coupling constant $g$ slowly decreases with increasing length scale $\ell$, which implies an increase of the magnitude of quantum fluctuations, and the theory reaches the strong-coupling regime $g \lesssim 1$ at a finite length scale $\ell\sim l_{\text{corr}}$, where
\begin{equation}
l_{\text{corr}} \sim l_{0}\exp\left(4\pi g_{0}\right)
\label{l-corr}
\end{equation}
is the correlation length.

At this stage, an analogy with the problem of Anderson localization is useful. It is known that field theories of Anderson localization are NLSMs \cite{evers08}. In particular, the $U(R)$ NLSM describes systems with quenched disorder which belong to the chiral unitary symmetry class AIII. The crucial difference between the measurement problem and Anderson localization is that in the latter case the relevant replica limit is $R\to 0$. In this limit, the perturbative beta-function for the $U(R)$ NLSM vanishes in all loops \cite{gade1991the}. 
On the other hand, in the replica limit $R\to 1$ relevant to the measurement problem, the one-loop RG flow given by Eq.~(\ref{eq:RG1}) is non-trivial. Interestingly, it is
analogous to the $R\to 0$ flow for the NLSM describing Anderson localization in two-dimensional systems with quenched disorder in the more conventional orthogonal symmetry class (class AI). There, the coupling constant $g$ has a meaning of conductance, and the flow corresponds to the well-known weak-localization phenomenon, leading to a negative quantum correction to the conductance. Thus, the length $l_{\text{corr}}$ identified above for the measurement problem is analogous to the localization length in the problem of two-dimensional Anderson localization.
At this scale, the dimensionless conductance becomes smaller than unity and the weak localization crosses over to strong Anderson localization. 

After this short detour, we return to the measurement problem.
On length scales smaller than $l_\text{corr}$, the Gaussian theory can still be applied to calculate $C(\boldsymbol{q})$ at a momentum $q$ but one should replace the bare coupling constant $g_0$ with the renormalized one $g(q)$, and introduce factor $Z^2(q)$ taking into account the source renormalization. To obtain $g(q)$ and $Z(q)$, one should integrate the RG flow equation
\eqref{eq:RG1} from the ultraviolet cutoff given by the mean free path $l_0$ up to the infrared length scale determined by external momentum, $\ell \sim q^{-1}$:
\begin{equation}
\label{eq:RunningG}
g(q)\approx g_{0}-\frac{1}{4\pi}\ln\frac{1}{ql_{0}},\quad Z(q) \approx 1
\end{equation}
The density correlation function $C(q)$ takes then the form
\begin{equation}
\label{eq:DensityCorrelator:Corrections1}
C(q)=Z^2(q)g(q)|q|.
\end{equation}

The perturbative one-loop RG result \eqref{eq:RunningG} for $q \gg l_{\text{corr}}^{-1}$ gives rise to a correction to the Gaussian-approximation result \eqref{cumulant-log} for the particle-number cumulant \eqref{eq:SecondCumulant} in a subsystem of length $l$ satisfying $l_0 < l < l_{\rm corr}$:
\begin{equation}
{\cal C}_{l}^{(2)}=\frac{1}{\pi}\int_{0}^{l_{0}^{-1}}\frac{dq}{|q|}g(q)(1-\cos ql)\approx\frac{2 g_{0}}{\pi}\ln\frac{l}{l_{0}}-\frac{1}{4\pi}\ln^{2}\frac{l}{l_{0}}.
\label{cumulant-with-corr}
\end{equation}
The one-loop correction thus leads to a reduction of the cumulant. 

Let us discuss now the behavior of the cumulant at largest scales, $l > l_{\rm corr}$. In this connection, it is instructive to recall general relations between the behavior of the correlator $C(q \to 0)$ at $t = 0$ and that of the cumulant, which follow from Eq. \eqref{eq:SecondCumulant}.
Specifically, the volume-law, logarithmic, and area-law scaling of the cumulant with $l$ are associated with the following types of the limiting behavior of $C(q \to 0)$ at $t = 0$:
\begin{itemize}
    \item volume law: $C(q)\to\const$;
    \item logarithmic law: $C(q) / |q| \to \const$;
    \item area law: $C(q) / |q| \to 0$.
\end{itemize}
In analogy with the two-dimensional Anderson localization (and, more generally, with conventional statistical-mechanics models), we expect that at $l \gg l_{\text{corr}}$ the system is ``strongly localized'' and exhibits an exponential decay of correlations. This implies that $C(q) / |q| = Z^2(q) g(q) \to 0$ as $q \to 0$, indicating the area law. Furthermore, the power-law decay of the density correlation function $\sim 1/x^2$  is superseded at $ x > l_{\rm corr}$ by the exponential decay $\sim \exp(-x / l_\text{corr})$.  Thus, the logarithmic growth \eqref{cumulant-log} of the particle-number cumulant obtained within the Gaussian approximation in Sec. \ref{sec:Gaussian} eventually saturates at the scale $l \sim l_{\text{corr}}$ giving rise to the area law behavior, with the saturation value estimated as
\begin{equation}
{\cal C}_{l}^{(2)}\sim g_{0}^{2},\quad l\gg l_{\text{corr}}.
\end{equation}

This behavior should be contrasted to  results of Ref. \cite{Fava2023}, where $SO(R)$ NLSM was derived and studied for the problem of Majorana fermion quantum random circuit. The sign of the one-loop RG term obtained in Ref. \cite{Fava2023} is opposite to 
that in our formula \eqref{eq:RG1}.
Before the replica limit $R \to 1$ is taken, the coefficient in Ref. \cite{Fava2023} is $R-2$, which should be compared to $R$ in our Eq.~\eqref{eq:RG}. The flow of $g(q)$ in Ref. \cite{Fava2023} thus is of the weak-antilocalization type, at variance with the localizing behavior manifest in our Eq. \eqref{eq:RunningG}. As a result, 
sign of the $\ln^2$ term in  Eq.~\eqref{cumulant-with-corr} becomes positive in the Majorana model of Ref. \cite{Fava2023}, and the cumulant  scales as $\ln^2 l$ in the large-$l$ limit.  

\section{Entanglement entropy}
\label{sec:Entropy}

Our focus so far has been on the scaling of the second cumulant of number of particles ${\cal C}_l^{(2)}$ with the subsystem size $l$. We analyze now its relation to the entanglement entropy $\mathcal{S}_{E}(l)$. As was discussed in Sec. \ref{sec:Model}, for our measurement protocol, the entanglement entropy can be expressed as a  series over even cumulants, Eq. \eqref{eq:KlichLevitov}. 
We thus have to estimate the behavior of higher cumulants.

At ``ballistic'' length scales $l \lesssim l_0$, the system is essentially indistinguishable from the Fermi gas heated to the infinite temperature. It is then natural to expect the behavior to follow the corresponding ``volume law'' for such small systems:
\begin{equation}
\label{eq:Entropy:Ballistic}
\mathcal{S}_{E}(l)\approx-\left[n\ln n+(1-n)\ln(1-n)\right]l,\quad l\ll l_{0}.
\end{equation}
Although the second cumulant $C_l^{(2)}$ in this region is also shown to follow the ``volume law'', see Eq. \eqref{eq:Cumulant:VolumeLaw}, the prefactor differs (although both prefactors vanish for $n=0$ and $n=1$). This means that at ballistic scales, all cumulants are expected to be parametrically of the same order.

For the ``diffusive'' region $l_0 \ll l \ll l_\text{corr}$, the situation changes. In this regime, the $SU(R)$ NLSM description from Sec. \ref{sec:NLSM} holds. In the Gaussian approximation, higher cumulants are zero. To find them, one should take into account the non-linearity on the NLSM manifold, which  yields an additional smallness in parameter $1 / g$. Therefore, we argue that in the diffusive region, the series for the entropy is dominated by the second cumulant, and the entanglement entropy is given by
\begin{equation}
\mathcal{S}_{E}(l)\approx\frac{\pi^{2}}{3}{\cal C}_{l}^{(2)}\approx\frac{4\pi}{3}n(1-n)l_{0}\ln\frac{l}{l_{0}},\quad l_{0}\ll l\ll l_{\text{corr}}.
\label{eq:Entanglement:log}
\end{equation}

Finally, as the system approaches correlation length $l_\text{corr}$, the role of the quantum fluctuations becomes more and more prominent. At $l \sim l_\text{corr}$, the conductance $g(l)$ is of order unity and higher cumulants are of the same order as the second cumulant. 
We thus have (up to unknown numerical coefficients)
\begin{equation}
\mathcal{S}_{E}(l)\sim{\cal C}_{l}^{(2)}\sim g_{0}^{2},\quad l \gtrsim l_{\text{corr}},
\end{equation}
which is the area-law behavior of the entanglement entropy. In the following Sec.~\ref{sec:Numerics}, we demonstrate numerically that the coefficient 
relating  $\mathcal{S}_{E}(l)$ and  ${\cal C}_{l}^{(2)}$ 
remains very close to $\pi^2 / 3$ 
(as in Eq.~\ref{eq:Entanglement:log})
even in the ``strong coupling regime''. Similar dominance of the second cumulant in the relation between the entanglement entropy and full counting statistics, Eq.~\eqref{eq:KlichLevitov}, is also known to hold for disordered systems in the vicinity of the Anderson metal-insulator transition, where the conductance (analogous to the coupling constant $g$ in our case) is of the order of unity \cite{Burmistrov2017}.

\section{Numerical analysis}
\label{sec:Numerics}

To verify our analytical predictions, we have performed numerical simulations for system sizes up to $L = 2000$. As the system is non-interacting and Gaussian, we are able to describe it in terms of the single-particle correlation matrix $G_{xy}=\left\langle \hat{\psi}^{\dagger}(x)\hat{\psi}(y)\right\rangle $, which fully characterizes the unitary time evolution together with jumps induced by random projective measurements. After performing a sufficient number of measurements for the system to reach the steady state, we have extracted the pair density correlation function 
\begin{equation}
C_{xy}\equiv G_{xy}\delta_{xy}-G_{xy}G_{yx}
\end{equation}
and averaged it over different runs of the simulation,
\begin{equation}
C(x-y) = \overline{C_{xy}}.
\end{equation}

\begin{figure}[ht!]
    \centering
    \includegraphics[width=\columnwidth]{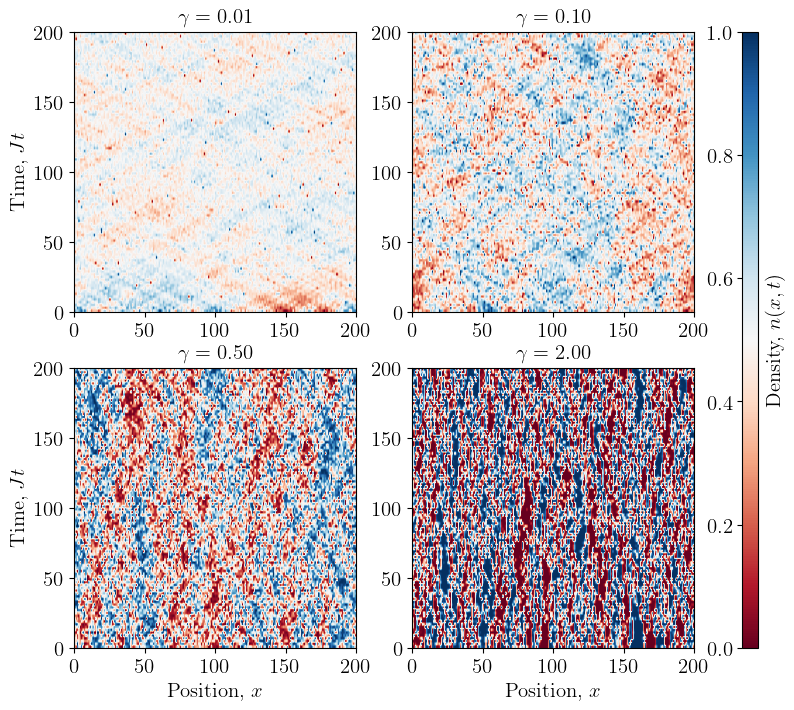}
    \caption{Typical time evolution of the density profile for system size $L = 200$ at half-filling $n=1/2$ for various values of measurement rate $\gamma$ with $J$ set to unity.}
    \label{fig:evolution}
\end{figure}

To illustrate the measurement-induced dynamics, we show in Fig. \ref{fig:evolution} the representative time evolution of the density profile for $L=200$ and several values of $\gamma$. For smaller values of $\gamma$, excitations created by rare measurements quickly relax and, as a consequence, the particle density fluctuates only weakly around its average value $n = 1/2$. The front of perturbation created by measurement moves with maximal group velocity $v_\text{max} = 2 J$, which can be seen as a pattern of tilted lines for $\gamma = 0.01$ and 0.1.
Note that the velocity $v_\text{max}$ is different from the root-mean-square velocity $v_{0} = \sqrt{2} J$ which defines the dynamic of $SU(R)$ NLSM fields at times $t \gg \gamma^{-1}$, see Eq.~\eqref{LPhi}. 

For larger values of the measurement rate, $\gamma= 0.5$ and 2,  the pattern changes dramatically. Specifically, we observe the quantum Zeno effect, which tends to pin the density on each site to values $n = 0, 1$, while the unitary evolution allows for rare ``jumps'' of pinned electrons between neighboring sites.

%%%%%%%%%%%%%%%%%%
%%%%%%%%%%%%%%%%%%

\begin{figure}[ht!]
    \centering
    \includegraphics[width=\columnwidth]{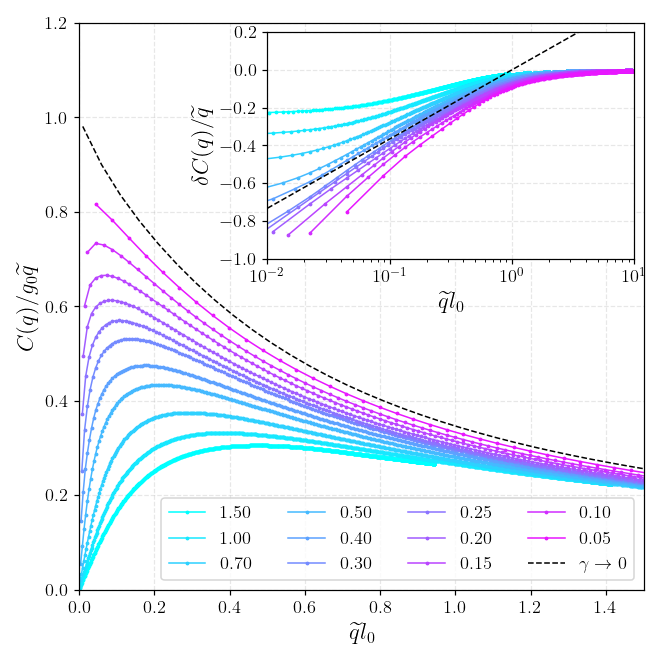}
    \caption{Trajectory-averaged equal-time density correlation function in the momentum space, $C(q)$, for several values of the measurement rate $\gamma / J$, as obtained by numerical simulations. The curves show
    $C(q) / g_0 \widetilde{q}$ as a function of $\widetilde{q} l_0$ (with $\widetilde{q} \equiv 2 \sin(q/2)$ taking into account a finite lattice spacing).
     Dashed line: limiting expression for $\gamma /J  \ll 1$ (Gaussian approximation) as given by Eq.~\eqref{eq:DensityCorrelator:BallisticCrossover}. The turndown of all curves at small $\widetilde{q} l_0$ is a manifestation of the area-law behavior in the thermodynamic limit. Inset: ``weak-localization correction'' defined as the difference between the corresponding curve and the dashed line in the main plot (without $g_0$ rescaling), in a semi-logarithmic plot. The dashed line corresponds to a logarithmic correction as predicted by Eq.~\eqref{eq:RunningG} but with a slope $-1/2\pi$ (i.e., two times larger). For larger $\gamma$, the ``localization'' becomes strong at the smallest momenta, so that a saturation of the correction is observed.
    Results were obtained for system size $L = 2000$ at half-filling $n=1/2$, averaged over $\sim 50$ measurement trajectories.}
    \label{fig:DensityNumerics}
\end{figure}

 In Fig.~\ref{fig:DensityNumerics},
 we present the numerical results for the equal-time density correlation function $C(q)$ obtained by averaging over $\sim 50$ quantum trajectories for various values of the dimensionless measurement rate $\gamma / J$. In the main panel, the ratio $C(q)/g_0 \widetilde{q}$ is displayed as a function of $\widetilde{q} l_0$, with $\widetilde{q} \equiv 2 \sin(q/2)$ being equal to $q$ in the long-wavelength limit and correctly taking into account a finite lattice spacing at large $q$.
In the Gaussian approximation,
all curves in this representation should collapse on a single curve, see Eq. \eqref{eq:DensityCorrelator:BallisticCrossover}, which is presented by a dashed line. The condition for this collapse is $q l_{\text{corr}} \gg 1$. Indeed, a very good collapse is observed for sufficiently large $q$. With decreasing $\gamma$ (and thus increasing $l_{\text{corr}}$) the numerical data follow the dashed line down to lower and lower values of $q$, as predicted. 

%%%%%%%%%%%%%%%%%%%
%%%%%%%%%%%%%%%
\begin{figure}
 \centering
 \includegraphics[width=\columnwidth]{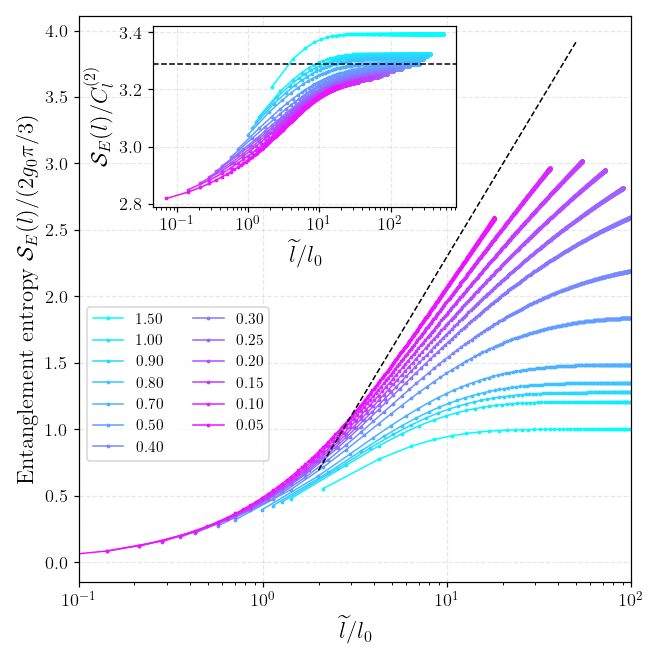}
 \caption{Trajectory-averaged entanglement entropy ${\cal S}_E(l)$ as a function of rescaled subsystem size $\widetilde{l} \equiv (L/\pi) \sin(\pi l / L)$ (to take into account finite-size effects) for different values of measurement rate $\gamma / J$.  Dashed line: logarithmic asymptotic, Eq. \eqref{eq:Entanglement:log}. Inset: ratio $\mathcal{S}_{E}(l)/ {\cal C}_{l}^{(2)}$ of the entanglement entropy and the second cumulant of number of particles;  dashed line: $\pi^2 / 3$ as given by the first term in the Eq. \eqref{eq:KlichLevitov}. In the ballistic regime, $l \ll l_0$, the ratio is in full agreement with the expected saturation at $-\left[n\ln n+(1-n)\ln(1-n)\right]/n(1-n) = 4 \ln 2 \approx 2.77$ (at half-filling), see Eq.~\eqref{eq:Entropy:Ballistic}}. 
 \label{fig:entropy}
 \end{figure}

 We recall that $C(q) / |q| \to \const$ at $q \to 0$, as found in Gaussian approximation and shown by dashed line, is
 responsible for the logarithmic behavior of the fluctuations of the number of particles.
 This behavior is, however, violated at the  smallest momenta $q$ and all the curves turn down, in consistency with our analytical prediction that, as a result of the $g(q)$ renormalization, $C(q) / |q| \to 0$ at $q \to 0$, implying the area law. For larger $\gamma$, the vanishing of  $C(q) / |q| \to 0$ is almost reached for our lowest $q$ since the correlation length $l_\text{corr}$ is smaller than the system size $L=2000$. At the same time, for smaller $\gamma$, the exponentially large correlation length \eqref{l-corr} strongly exceeds $L$, so that the ``strong localization'' cannot be observed. We can capture however its precursor---the perturbative ``weak localization'' correction, Eq. \eqref{eq:DensityCorrelator:Corrections1}. For the smallest $\gamma$, even the turndown is barely visible, as the mean free path becomes of the order of the system size.

In the inset of Fig.~\ref{fig:DensityNumerics}, we display the weak-localization correction. Specifically, we show the ratio $\delta C(q) / \widetilde{q}$, where $\delta C(q)$ is the difference between $C(q)$ and its Gaussian approximation. It is seen that $\delta C(q) / \widetilde{q}$ is proportional to $\ln \widetilde{q}$ and its (negative) slope is independent of $\gamma$, in agreement with our analytical prediction \eqref{eq:RunningG}.

 Finally, we have also calculated numerically the entanglement entropy and its dependence on the subsystem size for systems of size $L = 800$. Figure~\ref{fig:entropy} demonstrates the logarithmic behavior of the entanglement entropy for smaller values of $\gamma$ alongside the tendency towards saturation upon increasing $\gamma$, as predicted by the RG analysis.
 
 Our analytical predictions for the entanglement entropy were based on keeping only the first term in the Klich-Levitov formula, Eq.~\eqref{eq:KlichLevitov}. For rare measurements, this is parametrically justified by $g\gg 1$, as higher cumulants are suppressed by powers of $1/g$. It turns out that, even when the renormalized $g$ becomes of order unity (so that all cumulants might become important), the entanglement entropy is still dominated by the second cumulant. To support this statement, we plot in the inset of Fig.~\ref{fig:entropy}  the ratio $\mathcal{S}_{E}(l)/ {\cal C}_{l}^{(2)}$ of the entanglement entropy and the second particle-number cumulant as a function of the subsystem size. It is seen that this ratio saturates at a constant value at large $l$, thus demonstrating that the second cumulant and the entanglement entropy exhibit the same behavior. Furthermore,  for $\gamma / J \ll 1$, the saturation value is close to $\pi^2/3\approx 3.29$, in full consistency with the prediction that the right-hand side of Eq.~\eqref{eq:KlichLevitov} is dominated by the second cumulant. 
 Remarkably, even for larger values of $\gamma$ corresponding to the strong-coupling regime, $g_0\lesssim 1$ (for which the correlation length is of the order of several lattice spacings), the relation $\mathcal{S}_{E}(l)/ {\cal C}_{l}^{(2)} \approx \pi^2/3$ still holds with good numerical accuracy of several percent.

\section{Conclusions and outlook}
\label{sec:Conclusions}

We have studied dynamics of one-dimensional free fermions on a chain subject to random projective measurements of local site occupation number.
Our main focus has been on the scaling behavior of the second cumulant of particle number in a subsystem, as well as that of the entanglement entropy. We have developed an analytical approach  based on the Keldysh formalism and the replica trick. The replica for this problem has an unconventional form, $R\to 1$. 

In the limit of rare measurements, $\gamma / J \ll 1$, we have derived an effective field theory of the problem, which is the NLSM (Sec.~ \ref{sec:NLSM}). Its replica-symmetric sector lives on the $U(2) / U(1) \times U(1)$ manifold and describes conventional diffusion. The replica-asymmetric (replicon) sector, which describes quantities of main interest, is a two-dimensional NLSM defined on the $SU(R)$ manifold.  

On the Gaussian level, this model predicts 
a logarithmic behavior for the second cumulant of number of particles in a subsystem and for the entanglement entropy. 
However, the one-loop RG analysis demonstrates that the logarithmic growth of the second cumulant saturates at a finite value even in the limit of rare measurements. This saturation corresponds to the area-law phase and implies the absence of a measurement-induced entanglement phase transition for free fermions. The crossover between logarithmic growth and saturation happens at exponentially large scale $l_\text{corr}$, $\ln l_\text{corr} \sim J / \gamma$. 

Overall, the behavior of the second cumulant ${\cal C}_{l}^{(2)}$ depending on the subsystem size $l$ can be summarized as follows:
\begin{equation}
{\cal C}_{l}^{(2)}\simeq n(1-n)\cdot\begin{cases}
l, & l\ll l_0,\\
\frac{4}{\pi}l_{0}\ln\frac{l}{l_{0}}, & l_{0}\ll l\ll l_{\text{corr}},\\
\sim l_0 g_0 , & l \gtrsim l_{\text{corr}}.
\end{cases}
\end{equation}
Here, $l_{0}=J/\gamma\sqrt{2}$ is the mean-free path,  $l_{\text{corr}}\sim l_{0}e^{4 \pi g_{0}}$ is the correlation length, and $g_0 = 2 l_0 n(1-n) \gg 1$. 
This scaling of the second cumulant directly translates into the same scaling of the entanglement entropy $\mathcal{S}_E$, implying the area law in the thermodynamic limit, see Sec.~\ref{sec:Entropy}.

These findings were supported by numerical analysis of the equal-time density correlation function obtained by means of direct simulation of the system's time evolution in Sec.~\ref{sec:Numerics}. Although exponentially large systems, which are required for achieving the thermodynamic limit for rare measurements, are not computationally accessible, available system sizes were sufficient to clearly demonstrate the tendency towards ``localization'' responsible for the area-law scaling, in consistency with one-loop RG equations. Our analytical and numerical results are also in agreement with the numerical analysis performed in Ref.~\cite{Coppola2022}.

While our results were obtained for the model of projective measurements, a conceptually similar theory can be developed for weak measurements or continuous monitoring. For this reason, we argue that the problem of weak measurements will fall in the same universality class and the long-wavelength limit will be described by essentially the same $SU(R)$ NLSM. In other words, one-dimensional free fermions with weak measurements are expected to demonstrate a qualitatively similar behavior---the absence of the measurement-induced entanglement phase transition.

Yet another important prediction, directly following from the present consideration, can be made regarding the behavior of monitored free fermions in higher-dimensional systems $d>1$. Our derivation can be extended to systems of an arbitrary dimension $d$, leading to an analogy between $d$-dimensional monitored free fermions and localization phenomena in $(d+1)$-dimensional disordered systems. It is well known that 
disordered systems exhibit Anderson localization transition above two dimensions. As a consequence, higher-dimensional ($d>1$) free-fermion systems should demonstrate, with increasing measurement rate, a transition between a ``critical phase'' with logarithmic law, ${\cal S}_{E}(l)\sim l^{d-1}\ln l$, to an area-law phase, ${\cal S}_{E}(l)\sim l^{d-1}$.

The analytical approach developed in the present paper is not restricted to non-interacting systems and can be generalized to include interactions between fermions. Indeed, the NLSM for disordered systems can be generalized to include interactions (and to study the emergent quantum phase transitions) within the replica and Keldysh formalisms, cf. Refs.~\cite{Finkelstein1984,
BelitzKirkpatrick1994,
KamenevLevchenko, SchwieteFinkelstein2014, Finkelstein2023}.  Importantly, the interacting NLSM for Anderson localization inherits the key property of a non-interacting NLSM:  it is a theory of interacting diffusive modes that emerge from the presence of conserved quantities -- particle number and energy. Based on this analogy, 
the NLSM for monitored systems should also be capable of describing the Goldstone modes of the interacting problem corresponding to the symmetries of the model. The interaction between the sigma-model modes could be then analyzed within the RG approach, similar to the framework for interacting disordered systems. Thus, the sigma-model approach to studying the measurement-induced phases, as developed in this work for non-interacting fermions, is a powerful framework for a unified description of a wide range of related problems, including those for monitored \textit{interacting} particles.

One crucial modification to our theory for studying interacting models is the need to distinguish between the statistics of particle-number cumulants and the statistics of entanglement entropy. Indeed, the presence of interparticle interactions breaks down the Gaussianity of many-body states, rendering direct application of the Klich-Levitov identity \eqref{eq:KlichLevitov} impossible. As a result, in the interacting model, the measurement-induced entanglement dynamics cannot be directly captured by analyzing the particle-number cumulants. Instead, the entanglement entropy should be calculated by adopting the twisted boundary conditions connecting different replicas at $t=t_f$, cf. Ref.~\cite{Calabrese2004, Cardy2008, Calabrese2009}, which can be straightforwardly incorporated into the NLSM formalism.

As discussed in Sec.~\ref{sec:intro}, numerical modeling of monitored one-dimensional systems of interacting fermions (or related interacting models described by a time-independent Hamiltonian) suggests that a measurement-induced entanglement transition takes place in such systems, similarly to quantum circuits.
An important question thus arises whether or not the entanglement (``information'') transition is accompanied by the particle-number-fluctuation (``charge'') transition, and, if yes, whether the two coincide or remain distinct for realistic interacting fermions. Furthermore, even if the two transitions occur concurrently, they might correspond to different types of behavior of the entanglement entropy and charge fluctuations in the corresponding phases (say, volume-to-area vs. logarithm-to-area phase transitions).
Essentially, this is a question as to whether the ``charge-information separation'' takes place in this class of systems (and, if yes, what are its implications). We foresee that it should be possible to describe the entanglement transition and behavior of charge fluctuations consistently within the unifying approach of the non-linear sigma model.

Incorporation of interactions in our formalism will produce additional terms in the fermionic Lagrangian, Eq.~\eqref{eq:Action:Fermionic}, of the form $\sum_{r}\left({\cal L}_{\text{int}}[\bar{\psi}_{r}^{+},\psi_{r}^{+}]-{\cal L}_{\text{int}}[\bar{\psi}_{r}^{-},\psi_{r}^{-}]\right)$, as interactions are directly included in the unitary evolution.  A preliminary analysis suggests that the terms generated in the NLSM are of the type $(U^{-1})_{rr} U_{rr}$, with a summation over the replica indices $r$. Importantly, these interaction terms in the full action of the monitored fermions can (assuming their RG relevance) partly break down the $SU(R)$ replica symmetry of our non-interacting theory. This would introduce a mass to some of the replicon modes $\hat{\Phi}$, which would correspond to a reduced symmetry that is expected to be $\mathtt{S}_R \times [U(1)]^{R-1}$, with $\mathtt{S}_R$ being a discrete group of replica permutations.
One may anticipate that, in the symmetry-broken phase, the charge fluctuations would behave similarly to Eq.~\eqref{eq:DensityCorrelator:GaussianResult:Cr}. However, the renormalization of the constant $g$ determining the prefactor in this correlator will be governed by the modified RG involving the interactions. This may lead, for instance, to the stabilization of the ``delocalized'' behavior of charge fluctuations in 1D.
If the measurement rate is sufficiently large while the interaction is weak, we expect that the effect of interaction can be studied perturbatively within the same framework to show the stability of the area-law phase for both entropy and charge fluctuations.

In this context, it is worth mentioning that, in specific classes of random quantum circuits involving Haar-random gates and qudits with a divergent number of states, which can be thought of as resembling interacting systems with conserved particle numbers, a so-called ``charge sharpening'' transition was predicted~\cite{Agrawal2022, Barratt2022}. 
Specifically, starting with a mixed state with no definite ``charge,'' repeated measurements yield a well-defined value of the charge, but the needed number of measurements scales differently with the system size in the charge-fuzzy and charge-sharp phases. It was found that this transition is distinct from the measurement-induced entanglement transition and occurs within the volume-law phase, being of the Berezinskii-Kosterlitz-Thouless type in 1D geometry.  It was also argued in Ref.~\cite{Barratt2022} that charge sharpening can be probed by the analysis of density correlation functions analogous to those studied in our work.

While connections between the physics of charge fluctuations in our model and that of charge sharpening in  Refs.~\cite{Agrawal2022, Barratt2022} appears to be very suggestive, it should be emphasized that the models are very different. 
Understanding the influence of entanglement and density correlations on each other's scaling behaviors in a system of monitored interacting particles (i.e., as considered in our work but with ``switched on'' interaction)
remains a challenging open question. In particular, it is important to understand the behavior of entanglement and of charge correlations as functions of measurement rate, interaction strength, and system size, for various spatial dimensionalities.
The analytical NLSM approach developed in this work, capable of handling particle fluctuations, entanglement, and interactions, represents a promising framework for exploring this class of measurement-induced phenomena in interacting systems.

Another challenging direction for further development of this theory is incorporation of static random potential (also in combination with interaction between fermions). This will, in particular, shed light on the interplay between measurements and Anderson localization (or many-body localization in the presence of interaction). In connection with symmetry classification of non-linear sigma models, it would also be very interesting to study possible physical realizations which would fall into different universality classes and be described by NLSMs with different symmetries. Another intriguing question is whether topological effects may be of relevance in the context of measurement problems. Finally, investigation of measurements that correspond to $R \ne 1$ (and thus do not satisfy Born's rules) is an interesting task, see a comment at the end of Appendix \ref{App:NLSM}.

We are grateful to A. Altland, D. Bernard, M. Buchhold, I. Burmistrov, S. Diehl, Y. Gefen, A. Lunkin, P. Ostrovsky, M. Szyniszewski, and R. Vasseur for fruitful discussions. We thank E. Doggen for comments on the manuscript. We acknowledge support by the Deutsche Forschungsgemeinschaft (DFG) via the grants MI 658/14-1 and GO 1405/6-1. IVG and ADM are grateful to Departamento de F{\'i}sica, FCFM, University of Chile (Santiago) for hospitality during the final stage of this work.

\appendix
\section{Derivation of the Keldysh action for measured fermions} 
\label{sec:appendix:Keldysh}

In this appendix, we present further details
of the derivation of Keldysh action, Sec.~\ref{sec:replica-trick}. Within the replica approach,
we should average $R$ copies of the unnormalized density matrix $\hat{D}$ over the measurement trajectories. For a fixed number of measurements $M$, this average can be written explicitly:
\begin{equation}
\hat{\rho}_{R}=\prod_{m=1}^{M}\left(\int\frac{d^{2}\textbf{x}_m}{LT}\sum_{n_{m}=0,1}\right)\otimes_{r=1}^{R}\hat{D}_{r},
\label{eq:Keldysh-appendix-1}
\end{equation}
where $\textbf{x}=(x,t)$ and $\int \!d^{2}\textbf{x}=\sum_{x=1}^{L}\int_{t_{i}}^{t_{f}}dt$.

For each replica $\hat{D}_{r}$ of the matrix $\hat{D}$, we introduce a separate replica of the Keldysh time contour $C_r = C_r^+ \cap C_r^- = (-\infty, +\infty) \cap (+\infty, -\infty)$. The expression for each replica of $\hat{D}$-matrix can be written explicitly utilizing the standard Keldysh-contour time-ordering symbol ${\cal T}_C$ as follows:
\begin{equation}
\hat{D} = {\cal T}_C \left\{\hat{\rho}_0 \hat{U}_C \prod_{m=1}^{M} \hat{\mathbb{P}}_{n_m}^{+}(x_{m}, t_{m}) \hat{\mathbb{P}}^{-}_{n_m}(x_m, t_{m})\right\},
\label{eq:Keldysh-appendix-2}
\end{equation}
where superscripts $+$ and $-$ refer to the forward and backward branches of the Keldysh contour, respectively,  and $\hat{U}_C$ denotes the unitary evolution over the full contour. Combining Eqs.~\eqref{eq:Keldysh-appendix-1} and \eqref{eq:Keldysh-appendix-2}
 and performing averaging over Poisson distribution of the number of measurements $M$ with the mean value $\overline{M} = \gamma L T$, we observe that the combination of projection operators gets exponentiated.  Finally, introducing the standard fermionic path integral representation, we arrive at the following replicated Keldysh action:
\begin{equation}
i S[\bar{\psi}, \psi] = i \sum_{r=1}^{R} \bar\psi_r \hat{G}_0^{-1} \psi_r + i \gamma\int d^2 \textbf{x}{\cal L}_{M}[\bar{\psi},\psi],
\end{equation}
with the bare free-fermion Green's function $\hat{G}_0$ being $2\times2$ matrix in the Keldysh space. The measurements produce an additional local contribution to the action with the following Lagrangian density:
\begin{align}
\label{eq:KeldyshLMpm}
i{\cal L}_{M}[\bar{\psi},\psi] &= \sum_{n=0,1}\prod_{r=1}^{R}V_{n}[\bar{\psi}_{r},\psi_{r}]-1,\\
V_{0}[\bar\psi, \psi] &= (1 - \bar{\psi}^{+} \psi^{+})(1 - \bar{\psi}^{-} \psi^{-}),\\
V_{1}[\bar\psi, \psi] &= \bar\psi^{+} \psi^{+} \bar{\psi}^{-} \psi^{-}.
\end{align}
The measurements thus give rise to an effective local ``interaction'' of fermionic fields between different branches of the Keldysh contour (despite the original problem being a non-interacting one), with the interaction vertices containing up to $4R$ fermionic fields.

As a next step, we perform a standard Larkin-Ovchinnikov rotation \cite{KamenevLevchenko} defined  via the relations (note opposite signs for $\psi$ and $\bar{\psi}$)
\begin{equation}
\begin{cases}
\psi_{1,2}=(\psi_{+}\pm\psi_{-})/\sqrt{2},\\
\bar{\psi}_{1,2}=(\bar{\psi}_{+}\mp\bar{\psi}_{-})/\sqrt{2}.
\end{cases}
\label{eq:Keldysh-rotation}
\end{equation}
In the new basis, the Green function acquires the following structure in the Keldysh space:
\begin{equation}
-i\left\langle \psi\bar{\psi}\right\rangle = \hat{G} = \begin{pmatrix}G_{R} & G_{K}\\
G_{\bar{K}} & G_{A}
\end{pmatrix}_{K} \,,
\end{equation}
where subscript $R,A,K,\bar{K}$ stands for retarded, advanced, Keldysh and anti-Keldysh components of Green function, respectively, with the later being zero in the conventional Keldysh technique. The inverse of the bare free fermion Green function in this basis can be written as:
\begin{equation}
\hat{G}_{0}^{-1}=i\partial_{t}-\hat{H}_{0}+i\delta\hat{\Lambda}_{0},\quad \hat{\Lambda}_{0}(\epsilon)=\begin{pmatrix}1 & 2F_{0}(\epsilon)\\
0 & -1
\end{pmatrix}_{K}\,,
\end{equation}
where the term proportional to infinitesimal $\delta \to +0$ fixes the correct causality properties of retarded and advanced Green functions and carries the information about the initial distribution function $f_0(\epsilon)$ via $F_0(\epsilon) = 1 - 2 f_0(\epsilon)$.

An important technical detail should be noted at this point. The interaction vertices $V_i[\bar{\psi}, \psi]$ consist of multiple fermionic fields taken at exactly same point in space and time, and Greens functions with coinciding arguments require special regularization. 
The general rule that follows from the derivation of the path integral representation is that, since projection operators were normal ordered, the time (anti-)ordering should also reduce to normal ordering for coinciding temporal arguments. Such a convention is, however, somewhat inconvenient; in particular, since it corresponds to a non-zero anti-Keldysh component of the local Green function. Furthermore, the Keldysh component, which is usually continuous in a sense $G_{K}(t \to +0) = G_{K}(t \to -0)$, actually contains a single point discontinuity $G_{K}(t \to \pm 0) \neq G_{K}(t \equiv 0)$. Because this discontinuity affects only a  set of measure zero, it is usually discarded. However, it should be treated carefully when working with local-in-time interactions.
In the present paper, we adopt an alternative, ``principal-value'' regularization: $G^{(\text{reg})}(t=0)=\lim_{t\to0}(G(t)+G(-t))/2$, which does not suffer from the above discontinuities and is related to the original Green function via $i\hat{G}^{(\text{reg})}_{ij}(t,t^\prime)=i\hat{G}_{ij}(t,t^\prime)+\delta_{tt^\prime} \delta_{ij}\hat{\tau}_{x}/2$, where indices $i,j$ incorporate real-space and replica structure. Note that $\delta_{t,t^\prime}$ is a Kronecker delta-symbol $\delta_{t,t^\prime}$ equal to unity for coinciding times and to zero otherwise; it should not be confused with the Dirac delta-function. 

Switching between different regularizations requires introduction of counter-terms in the action, $\delta S=S_M^{(\text{reg})}-S_M$, so that arbitrary observable quantities remain unchanged:
\begin{multline}    
\label{eq:RegularizationIdentity}
\int{\cal D}\bar{\psi}{\cal D}\psi\exp\left(i\bar{\psi}\hat{G}_{0}^{-1}\psi+i \gamma S_{M}\right)\\
=\int{\cal D}\bar{\psi}{\cal D}\psi\exp\left(i\bar{\psi}\hat{G}_{0}^{(\text{reg})-1}\psi+i \gamma S_{M}^{(\text{reg})}\right).
\end{multline}
One can consider a standard diagrammatic expansion in $\gamma$ to arbitrary order of perturbation theory, and explicitly build $S_{M}^{(\text{reg})}$ such that these expansions coincide. 

Consider an arbitrary Feynman diagram in the expansion of the left-hand side of Eq. \eqref{eq:RegularizationIdentity}, and substitute $iG_{0,ij}(t,t^{\prime})=iG^{\text{(reg)}}_{0,ij}(t,t^{\prime})-\delta_{tt^{\prime}}\delta_{ij}\hat{\tau}_{x}/2$. In order for the identity \eqref{eq:RegularizationIdentity} to be fulfilled, the difference between these Green functions should be produced by the counter-terms in the right-hand side.
From the structure of diagrammatic expansion we then deduce that counter-terms are given by the sum over all \textit{partial} Wick contractions of the original action $S_M$, with each contraction replaced by $-\delta_{t t^\prime} \delta_{ij} \hat{\tau}_x / 2$. As expected, only terms local in time, in real space, and in replica space  give non-zero counter-terms. Applying this procedure to the action \eqref{eq:KeldyshLMpm} and performing the rotation \eqref{eq:Keldysh-rotation} in Keldysh space, we find that the regularized action keeps the same product form \eqref{eq:KeldyshLMpm} but with regularized interaction vertices:
\begin{align}
V_{0}^{(\text{reg})}[\bar{\psi},\psi]&=\frac{1}{4}-\frac{1}{2}\left(\bar{\psi}_{2}\psi_{1}+\bar{\psi}_{1}\psi_{2}\right)-\bar{\psi}_{1}\psi_{1}\bar{\psi}_{2}\psi_{2},\\
V_{1}^{(\text{reg})}[\bar{\psi},\psi]&=\frac{1}{4}+\frac{1}{2}\left(\bar{\psi}_{1}\psi_{2}+\bar{\psi}_{2}\psi_{1}\right)-\bar{\psi}_{1}\psi_{1}\bar{\psi}_{2}\psi_{2}.
\end{align}
which is Eq.~\eqref{eq17} of the main text. 
We work with the regularized action, dropping the superscript ``(reg)'' for brevity.

\section{Generalized Hubbard-Stratonovich transformation}
\label{sec:appendix:HubbardStratonovich}

In this appendix, we provide details of  derivation of the generalized Hubbard-Stratonovich transformation (Sec.~\eqref{sec:HS-transformation} of the main text) for the $4R$-fermion interaction of the form \eqref{eq:Action:Measurement:Grassman}.

We begin with the following identity valid for arbitrary positive parameter $\epsilon > 0$:
\begin{equation}
\label{eq:HubbardStratonovich:1}
1=\int{\cal D}\hat{{\cal G}}{\cal D}\hat{\Sigma}\exp\left(-\frac{1}{2\epsilon}\Tr(\hat{{\cal G}}+i\psi\bar{\psi})^{2}-\frac{\epsilon}{2}\Tr\hat{\Sigma}^{2}\right),
\end{equation}
where the integration is performed over $2R \times 2R$ time- and space-dependent Hermitian matrices $\hat{{\cal G}}(x,t)$ and $\hat{\Sigma}(x,t)$ with a flat integration measure. The $\Tr$ symbol here includes  the trace over replica and Keldysh space as well as integration over time and summation over lattice sites. In the limit $\epsilon \to +0$, the first term in the exponential 
in Eq.~\eqref{eq:HubbardStratonovich:1}
acts as a delta-function, which imposes $\hat{{\cal G}}=-i\psi\bar{\psi}$. This property will be used to rewrite the interaction in terms of the ${\cal G}$-matrix.
Performing a shift $\hat{\Sigma}\mapsto\hat{\Sigma}+i\left(\hat{{\cal G}}+i\psi\bar{\psi}\right) / \epsilon$, we arrive at another form of Eq. \eqref{eq:HubbardStratonovich:1}:
\begin{equation}
\label{eq:HubbardStratonovich}
1=\int{\cal D}\hat{{\cal G}}{\cal D}\hat{\Sigma}\exp\left(-\frac{\epsilon}{2}\Tr\hat{\Sigma}^{2}-i\Tr(\hat{\Sigma}\hat{{\cal G}})-\bar{\psi}\hat{\Sigma}\psi\right).
\end{equation}
The first term in the exponential 
in Eq.~\eqref{eq:HubbardStratonovich}
is required only to enforce the convergence of the integral over $\hat{\Sigma}$; we will omit it in what follows for brevity.

Now that we have the identification $\psi\bar{\psi} = - i \hat{{\cal G}}$, the we can rewrite the interaction in terms of ${{\cal G}}$-matrix. Formally one then can consider an arbitrary decoupling of the non-linear interaction $S_M$ in bilinears $\psi\bar{\psi}$ 
(which can be viewed as a single pattern of Wick contractions) and replace the corresponding pair products of fermionic operators by ${\cal G}_{ij}$.
 Although this would be mathematically correct, physically it would correspond to decoupling of the interaction in a single \textit{channel}. Indeed, we want to consider matrix ${\cal G}$ in what follows as a slow mode. It is thus crucial to consider decoupling in all possible channels.
The procedure bears similarity with the one discussed in the context of Anderson localization in the orthogonal symmetry class, where slow modes include diffusons and cooperons, and the quartic interaction coming from averaging over quenched disorder is decoupled in two different channels simultaneously, see Ref. \cite{EfetovBook}. 

To implement this technically, we switch to the Fourier space and introduce an energy and momentum cutoff $\Lambda$ which should be smaller than the size of the Brillouin zone $\pi$, but larger then any other characteristic scale arising in our problem:
\begin{equation}
\label{eq:Gmomentum}
\hat{{\cal G}}_{ij}(\boldsymbol{q})\simeq-i \theta(\Lambda-|\boldsymbol{q}|)\sum_{\boldsymbol{k}}\psi_{i}(\boldsymbol{k}+\boldsymbol{q}/2)\bar{\psi}_{j}(\boldsymbol{k}-\boldsymbol{q}/2),
\end{equation}
with indices $i, j$ corresponding to the Keldysh and replica structure of fermionic fields. Considering an arbitrary local interaction vertex of order $2N$ that we want to decouple using the generalized Hubbard-Stratonovich transformation, we rewrite it in the momentum representation:
\begin{multline}
V_{2N}[\bar{\psi},\psi]=\int d^2\textbf{x}\prod_{i=1}^{N}\bar{\psi}_{b_{i}}(\textbf{x})\psi_{a_{i}}(\textbf{x})\\
=\sum_{\begin{array}{c}
\boldsymbol{k}_{1},\dots,\boldsymbol{k}_{N}\\
\boldsymbol{k}_{1}^{\prime},\dots,\boldsymbol{k}_{N}^{\prime}
\end{array}}\delta\left(\sum_{i=1}^{N}\boldsymbol{k}_{i}=\sum_{i=1}^{N}\boldsymbol{k}_{i}^{\prime}\right)\prod_{i=1}^{N}\bar{\psi}_{b_{i}}(\boldsymbol{k}_{i}^\prime)\psi_{a_{i}}(\boldsymbol{k}_{i}).
\end{multline}
Within the whole $2N$-dimensional momentum space, there is $N!$ sectors where momentums are grouped into $N$ pairs $\{\boldsymbol{k}_{\alpha_{i}}, \boldsymbol{k}_{\beta_{i}}^{\prime}\}$  with small momentum difference in each pair, $|\boldsymbol{k}_{\alpha_{i}}-\boldsymbol{k}_{\beta_{i}}^{\prime}|\lesssim\Lambda$. 
These sectors are nearly non-overlapping: the overlap would correspond to more than two momenta being close to each other, and the phase volume of such region in momentum space contains an additional smallness in parameter $\Lambda \ll 1$. As the long-wavelength fluctuations of matrix ${\cal G}$ are expected to dominate the physical behavior of the system, such ``pairings'' of fermionic fileds into low-momenta bilinears should dominate the original fermionic path integral.

Introducing for each pair ``center-of-mass'' and ``relative-motion'' momenta defined as $\boldsymbol{K}_{i}=(\boldsymbol{k}_{\alpha_{i}}+\boldsymbol{k}_{\beta_{i}}^{\prime})/2$ and $\boldsymbol{q}_{i}=\boldsymbol{k}_{\alpha_{i}}-\boldsymbol{k}_{\beta_{i}}^{\prime}$, we see that summation over the ``center-of-mass'' momentum reduces to the corresponding matrix element of matrix ${\cal G}$, see Eq. \eqref{eq:Gmomentum}, which allows us to rewrite the interaction vertex as:
\begin{multline*}
V_{2N}\approx\sum_{\{\alpha_{i},\beta_{i}\}\in{\cal P}_{2N}}(-1)^{F}\sum_{|\boldsymbol{q}_{i}|<\Lambda}\delta\left(\sum_{i=1}^{N}\boldsymbol{q}_{i}=0\right)\prod_{i=1}^{N}i{\cal G}_{\alpha_{i}\beta_{i}}(\boldsymbol{q}_{i})
\\= \int d^2\textbf{x}\left(\sum_{\{\alpha_{i},\beta_{i}\}\in{\cal P}_{2N}}(-1)^{F}\prod_{i=1}^{N}i {\cal G}_{\alpha_{i}\beta_{i}}(\textbf{x})\right).
\end{multline*}
Here the outermost sum runs over the $N!$ sectors ($N!$ pairings of the set $\{a_i, b_i\}$) denoted as ${\cal P}_{2N}$, and the $(-1)^F$ factor accounts for sign changes that arise when Grassmann fields belonging to each pair are brought together. As the last step, we note that this expression is formally equivalent to the result of application of the Wick theorem to the following Gaussian Grassmann integral:
\begin{equation}
V_{2N}[{\cal G}]=\int\frac{{\cal D}\bar{\psi}{\cal D}\psi}{\det\left(-i{\cal G}^{-1}\right)}\exp\left(i\bar{\psi}{\cal G}^{-1}\psi\right)V_{2N}[\bar{\psi},\psi].
\label{eq:V2NG-gaussian-int}
\end{equation}
This equation is the main result of the present derivation: decoupling a given interaction vertex in all possible slow channels is equivalent to calculating the Grassmann Gaussian average of this interaction vertex. We reiterate that the matrix ${\cal G}$ is  assumed to be a slow field in this derivation; without this restriction, one would formally get a multiple counting (each term of the form $V_{2N}$ would be counted $N!$ times).

We are now ready to apply this scheme to the interaction in our problem. Substituting  the exponential form of the interaction \eqref{eq:Action:Measurement:Exponential}
in Eq.~\eqref{eq:V2NG-gaussian-int}, we are left with Gaussian integrals over 
$\bar{\psi},\psi$, which can be readily calculated. This finally brings us to the following form of the interaction rewritten now in terms of matrix ${\cal G}$:
\begin{equation}
i{\cal L}_{M}[{\cal G}]=\det\left(\frac{1}{2}+i\hat{{\cal G}}\hat{\tau}_{x}\right)+\det\left(\frac{1}{2}-i\hat{{\cal G}}\hat{\tau}_{x}\right)-1,
\end{equation}
which is
Eq.~\eqref{eq:Action:Measurement:Matrix}
of the main text.

\section{Matrix field theory: Saddle points and Gaussian fluctuations}
\label{App:NLSM}

In this appendix, we provide additional details to the analysis of saddle points and Gaussian fluctuations in Sections
\ref{sec:saddle-point-analysis} and
\ref{sec:quadratic-fluctuations} of the main text. 

To determine spatially homogeneous saddle points of the matrix action \eqref{eq:S-S0-LM}, we consider a variation of the action with respect to $\hat{\Sigma}$, which yields the following saddle-point equation:
\begin{equation}
-i\hat{{\cal G}}_{0}+i~\mathrm{v.p.}\int\frac{d\epsilon}{2\pi}\int_{-\pi}^{\pi}\frac{dk}{2\pi}(\epsilon-\xi_{k}+i\hat{\Sigma}_0)^{-1} = 0 \,.
\end{equation}
This equation can be solved for $\hat{{\cal G}}_0$ in the basis where $\hat{\Sigma}$ is diagonal. Let us write $\hat{\Sigma}_{0}=\hat{{\cal R}}\hat{\lambda}_{0}\hat{{\cal R}}^{-1}$ with a diagonal matrix $\hat{\lambda}_0$; then the solution reads:
\begin{equation}
\hat{{\cal G}}_{0}=-i\hat{Q}_{0}/2,\quad \hat{Q}_{0}\equiv\hat{{\cal R}} (\sign{\rm Re}\hat{\lambda}_{0})\hat{{\cal R}}^{-1}.
\end{equation}
By construction, matrix $\hat{Q}_0$ satisfies the NLSM constraint $\hat{Q}_0^2 = 1$. 

The ``quantum'' Keldysh component of the fermionic density on this solution is given by
\begin{equation}
\rho_{0}^{(\text{q})}=-\frac{1}{4}\Tr\hat{Q}_{0}.
\end{equation}
Since eigenvalues of $\hat{Q}_0$ are $\pm 1$, this quantity has a discrete set of possible values. On physical grounds, we request that the quantum component is zero on the saddle point, i.e., $\Tr \hat{Q}_0 = 0$.

We focus first on the replica-symmetric saddle points $\hat{Q}_{0}=(\hat{Q}_{0})_{K}\otimes\hat{1}_{R}$. Consider arbitrary fluctuations (including those with a non-trivial structure in the replica space) around this saddle point, $\hat{{\cal G}}=-i(\hat{Q}_{0}+\delta\hat{Q}_{{\cal G}})/2$. Properties of matrix $\hat{Q}_0$ allow us to rewrite the measurement action \eqref{eq:Action:Measurement:Matrix} in the following form convenient for an expansion in
$\delta\hat{Q}_{{\cal G}}$:
\begin{multline}
i{\cal L}_{M}[{\cal G}]=\rho_{0}^{R}\det\left(1+\frac{\left(\hat{Q}_{0}-\hat{\tau}_{x}\right)\delta\hat{Q}_{{\cal G}}}{4\rho_{0}}\right)\\
+(1-\rho_{0})^{R}\det\left(1+\frac{\left(\hat{Q}_{0}+\hat{\tau}_{x}\right)\delta\hat{Q}_{{\cal G}}}{4(1-\rho_{0})}\right)-1 \,,
\end{multline}
where we have introduced the ``classical'' Keldysh component of the density defined as
\begin{equation}
\rho_{0}=\frac{1}{4}\Tr\left(1-\hat{Q}_{0}\hat{\tau}_{x}\right).
\end{equation}
To perform the expansion, we use the formula 
\begin{multline}
\det(1+\epsilon \hat{X})=\exp\left(\Tr\ln(1+\epsilon\hat{X})\right)\\
\approx1+\epsilon\Tr\hat{X}-\frac{\epsilon^2}{2}\left(\Tr\hat{X}^{2}-\Tr^{2}\hat{X}\right) + O(\epsilon^3).
\end{multline}
This yields the following results for the terms of zeroth and first order:
\begin{equation}
i{\cal L}_{M}^{(0)}=\rho_{0}^{R}+(1-\rho_{0})^{R}-1 \,,
\end{equation}
\begin{multline}
\label{eq:Action:Measurement:Expansion:Linear}
i{\cal L}_{M}^{(1)}=\frac{1}{4}\Tr\Big(\delta\hat{Q}_{{\cal G}}\Big[\left(\rho_{0}^{R-1}+(1-\rho_{0})^{R-1}\right)\hat{Q}_{0}\\
+\left((1-\rho_{0})^{R-1}-\rho_{0}^{R-1}\right)\hat{\tau}_{x}\Big]\Big),
\end{multline}
and the following two quadratic terms:
\begin{multline}
\label{eq:Action:Measurement:Expansion:Quadratic:1}
i{\cal L}_{M}^{(2,1)}=-\frac{1}{32}\Tr\Big[\rho_{0}^{R-2}\left(\left(\hat{Q}_{0}-\hat{\tau}_{x}\right)\delta\hat{Q}_{{\cal G}}\right)^{2}\\
+(1-\rho_{0})^{R-2}\left((\hat{Q}_{0}+\hat{\tau}_{x})\delta\hat{Q}_{{\cal G}}\right)^{2}\Big],
\end{multline}
\begin{multline}
\label{eq:Action:Measurement:Expansion:Quadratic:2}
i{\cal L}_{M}^{(2,2)}=\frac{\rho_{0}^{R-2}}{32}\Tr^{2}\left[(\hat{Q}_{0}-\hat{\tau}_{x})\delta\hat{Q}_{{\cal G}}\right]\\
+\frac{(1-\rho_{0})^{R-2}}{32}\Tr^{2}\left[(\hat{Q}_{0}+\hat{\tau}_{x})\delta\hat{Q}_{{\cal G}}\right].
\end{multline}

Equation \eqref{eq:Action:Measurement:Expansion:Linear} allows us to write the second saddle point equation, which is obtained by varying the full action \eqref{eq:S-S0-LM} with respect to ${\cal G}$:
\begin{multline}
-i\hat{\Sigma}_{0}+i\gamma\Big[\frac{1}{2}\left(\rho_{0}^{R-1}+(1-\rho_{0})^{R-1}\right)\hat{Q}_{0}\\
+\frac{1}{2}\left((1-\rho_{0})^{R-1}-\rho_{0}^{R-1}\right)\hat{\tau}_{x}\Big] = 0 \,.
\end{multline}
The term proportional to $\hat{\tau}_x$ vanishes in two cases: (i) in the replica limit $R \to 1$ for arbitrary density $\rho_0$, and (ii) for half-filling $\rho_0 = 1/2$ and for arbitrary number of replicas $R$.  The physics that we are interested in is expected to be independent on $\rho_0$, so that the case $\rho_0 = 1/2$ should be representative. We thus retain the saddle-point manifold  $\hat{\Sigma}_{0}=\gamma\hat{Q}_{0}/2^{R-1}$.

As the last step, we parametrize fluctuations of $\Sigma$ as $\hat{\Sigma}=\gamma_{R}\left(\hat{Q}_{0}+\delta\hat{Q}_{\Sigma}\right)$ with $\gamma_R = \gamma / 2^{R-1}$, and perform a quadratic expansion of action \eqref{eq:Action:Matrix} in $\delta{Q}_\Sigma$. The zeroth order term vanishes, and the result for the second-order term reads
\begin{equation}
\label{eq:S02}
iS_{0}^{(2)}=\frac{1}{2}\Tr\left(\gamma_{R}^{2}\hat{G}\delta\hat{Q}_{\Sigma}\hat{G}\delta\hat{Q}_{\Sigma}-\gamma_R \delta\hat{Q}_{\Sigma}\delta\hat{Q}_{{\cal G}}\right).
\end{equation}
Here $\hat{G}$ is a dressed Green function that has the form
\begin{multline}
\hat{G}(\boldsymbol{k})=(\epsilon-\xi_{k}+i\gamma_{R}\hat{Q}_{0})^{-1}\\
=\frac{1}{2}G_{R}(\boldsymbol{k})(1+\hat{Q}_{0})+\frac{1}{2}G_{A}(\boldsymbol{k})(1-\hat{Q}_{0}),
\end{multline}
with SCBA-dressed retarded and advanced Green functions defined as 
\begin{equation}
G_{R/A}^{-1}(\boldsymbol{k})=\epsilon-\xi_{k}\pm i\gamma_{R}.
\end{equation}
Due to causality properties of $G_{R/A}$,  only the cross term proportional to the elementary ``diffuson'' block ${\cal B}(\textbf{x})=G_{R}(\textbf{x})G_{A}(-\textbf{x})$ survives
in the first term of Eq. \eqref{eq:S02}.
Writing explicitly the space and time integration included in symbol $\Tr$ in Eq. \eqref{eq:S02}, we obtain
\begin{multline}
iS_{0}^{(2)}=\frac{\gamma_{R}^{2}}{4}\int d^2\textbf{x}_{1}d^2\textbf{x}_{2}{\cal B}(\textbf{x}_{1}-\textbf{x}_{2})\\
\times\Tr\left[\delta\hat{Q}_{\Sigma}(\textbf{x}_{1})(1+\hat{Q}_{0})\delta\hat{Q}_{\Sigma}(\textbf{x}_{2})(1-\hat{Q}_{0})\right]\\
-\frac{\gamma_R}{2}\int d\textbf{x}\Tr(\delta\hat{Q}_{\Sigma}(\textbf{x})\delta\hat{Q}_{{\cal G}}(\textbf{x})).
\end{multline}
This is Eq.~\eqref{eq:Action:MatrixGaussian} of the main text.

It is worth noting that, for arbitrary $R$ and $n$, there are exact saddle points of the action of the form $\hat{Q}_0 = \pm \hat{\tau}_x$ and $\hat\tau_z$; for $n=1/2$ the latter coincides with $\hat\Lambda$. 
They correspond to densities $\rho_0 = 0$,  $\rho_0 = 1$, and $\rho_0 = 1/2$, respectively.
We conjecture that these saddle points may correspond to breaking of the system into domains for the case of $R>1$, i.e., for measurements with probabilities not satisfying Born's rules. Indeed, our preliminary numerical results for such unconventional measurements indicate a trend towards formation of domains. We relegate a systematic investigation of this issue to future work.

\section{Crossover between the ballistic and diffusive regimes in the Gaussian approximation}
\label{sec:appendix:BallisticCrossover}

In this appendix, we present details of an exact calculation of the density correlation function
$C(q)$,
Eq.~\eqref{eq:DensityCorrelator:GaussianResult:C}, and the second cumulant ${\cal C}_{l}^{(2)}$,
Eq.~\eqref{eq:SecondCumulant}, within the Gaussian theory. The result includes the ballistic and diffusive regimes and a crossover between them. The only assumption is that the mean free path is large, $l_0 \gg 1$; a relation between $l_0$ and the length scale $l$ can be arbitrary. 

The inversion of the quadratic operator that enters the Gaussian action (Eqs.~\ref{eq:Action:MatrixGaussian}-\ref{eq:LM2}) in the presence of an absorbing boundary in the time domain at $t = t_f$ and the calculation of the density-correlation function \eqref{eq:ReplicaOffdiagonalCorrelationFunction} is equivalent to the solution of the Wiener-Hopf integral equation. Without loss of generality, we put $t_i \to -\infty$ and $t_f = 0$, and obtain:
\begin{equation}
L(t,t^{\prime})-\frac{1}{2\tau_{0}}\int_{-\infty}^{0} dt^{\prime\prime}{\cal B}(q,|t-t^{\prime\prime}|)L(t^{\prime\prime},t^{\prime})=\delta(t-t^{\prime}),
\end{equation}
with ${\cal B}(q,t)$ being the Fourier transform, with respect to time, of the ``diffuson block'', Eq. \eqref{eq:DiffusiveBlock}:
\begin{equation}
{\cal B}(q,t)=\theta(t)e^{-t/\tau_{0}}J_{0}\left(4Jt\sin\frac{q}{2}\right).
\end{equation}
The replica-off-diagonal density correlation function is related to the kernel $L$ via
\begin{multline}
\frac{C_{\text{repl}}(q,t,t^{\prime})}{n(1-n)}={\cal B}(q,|t-t^{\prime}|)\\
+\frac{1}{2\tau_0}\int_{-\infty}^{0}dt_{1}dt_{2}{\cal B}(q,|t-t_{1}|)L(t_{1},t_{2}){\cal B}(q,|t_{2}-t^{\prime}|)\\
-\frac{1}{\tau_0}\int_{-\infty}^{0}dt_{1}dt_{2}L(t_{1},t_{2})\Big[{\cal B}(q,t-t_{1}){\cal B}(q,t^{\prime}-t_{2})\\
+{\cal B}(q,t_{1}-t){\cal B}(t_{2}-t^{\prime})\Big].
\end{multline}

As we are interested only in the equal-time density correlation function $C(q) = C_{\text{repl}}(q, t=t^\prime = t_f = 0)$, the problem can be slightly simplified by introducing an auxiliary function
\begin{equation}
F(q,t)\equiv\frac{1}{2}\int_{-\infty}^{0}L(t,t^{\prime}){\cal B}(q,-t^{\prime})dt^{\prime},
\end{equation}
which satisfies the integral equation:
\begin{equation}
F(q,t)-\frac{1}{2\tau_{0}}\int_{-\infty}^{0}dt^{\prime}{\cal B}(q,|t-t^{\prime}|)F(q,t^{\prime})=\frac{1}{2}{\cal B}(q,-t),
\label{Fqt}
\end{equation}
and determines the density correlation function through
\begin{equation}
\label{eq:CqFq}
C(q)=n(1-n) 2\left[1-F(q,0)\right]
\end{equation}
The integral equation for $F(q,t)$ depends on a single parameter
\begin{equation}
u = 2 l_0 \sin(q/2) \approx q l_0 \,.
\end{equation}

%%%%%%%%%%%%%%%%
\begin{figure}[ht]
    \centering
    \includegraphics[width=\columnwidth]{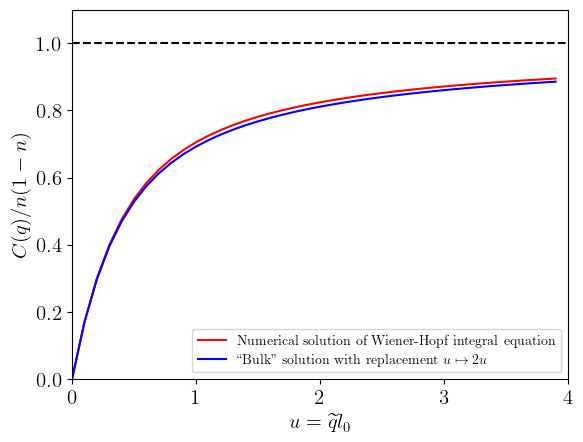}
    \caption{Comparison between numerical solution of the Wiener-Hopf integral equation as given by Eqs. (\ref{Fqt},\ref{eq:CqFq}) (red curve) with the ``bulk'' solution with replacement $u \mapsto 2u$ as given by Eq. \eqref{eq:DensityCorrelator:BallisticCrossover} (blue curve). Although the discrepancy is visible, it of the order of one percent and is comparable to the numerical error.}
    \label{fig:WienerHopf}
\end{figure}

We have solved Eq.~\eqref{Fqt} and calculated $C(q)$  numerically in  a broad range of values of $u$, the result is presented in Fig. \ref{fig:WienerHopf}.
This revealed an interesting property: within numerical accuracy, the solution $C(q)$ coincides with the solution  $\widetilde{C}(q)$  of the corresponding  \textit{bulk} equation (i.e., the one with upper limit in Eq.~\eqref{Fqt} replaced by $+\infty$), but with parameter $u$ being exactly twice larger:
\begin{align}
\label{eq:DensityCorrelator:BallisticCrossover}
C(q; u)&\simeq\widetilde{C}(q; 2u)=n(1-n)\cdot\widetilde{c}(u),\\
\widetilde{c}(u)&=\int_{0}^{\infty}\frac{2dv}{\pi}\frac{{\rm Re}b(v,2u)-|b(v,2u)|^{2}}{1-{\rm Re}b(v,2u)}.
\end{align}
Here, $v = \omega \tau_0$ is the dimensionless frequency, and $b(v,u)$ is the dimensionless block of the ladder, \eqref{eq:DiffusiveBlock}:
\begin{equation}
b(v,u)=\frac{1}{\sqrt{(1-iv)^{2}+2u^{2}}}
\end{equation}
While the above simple relation between the bulk and boundary correlation functions has not been demonstrated analytically in the whole range of parameters, it is straightforward to see that it correctly reproduces the asymptotic behavior in both limits of small and large $u$:
\begin{equation}
\widetilde{c}(u)\approx\begin{cases}
2u, & u\ll1\\
1 - \ln u / 2 \pi \sqrt{2} u, & u\gg1
\end{cases}
\end{equation}

Performing the Fourier transformation, we obtain a universal scaling form for the (equal-time) density correlation function in real space,
\begin{align}
C(x) &=n(1-n)\left[\delta_{x,0}-\frac{1}{l_{0}}c\left(\frac{x}{l_{0}}\right)\right],\\
c\left(y = \frac{x}{l_0}\right) &=\frac{1}{\pi}\int_{0}^{\infty}du\, 
[1-\tilde{c}(u)]\cos(uy)
\end{align}
with asymptotic behavior:
\begin{equation}
c(y)\approx\begin{cases}
2/\pi y^{2}, & y\gg1\\
\ln^{2}(1/y)/4\sqrt{2}\pi^{2}, & y\ll1
\end{cases}
\end{equation}
Substituting this result 
into Eq.~\eqref{eq:SecondCumulant}, we obtain the universal scaling form of the second cumulant,
\begin{align}
{\cal C}_{l}^{(2)} &=n(1-n)\, l_{0}\, c_2\left(\frac{l}{l_{0}}\right),\\
c_{2}\left(y=\frac{l}{l_{0}}\right)	&=\frac{2}{\pi}\int_{0}^{\infty}\frac{du}{u^2}\widetilde{c}(u)(1 - \cos u y),
\end{align}
with the asymptotic behavior given by
\begin{equation}
c_2(y) \approx \begin{cases}
y, & y\ll1,\\
(4/\pi)\ln y, & y\gg1.
\end{cases}
\end{equation}
This scaling function is plotted in   Fig.~\ref{fig:DensityCorrelator:Analytics} of the main text. 

\section{Derivation of the \textit{SU(R)} NLSM: Effective action for the replicon modes}
\label{sec:appendix:RepliconAction}

In this appendix we present details of the derivation of the $SU(R)$ NLSM action, Sec.~\ref{Sec:VD}.
As a starting point, we use Eqs. \eqref{eq:Action:NLSM} and \eqref{eq:Action:Measurement:NLSM} and utilize the parametrization \eqref{eq:QMatrix:Replicon}.
We then expand the action to quadratic order in the massive $\hat{\Theta}$-modes (however keeping it exact in $\hat{Q}_0$ and  $\hat{\Phi}$) and then integrate over $\hat{\Theta}$. In all prefactors, we take the limit $R \to 1$.

\paragraph{Measurement action.}
We start with Eq.~\eqref{eq:Action:Measurement:NLSM}, which is manifestly independent of $\hat{{\cal R}}_\Phi$ because $\hat{{\cal R}}_\Phi$ commutes with $\hat{\tau}_x$. To perform the expansion in $\hat{\Theta}$-modes, we use formulas from Appendix \ref{App:NLSM}.
Since $\hat{\Theta}$ is traceless in replica space, we can directly use Eqs.~\eqref{eq:Action:Measurement:Expansion:Linear} and \eqref{eq:Action:Measurement:Expansion:Quadratic:1}, with the replacement $Q_{\cal G} \mapsto Q_\Theta$, where
\begin{multline}
\label{eq:QTheta}
\hat{Q}_{\Theta}=\hat{{\cal R}}_{\Theta}\hat{Q}_{0}\hat{{\cal R}}_{\Theta}^{-1}\approx\hat{Q}_{0}+i\hat{\Theta}[\hat{\tau}_{y},\hat{Q}_{0}]/2\\
-\hat{\Theta}^{2}(\hat{Q}_{0}-\hat{\tau}_{y}\hat{Q}_{0}\hat{\tau}_{y})/4.
\end{multline}
After separating trace in Keldysh and replica space yields:
\begin{equation}
\label{eq:Action:Measurement:ThetaExpansion}
i{\cal L}_{M}[\hat{\Theta},\hat{Q}_{0}]=- \frac{\tr_{K}^{2}(\hat{Q}_{0}\hat{\tau}_{z})}{32\rho_{0}(1-\rho_{0})}\tr_{R}\hat{\Theta}^{2}
\end{equation}
At the saddle point $\hat{Q}_0 = \hat{\Lambda}$, the prefactor, which gives a mass of $\hat{\Theta}$-mode, is finite and equal to $1 / 8 n(1-n)$.

\paragraph{Dynamic term.}
We proceed with the time-derivative term from Eq.~\eqref{eq:Action:NLSM}, which we  denote as ${\cal L}_{\text{dyn}}$. The parametrization \eqref{eq:QMatrix:Replicon} corresponds to rotation matrices $\hat{{\cal R}}=\hat{{\cal R}}_{\Phi}\hat{{\cal R}}_{\Theta}\hat{{\cal R}}_{0}$, where matrix $\hat{{\cal R}}_{0}$ generates the replica-symmetric part $\hat{Q}_{0}=\hat{{\cal R}}_{0}\hat{\Lambda}\hat{{\cal R}}_{0}^{-1}$. The direct substitution generates the following terms:
\begin{multline}
\label{eq:Action:Dynamic:Expansion}
i{\cal L}_{\text{dyn}}[\hat{Q}]=i{\cal L}_{\text{dyn}}[\hat{Q}_{0}]+\frac{1}{2}\Tr(\hat{Q}_{0}\hat{{\cal R}}_{\Theta}^{-1}\partial_{t}\hat{{\cal R}}_{\Theta})\\
+\frac{1}{2}\Tr(\hat{Q}_{\Theta}\hat{{\cal R}}_{\Phi}^{-1}\partial_{t}\hat{{\cal R}}_{\Phi}).
\end{multline}

The second term in Eq. \eqref{eq:Action:Dynamic:Expansion} vanishes exactly since $\hat{\Theta}$ is traceless in replica space. The last one, however, is very important as it generates interaction between massive mode $\hat{\Theta}$ and massless $\hat{\Phi}$. Separating explicitly the trace over Keldysh space, we arrive at
\begin{equation}
\label{eq:Action:Dynamic:ThetaExpansion}
i\delta{\cal L}_{\text{dyn}}[\hat{\Theta}, \hat{\Phi}, \hat{Q}_0] =\frac{1}{4}\tr_{K}(\hat{Q}_{0}\hat{\tau}_{z})\cdot\tr_{R}\left(\hat{\Theta}\hat{U}^{-1/2}\partial_{t}\hat{U}\hat{U}^{-1/2}\right)
\end{equation}
with $\hat{U} = \exp(i \hat{\Phi})$.

\paragraph{Spatial gradient term.}
Next, we consider the spatial-gradient term from Eq.~\eqref{eq:Action:NLSM}. The derivative of the $\hat{Q}$-matrix can be written in the following form:
\begin{multline*}
\partial_{x}\hat{Q}=\hat{{\cal R}}_{\Phi}\hat{{\cal R}}_{\Theta}\partial_{x}\hat{Q}_{0}\hat{{\cal R}}_{\Theta}^{-1}\hat{{\cal R}}_{\Phi}^{-1}\\
+\hat{{\cal R}}_{\Phi}\hat{{\cal R}}_{\Theta}\left[\hat{{\cal R}}_{\Theta}^{-1}\partial_{x}\hat{{\cal R}}_{\Theta},\hat{Q}_{0}\right]\hat{{\cal R}}_{\Theta}^{-1}\hat{{\cal R}}_{\Phi}^{-1}\\
+\hat{{\cal R}}_{\Phi}\left[\hat{{\cal R}}_{\Phi}^{-1}\partial_{x}\hat{{\cal R}}_{\Phi},\hat{Q}_{\Theta}\right]\hat{{\cal R}}_{\Phi}^{-1}.
\end{multline*}
Upon squaring this expression, some of the terms vanish after taking the trace over replicas.
Integrating over $\hat{\Theta}$ and keeping only terms with two gradients (i.e., discarding terms with higher gradients) we left with the bare replica-symmetric term for $\hat{Q}_0$-matrix and the following term containing gradients of the massless field $\hat{\Phi}$:
\begin{equation}
i\delta{\cal L}_{\text{grad}}[\hat{\Phi}, \hat{Q}_0]=-D\rho_{0}(1-\rho_{0})\tr_{R}\left(\partial_{x}\hat{U}^{-1}\partial_{x}\hat{U}\right).
\label{eq:E5}
\end{equation}

\paragraph{Density source term.}
Last but not least---replicon modes also couple to the density source term defined by Eq.~\eqref{eq:Action:DensitySource}. In our parametrization, it acquires the following form:
\begin{equation}
i{\cal L}_{\text{source}}[\hat{Q}, \hat{\xi}]=\frac{i}{4}\Tr\left(\hat{\xi}_{\Phi}\left(1-\hat{Q}_{\Theta}\hat{\tau}_{x}\right)\right)
\label{eq:E6}
\end{equation}
with the $\Phi$-rotated source $\hat{\xi}_{\Phi}=\hat{{\cal R}}_{\Phi}^{-1}\hat{\xi}\hat{{\cal R}}_{\Phi}$, and $\hat{Q}_{\Theta}$ defined in Eq.~\eqref{eq:QTheta}. Expanding Eq.~\eqref{eq:E6} in $\Theta$, we arrive at the following expression:
\begin{multline}
\label{eq:Action:Source:ThetaExpansion}
i{\cal L}_{\text{source}}[\hat{\Theta}, \hat{\Phi}, \hat{Q}_0, \hat{\xi}]=i\rho_{0}\xi_0+\frac{i}{8}\tr_{K}(\hat{Q}_{0}\hat{\tau}_{z})\\
\times \tr_{R}\left[\hat{\Theta}\left(\hat{U}^{-1/2}\hat{\Xi}\hat{U}^{1/2}+\hat{U}^{1/2}\hat{\Xi}\hat{U}^{-1/2}\right)\right].
\end{multline}
As expected, the replica-symmetric part of the source $\xi_0 \equiv \tr_R \hat{\xi}$ couples to the replica-symmetric density, while the replicon part of the source $\hat{\Xi} = \hat{\xi} - \xi_0$ couples to the replicon modes.

\paragraph{Gaussian integration.}
As the final step of the derivation, we collect all $\Theta$-dependent terms, Eqs.~\eqref{eq:Action:Measurement:ThetaExpansion}, \eqref{eq:Action:Dynamic:ThetaExpansion}, and \eqref{eq:Action:Source:ThetaExpansion}, and perform Gaussian integration over $\hat{\Theta}$ mode to obtain the effective action for $\hat{\Phi}$ fields. We see that $\tr_K (\hat{Q}_0 \hat{\tau}_z)$ factors cancel out, yielding:
\begin{multline}
\int{\cal D}\Theta\exp\left[i\left(\gamma{\cal L}_{M}+\delta{\cal L}_{\text{dyn}}+\delta{\cal L}_{\text{source}}\right)\right]\\
=\exp\left[\frac{\rho_{0}(1-\rho_{0})}{2\gamma}\tr_{R}\left(\hat{U}^{-1}\partial_{t}\hat{U}+\frac{i}{2}\left(\hat{\Xi}+U^{-1}\hat{\Xi}\hat{U}\right)\right)^{2}\right]\\
=\exp\left[-\rho_{0} (1-\rho_{0}) \tau_0\tr_{R}\left(\partial^{\Xi}_{t}\hat{U}(\partial^{\Xi}_{t}\hat{U})^{\dagger}\right)\right],
\label{eq:E8}
\end{multline}
with
\begin{align}
\partial^\Xi_t \hat{U} &= \partial_{t}\hat{U}+\frac{i}{2}\{\hat{U},\hat{\Xi}\},\\
(\partial^\Xi_t \hat{U})^\dagger &= \partial_{t}\hat{U}^{-1}-\frac{i}{2}\left\{ \hat{U}^{-1},\hat{\Xi}\right\}. 
\end{align}
Combining Eqs.~(\ref{eq:E5}) and (\ref{eq:E8}), we arrive at Eq.~(\ref{LPhi}) of the main text.

\section{Renormalization-group equations for \textit{SU(R)} NLSM with boundary}
\label{sec:appendix:RG}

The renormalized $SU(R)$ NLSM can be parametrized by two running coupling constants, $g$ and $Z_s$, as follows:
\begin{multline}
\label{eq:RenormalizedAction}
iS[\hat{U}]=-\frac{g}{2}\int d^{2}\boldsymbol{x}\tr\left[\partial_{\mu}\hat{U}\partial_{\mu}\hat{U}^{\dagger}\right]\\
+gZ_{s}\int dx\tr\left[\partial_{t}\hat{\Phi}(x,t=0)\hat{\Xi}(x)\right].
\end{multline}
Here, we use the dimensionless units $t \mapsto t/\tau_0$ and $x \mapsto x/l_0$, with the absorbing boundary in the time domain fixed, for simplicity, at $t_f = 0$. Utilizing the background-field method, for a single RG step, we perform a splitting of ``fast'' and ``slow'' modes as $\hat{U}=\hat{U}_{f}\hat{U}_{0}$, so that interaction vertices that couple fast and slow modes read:
\begin{equation}
{\cal L}_{\text{int}}=-g\tr\left[\hat{W}_{\mu}\partial_{\mu}\hat{U}_{0}\hat{U}_{0}^{\dagger}\right],\quad \hat{W}_{\mu}\equiv-i\hat{U}_{f}^{\dagger}\partial_{\mu}\hat{U}_{f},
\end{equation}
with the following perturbative expansion:
\begin{equation}
\hat{W}_{\mu}\approx\partial_{\mu}\hat{\Phi}_{f}-\frac{i}{2}\left[\hat{\Phi}_{f},\partial_{\mu}\hat{\Phi}_{f}\right]-\frac{1}{6}\left[\hat{\Phi}_{f}, \left[\hat{\Phi}_{f}, \partial_\mu \hat{\Phi}_{f}\right]\right].
\end{equation}

\paragraph{Bulk renormalization.}
The coupling constant $g$ is defined in the bulk (in the time domain). The renormalization then comes from the second-order perturbation with two interaction vertices quadratic in fast modes. Neglecting the boundary, the effective action reads:
\begin{multline*}
    iS_{\text{eff}}^{(1)}=\frac{g^{2}}{8}\int d\boldsymbol{r}_{1}d\boldsymbol{r}_{2}\Big\llangle \tr\left(\left[\hat{\Phi}_{f},\partial_{\mu}\hat{\Phi}_{f}\right]\partial_{\mu}\hat{U}_{0}\hat{U}_{0}^{\dagger}\right)_{\boldsymbol{r}_{1}}\\
    \times \tr\left(\left[\hat{\Phi}_{f},\partial_{\nu}\hat{\Phi}_{f}\right]\partial_{\nu}\hat{U}_{0}\hat{U}_{0}^{\dagger}\right)_{\boldsymbol{r}_{2}}\Big\rrangle.
\end{multline*}
Performing the Wick contraction, and switching the integration to the ``center of mass'' $\boldsymbol{R} = (\boldsymbol{r}_1 + \boldsymbol{r}_2) / 2$ and relative motion $\boldsymbol{\rho} = \boldsymbol{r}_1 - \boldsymbol{r}_2$, we obtain:
\begin{multline}
iS^{(1)}_{\text{eff}}\approx\frac{g^{2}R}{4}\int d\boldsymbol{\rho}(\partial_{\mu}G_{f}(\boldsymbol{\rho}))^{2}\int d\boldsymbol{R}\tr\left(\partial_{\mu}\hat{U}_{0}\partial_{\mu}\hat{U}_{0}^{\dagger}\right)\\
\approx\frac{R}{8\pi}\ln\frac{\Lambda}{\Lambda^{\prime}}\int d\boldsymbol{R}\tr\left(\partial_{\mu}\hat{U}_{0}\partial_{\mu}\hat{U}_{0}^{\dagger}\right),
\end{multline}
where $\Lambda$ and $\Lambda^\prime$ are the ultraviolet cutoffs before and after the renormalization step. This gives the RG equation for $g$:
\begin{equation}
-\frac{\partial g}{\partial\ln\Lambda}=-\frac{R}{4\pi} + O(1/g)
\end{equation}

\paragraph{Boundary renormalization.}
The renormalization of the density source $\Xi$ at the boundary $t=0$ comes from the second order of perturbation theory, with one fast field put at the boundary and a single cubic interaction vertex:
\begin{multline}
    i S^{(2)}_{\text{eff}}=-\frac{i}{6}g^{2}Z_{s}\int d\boldsymbol{r}_{1}dx_{2}\\
    \left\llangle \tr\left(\left[\hat{\Phi}_{f}, \left[\hat{\Phi}_{f}, \partial_\mu \hat{\Phi}_{f}\right]\right]\partial_{t}\hat{U}_{0}\hat{U}_{0}^{\dagger}\right)_{\boldsymbol{r}_{1}}\tr\left(\partial_{t}\hat{\Phi}_{f}\hat{\Xi}\right)_{x_{2}}\right\rrangle. 
\end{multline}
After performing the Wick contraction and introducing ``center-of-mass'' coordinate $X = (x_1 + x_2) / 2$ and relative motion coordinates $\boldsymbol{\rho} = (\rho = x_1 - x_2, t_1)$, this becomes
\begin{multline}
iS^{(2)}_{\text{eff}}=-g^{2}Z_{s}R\int d\boldsymbol{\rho}\left[\partial_{t^{\prime}}G_{f}(0,t,t^{\prime})\partial_{t^{\prime\prime}}G(\rho,t,t^{\prime\prime})\right]_{t^{\prime\prime}\to0}^{t^{\prime}\to t}\\
\times \int dX\tr\left(\partial_{t}\hat{\Phi}\hat{\Xi}\right).
\end{multline}
Such an integral would be zero without the boundary, as it would be impossible to satisfy the frequency conservation laws; however, in the presence of the boundary this term also contains the logarithmic divergence, yielding
\begin{equation}
iS^{(2)}_{\text{eff}}=-\frac{Z_{s}R}{4\pi}\ln\frac{\Lambda}{\Lambda^{\prime}}\int dX\tr\left(\partial_{t}\hat{\Phi}\hat{\Xi}\right)
\end{equation}
Interestingly, within the chosen parametrization of the renormalized action, Eq.~\eqref{eq:RenormalizedAction}, this correction is completely absorbed into the renormalization of $g$ and, thus, within the one-loop approximation, the renormalization of $Z_s$ is absent:
\begin{equation}
-\frac{d\ln Z_{s}}{d\ln\Lambda}=0 + O(1/g^2).
\end{equation}

\bibliography{refs}

\begin{thebibliography}{96}%
\makeatletter
\providecommand \@ifxundefined [1]{%
 \@ifx{#1\undefined}
}%
\providecommand \@ifnum [1]{%
 \ifnum #1\expandafter \@firstoftwo
 \else \expandafter \@secondoftwo
 \fi
}%
\providecommand \@ifx [1]{%
 \ifx #1\expandafter \@firstoftwo
 \else \expandafter \@secondoftwo
 \fi
}%
\providecommand \natexlab [1]{#1}%
\providecommand \enquote  [1]{``#1''}%
\providecommand \bibnamefont  [1]{#1}%
\providecommand \bibfnamefont [1]{#1}%
\providecommand \citenamefont [1]{#1}%
\providecommand \href@noop [0]{\@secondoftwo}%
\providecommand \href [0]{\begingroup \@sanitize@url \@href}%
\providecommand \@href[1]{\@@startlink{#1}\@@href}%
\providecommand \@@href[1]{\endgroup#1\@@endlink}%
\providecommand \@sanitize@url [0]{\catcode `\\12\catcode `\$12\catcode
  `\&12\catcode `\#12\catcode `\^12\catcode `\_12\catcode `\%12\relax}%
\providecommand \@@startlink[1]{}%
\providecommand \@@endlink[0]{}%
\providecommand \url  [0]{\begingroup\@sanitize@url \@url }%
\providecommand \@url [1]{\endgroup\@href {#1}{\urlprefix }}%
\providecommand \urlprefix  [0]{URL }%
\providecommand \Eprint [0]{\href }%
\providecommand \doibase [0]{https://doi.org/}%
\providecommand \selectlanguage [0]{\@gobble}%
\providecommand \bibinfo  [0]{\@secondoftwo}%
\providecommand \bibfield  [0]{\@secondoftwo}%
\providecommand \translation [1]{[#1]}%
\providecommand \BibitemOpen [0]{}%
\providecommand \bibitemStop [0]{}%
\providecommand \bibitemNoStop [0]{.\EOS\space}%
\providecommand \EOS [0]{\spacefactor3000\relax}%
\providecommand \BibitemShut  [1]{\csname bibitem#1\endcsname}%
\let\auto@bib@innerbib\@empty
%</preamble>
\bibitem [{\citenamefont {Aharonov}(2000)}]{Aharonov2000a}%
  \BibitemOpen
  \bibfield  {author} {\bibinfo {author} {\bibfnamefont {D.}~\bibnamefont
  {Aharonov}},\ }\bibfield  {title} {\bibinfo {title} {Quantum to classical
  phase transition in noisy quantum computers},\ }\href
  {https://doi.org/10.1103/PhysRevA.62.062311} {\bibfield  {journal} {\bibinfo
  {journal} {Phys. Rev. A}\ }\textbf {\bibinfo {volume} {62}},\ \bibinfo
  {pages} {062311} (\bibinfo {year} {2000})}\BibitemShut {NoStop}%
\bibitem [{\citenamefont {Preskill}(2018)}]{Preskill2018a}%
  \BibitemOpen
  \bibfield  {author} {\bibinfo {author} {\bibfnamefont {J.}~\bibnamefont
  {Preskill}},\ }\bibfield  {title} {\bibinfo {title} {Quantum computing in the
  {NISQ} era and beyond},\ }\href {https://doi.org/10.22331/q-2018-08-06-79}
  {\bibfield  {journal} {\bibinfo  {journal} {{Quantum}}\ }\textbf {\bibinfo
  {volume} {2}},\ \bibinfo {pages} {79} (\bibinfo {year} {2018})}\BibitemShut
  {NoStop}%
\bibitem [{\citenamefont {Bharti~\textit{et~al.}}(2022)}]{Bharti2022}%
  \BibitemOpen
  \bibfield  {author} {\bibinfo {author} {\bibfnamefont {K.}~\bibnamefont
  {Bharti~\textit{et~al.}}},\ }\bibfield  {title} {\bibinfo {title} {Noisy
  intermediate-scale quantum algorithms},\ }\href
  {https://doi.org/10.1103/RevModPhys.94.015004} {\bibfield  {journal}
  {\bibinfo  {journal} {Reviews of Modern Physics}\ }\textbf {\bibinfo {volume}
  {94}},\ \bibinfo {pages} {015004} (\bibinfo {year} {2022})}\BibitemShut
  {NoStop}%
\bibitem [{\citenamefont {Li}\ \emph {et~al.}(2018)\citenamefont {Li},
  \citenamefont {Chen},\ and\ \citenamefont {Fisher}}]{Li2018a}%
  \BibitemOpen
  \bibfield  {author} {\bibinfo {author} {\bibfnamefont {Y.}~\bibnamefont
  {Li}}, \bibinfo {author} {\bibfnamefont {X.}~\bibnamefont {Chen}},\ and\
  \bibinfo {author} {\bibfnamefont {M.~P.~A.}\ \bibnamefont {Fisher}},\
  }\bibfield  {title} {\bibinfo {title} {Quantum {Z}eno effect and the
  many-body entanglement transition},\ }\href
  {https://doi.org/10.1103/PhysRevB.98.205136} {\bibfield  {journal} {\bibinfo
  {journal} {Phys. Rev. B}\ }\textbf {\bibinfo {volume} {98}},\ \bibinfo
  {pages} {205136} (\bibinfo {year} {2018})}\BibitemShut {NoStop}%
\bibitem [{\citenamefont {Skinner}\ \emph {et~al.}(2019)\citenamefont
  {Skinner}, \citenamefont {Ruhman},\ and\ \citenamefont
  {Nahum}}]{Skinner2019a}%
  \BibitemOpen
  \bibfield  {author} {\bibinfo {author} {\bibfnamefont {B.}~\bibnamefont
  {Skinner}}, \bibinfo {author} {\bibfnamefont {J.}~\bibnamefont {Ruhman}},\
  and\ \bibinfo {author} {\bibfnamefont {A.}~\bibnamefont {Nahum}},\ }\bibfield
   {title} {\bibinfo {title} {Measurement-induced phase transitions in the
  dynamics of entanglement},\ }\href
  {https://doi.org/10.1103/PhysRevX.9.031009} {\bibfield  {journal} {\bibinfo
  {journal} {Phys. Rev. X}\ }\textbf {\bibinfo {volume} {9}},\ \bibinfo {pages}
  {031009} (\bibinfo {year} {2019})}\BibitemShut {NoStop}%
\bibitem [{\citenamefont {Chan}\ \emph {et~al.}(2019)\citenamefont {Chan},
  \citenamefont {Nandkishore}, \citenamefont {Pretko},\ and\ \citenamefont
  {Smith}}]{Chan2019a}%
  \BibitemOpen
  \bibfield  {author} {\bibinfo {author} {\bibfnamefont {A.}~\bibnamefont
  {Chan}}, \bibinfo {author} {\bibfnamefont {R.~M.}\ \bibnamefont
  {Nandkishore}}, \bibinfo {author} {\bibfnamefont {M.}~\bibnamefont
  {Pretko}},\ and\ \bibinfo {author} {\bibfnamefont {G.}~\bibnamefont
  {Smith}},\ }\bibfield  {title} {\bibinfo {title} {Unitary-projective
  entanglement dynamics},\ }\href {https://doi.org/10.1103/PhysRevB.99.224307}
  {\bibfield  {journal} {\bibinfo  {journal} {Phys. Rev. B}\ }\textbf {\bibinfo
  {volume} {99}},\ \bibinfo {pages} {224307} (\bibinfo {year}
  {2019})}\BibitemShut {NoStop}%
\bibitem [{\citenamefont {Szyniszewski}\ \emph {et~al.}(2019)\citenamefont
  {Szyniszewski}, \citenamefont {Romito},\ and\ \citenamefont
  {Schomerus}}]{Szyniszewski2019a}%
  \BibitemOpen
  \bibfield  {author} {\bibinfo {author} {\bibfnamefont {M.}~\bibnamefont
  {Szyniszewski}}, \bibinfo {author} {\bibfnamefont {A.}~\bibnamefont
  {Romito}},\ and\ \bibinfo {author} {\bibfnamefont {H.}~\bibnamefont
  {Schomerus}},\ }\bibfield  {title} {\bibinfo {title} {Entanglement transition
  from variable-strength weak measurements},\ }\href
  {https://doi.org/10.1103/PhysRevB.100.064204} {\bibfield  {journal} {\bibinfo
   {journal} {Phys. Rev. B}\ }\textbf {\bibinfo {volume} {100}},\ \bibinfo
  {pages} {064204} (\bibinfo {year} {2019})}\BibitemShut {NoStop}%
\bibitem [{\citenamefont {Li}\ \emph {et~al.}(2019)\citenamefont {Li},
  \citenamefont {Chen},\ and\ \citenamefont {Fisher}}]{Li2019a}%
  \BibitemOpen
  \bibfield  {author} {\bibinfo {author} {\bibfnamefont {Y.}~\bibnamefont
  {Li}}, \bibinfo {author} {\bibfnamefont {X.}~\bibnamefont {Chen}},\ and\
  \bibinfo {author} {\bibfnamefont {M.~P.~A.}\ \bibnamefont {Fisher}},\
  }\bibfield  {title} {\bibinfo {title} {Measurement-driven entanglement
  transition in hybrid quantum circuits},\ }\href
  {https://doi.org/10.1103/PhysRevB.100.134306} {\bibfield  {journal} {\bibinfo
   {journal} {Phys. Rev. B}\ }\textbf {\bibinfo {volume} {100}},\ \bibinfo
  {pages} {134306} (\bibinfo {year} {2019})}\BibitemShut {NoStop}%
\bibitem [{\citenamefont {Bao}\ \emph {et~al.}(2020)\citenamefont {Bao},
  \citenamefont {Choi},\ and\ \citenamefont {Altman}}]{Bao2020a}%
  \BibitemOpen
  \bibfield  {author} {\bibinfo {author} {\bibfnamefont {Y.}~\bibnamefont
  {Bao}}, \bibinfo {author} {\bibfnamefont {S.}~\bibnamefont {Choi}},\ and\
  \bibinfo {author} {\bibfnamefont {E.}~\bibnamefont {Altman}},\ }\bibfield
  {title} {\bibinfo {title} {Theory of the phase transition in random unitary
  circuits with measurements},\ }\href
  {https://doi.org/10.1103/PhysRevB.101.104301} {\bibfield  {journal} {\bibinfo
   {journal} {Phys. Rev. B}\ }\textbf {\bibinfo {volume} {101}},\ \bibinfo
  {pages} {104301} (\bibinfo {year} {2020})}\BibitemShut {NoStop}%
\bibitem [{\citenamefont {Choi}\ \emph {et~al.}(2020)\citenamefont {Choi},
  \citenamefont {Bao}, \citenamefont {Qi},\ and\ \citenamefont
  {Altman}}]{Choi2020a}%
  \BibitemOpen
  \bibfield  {author} {\bibinfo {author} {\bibfnamefont {S.}~\bibnamefont
  {Choi}}, \bibinfo {author} {\bibfnamefont {Y.}~\bibnamefont {Bao}}, \bibinfo
  {author} {\bibfnamefont {X.-L.}\ \bibnamefont {Qi}},\ and\ \bibinfo {author}
  {\bibfnamefont {E.}~\bibnamefont {Altman}},\ }\bibfield  {title} {\bibinfo
  {title} {Quantum error correction in scrambling dynamics and
  measurement-induced phase transition},\ }\href
  {https://doi.org/10.1103/PhysRevLett.125.030505} {\bibfield  {journal}
  {\bibinfo  {journal} {Phys. Rev. Lett.}\ }\textbf {\bibinfo {volume} {125}},\
  \bibinfo {pages} {030505} (\bibinfo {year} {2020})}\BibitemShut {NoStop}%
\bibitem [{\citenamefont {Gullans}\ and\ \citenamefont
  {Huse}(2020{\natexlab{a}})}]{Gullans2020a}%
  \BibitemOpen
  \bibfield  {author} {\bibinfo {author} {\bibfnamefont {M.~J.}\ \bibnamefont
  {Gullans}}\ and\ \bibinfo {author} {\bibfnamefont {D.~A.}\ \bibnamefont
  {Huse}},\ }\bibfield  {title} {\bibinfo {title} {Dynamical purification phase
  transition induced by quantum measurements},\ }\href
  {https://doi.org/10.1103/PhysRevX.10.041020} {\bibfield  {journal} {\bibinfo
  {journal} {Phys. Rev. X}\ }\textbf {\bibinfo {volume} {10}},\ \bibinfo
  {pages} {041020} (\bibinfo {year} {2020}{\natexlab{a}})}\BibitemShut
  {NoStop}%
\bibitem [{\citenamefont {Gullans}\ and\ \citenamefont
  {Huse}(2020{\natexlab{b}})}]{Gullans2020b}%
  \BibitemOpen
  \bibfield  {author} {\bibinfo {author} {\bibfnamefont {M.~J.}\ \bibnamefont
  {Gullans}}\ and\ \bibinfo {author} {\bibfnamefont {D.~A.}\ \bibnamefont
  {Huse}},\ }\bibfield  {title} {\bibinfo {title} {Scalable probes of
  measurement-induced criticality},\ }\href
  {https://doi.org/10.1103/PhysRevLett.125.070606} {\bibfield  {journal}
  {\bibinfo  {journal} {Phys. Rev. Lett.}\ }\textbf {\bibinfo {volume} {125}},\
  \bibinfo {pages} {070606} (\bibinfo {year} {2020}{\natexlab{b}})}\BibitemShut
  {NoStop}%
\bibitem [{\citenamefont {Jian}\ \emph {et~al.}(2020)\citenamefont {Jian},
  \citenamefont {You}, \citenamefont {Vasseur},\ and\ \citenamefont
  {Ludwig}}]{Jian2020a}%
  \BibitemOpen
  \bibfield  {author} {\bibinfo {author} {\bibfnamefont {C.-M.}\ \bibnamefont
  {Jian}}, \bibinfo {author} {\bibfnamefont {Y.-Z.}\ \bibnamefont {You}},
  \bibinfo {author} {\bibfnamefont {R.}~\bibnamefont {Vasseur}},\ and\ \bibinfo
  {author} {\bibfnamefont {A.~W.~W.}\ \bibnamefont {Ludwig}},\ }\bibfield
  {title} {\bibinfo {title} {Measurement-induced criticality in random quantum
  circuits},\ }\href {https://doi.org/10.1103/PhysRevB.101.104302} {\bibfield
  {journal} {\bibinfo  {journal} {Phys. Rev. B}\ }\textbf {\bibinfo {volume}
  {101}},\ \bibinfo {pages} {104302} (\bibinfo {year} {2020})}\BibitemShut
  {NoStop}%
\bibitem [{\citenamefont {Zabalo}\ \emph {et~al.}(2020)\citenamefont {Zabalo},
  \citenamefont {Gullans}, \citenamefont {Wilson}, \citenamefont
  {Gopalakrishnan}, \citenamefont {Huse},\ and\ \citenamefont
  {Pixley}}]{Zabalo2020a}%
  \BibitemOpen
  \bibfield  {author} {\bibinfo {author} {\bibfnamefont {A.}~\bibnamefont
  {Zabalo}}, \bibinfo {author} {\bibfnamefont {M.~J.}\ \bibnamefont {Gullans}},
  \bibinfo {author} {\bibfnamefont {J.~H.}\ \bibnamefont {Wilson}}, \bibinfo
  {author} {\bibfnamefont {S.}~\bibnamefont {Gopalakrishnan}}, \bibinfo
  {author} {\bibfnamefont {D.~A.}\ \bibnamefont {Huse}},\ and\ \bibinfo
  {author} {\bibfnamefont {J.~H.}\ \bibnamefont {Pixley}},\ }\bibfield  {title}
  {\bibinfo {title} {Critical properties of the measurement-induced transition
  in random quantum circuits},\ }\href
  {https://doi.org/10.1103/PhysRevB.101.060301} {\bibfield  {journal} {\bibinfo
   {journal} {Phys. Rev. B}\ }\textbf {\bibinfo {volume} {101}},\ \bibinfo
  {pages} {060301(R)} (\bibinfo {year} {2020})}\BibitemShut {NoStop}%
\bibitem [{\citenamefont {Iaconis}\ \emph {et~al.}(2020)\citenamefont
  {Iaconis}, \citenamefont {Lucas},\ and\ \citenamefont {Chen}}]{Iaconis2020a}%
  \BibitemOpen
  \bibfield  {author} {\bibinfo {author} {\bibfnamefont {J.}~\bibnamefont
  {Iaconis}}, \bibinfo {author} {\bibfnamefont {A.}~\bibnamefont {Lucas}},\
  and\ \bibinfo {author} {\bibfnamefont {X.}~\bibnamefont {Chen}},\ }\bibfield
  {title} {\bibinfo {title} {Measurement-induced phase transitions in quantum
  automaton circuits},\ }\href {https://doi.org/10.1103/PhysRevB.102.224311}
  {\bibfield  {journal} {\bibinfo  {journal} {Phys. Rev. B}\ }\textbf {\bibinfo
  {volume} {102}},\ \bibinfo {pages} {224311} (\bibinfo {year}
  {2020})}\BibitemShut {NoStop}%
\bibitem [{\citenamefont {Turkeshi}\ \emph {et~al.}(2020)\citenamefont
  {Turkeshi}, \citenamefont {Fazio},\ and\ \citenamefont
  {Dalmonte}}]{Turkeshi2020a}%
  \BibitemOpen
  \bibfield  {author} {\bibinfo {author} {\bibfnamefont {X.}~\bibnamefont
  {Turkeshi}}, \bibinfo {author} {\bibfnamefont {R.}~\bibnamefont {Fazio}},\
  and\ \bibinfo {author} {\bibfnamefont {M.}~\bibnamefont {Dalmonte}},\
  }\bibfield  {title} {\bibinfo {title} {Measurement-induced criticality in
  $(2+1)$-dimensional hybrid quantum circuits},\ }\href
  {https://doi.org/10.1103/PhysRevB.102.014315} {\bibfield  {journal} {\bibinfo
   {journal} {Phys. Rev. B}\ }\textbf {\bibinfo {volume} {102}},\ \bibinfo
  {pages} {014315} (\bibinfo {year} {2020})}\BibitemShut {NoStop}%
\bibitem [{\citenamefont {Zhang}\ \emph {et~al.}(2020)\citenamefont {Zhang},
  \citenamefont {Reyes}, \citenamefont {Kourtis}, \citenamefont {Chamon},
  \citenamefont {Mucciolo},\ and\ \citenamefont {Ruckenstein}}]{Zhang2020c}%
  \BibitemOpen
  \bibfield  {author} {\bibinfo {author} {\bibfnamefont {L.}~\bibnamefont
  {Zhang}}, \bibinfo {author} {\bibfnamefont {J.~A.}\ \bibnamefont {Reyes}},
  \bibinfo {author} {\bibfnamefont {S.}~\bibnamefont {Kourtis}}, \bibinfo
  {author} {\bibfnamefont {C.}~\bibnamefont {Chamon}}, \bibinfo {author}
  {\bibfnamefont {E.~R.}\ \bibnamefont {Mucciolo}},\ and\ \bibinfo {author}
  {\bibfnamefont {A.~E.}\ \bibnamefont {Ruckenstein}},\ }\bibfield  {title}
  {\bibinfo {title} {Nonuniversal entanglement level statistics in
  projection-driven quantum circuits},\ }\href
  {https://doi.org/10.1103/PhysRevB.101.235104} {\bibfield  {journal} {\bibinfo
   {journal} {Phys. Rev. B}\ }\textbf {\bibinfo {volume} {101}},\ \bibinfo
  {pages} {235104} (\bibinfo {year} {2020})}\BibitemShut {NoStop}%
\bibitem [{\citenamefont {Nahum}\ \emph {et~al.}(2021)\citenamefont {Nahum},
  \citenamefont {Roy}, \citenamefont {Skinner},\ and\ \citenamefont
  {Ruhman}}]{Nahum2021a}%
  \BibitemOpen
  \bibfield  {author} {\bibinfo {author} {\bibfnamefont {A.}~\bibnamefont
  {Nahum}}, \bibinfo {author} {\bibfnamefont {S.}~\bibnamefont {Roy}}, \bibinfo
  {author} {\bibfnamefont {B.}~\bibnamefont {Skinner}},\ and\ \bibinfo {author}
  {\bibfnamefont {J.}~\bibnamefont {Ruhman}},\ }\bibfield  {title} {\bibinfo
  {title} {Measurement and entanglement phase transitions in all-to-all quantum
  circuits, on quantum trees, and in {Landau-Ginsburg} theory},\ }\href
  {https://doi.org/10.1103/PRXQuantum.2.010352} {\bibfield  {journal} {\bibinfo
   {journal} {PRX Quantum}\ }\textbf {\bibinfo {volume} {2}},\ \bibinfo {pages}
  {010352} (\bibinfo {year} {2021})}\BibitemShut {NoStop}%
\bibitem [{\citenamefont {Ippoliti}\ \emph {et~al.}(2021)\citenamefont
  {Ippoliti}, \citenamefont {Gullans}, \citenamefont {Gopalakrishnan},
  \citenamefont {Huse},\ and\ \citenamefont {Khemani}}]{Ippoliti2021a}%
  \BibitemOpen
  \bibfield  {author} {\bibinfo {author} {\bibfnamefont {M.}~\bibnamefont
  {Ippoliti}}, \bibinfo {author} {\bibfnamefont {M.~J.}\ \bibnamefont
  {Gullans}}, \bibinfo {author} {\bibfnamefont {S.}~\bibnamefont
  {Gopalakrishnan}}, \bibinfo {author} {\bibfnamefont {D.~A.}\ \bibnamefont
  {Huse}},\ and\ \bibinfo {author} {\bibfnamefont {V.}~\bibnamefont
  {Khemani}},\ }\bibfield  {title} {\bibinfo {title} {Entanglement phase
  transitions in measurement-only dynamics},\ }\href
  {https://doi.org/10.1103/PhysRevX.11.011030} {\bibfield  {journal} {\bibinfo
  {journal} {Phys. Rev. X}\ }\textbf {\bibinfo {volume} {11}},\ \bibinfo
  {pages} {011030} (\bibinfo {year} {2021})}\BibitemShut {NoStop}%
\bibitem [{\citenamefont {Ippoliti}\ and\ \citenamefont
  {Khemani}(2021)}]{Ippoliti2021b}%
  \BibitemOpen
  \bibfield  {author} {\bibinfo {author} {\bibfnamefont {M.}~\bibnamefont
  {Ippoliti}}\ and\ \bibinfo {author} {\bibfnamefont {V.}~\bibnamefont
  {Khemani}},\ }\bibfield  {title} {\bibinfo {title} {Postselection-free
  entanglement dynamics via spacetime duality},\ }\href
  {https://doi.org/10.1103/PhysRevLett.126.060501} {\bibfield  {journal}
  {\bibinfo  {journal} {Phys. Rev. Lett.}\ }\textbf {\bibinfo {volume} {126}},\
  \bibinfo {pages} {060501} (\bibinfo {year} {2021})}\BibitemShut {NoStop}%
\bibitem [{\citenamefont {Lavasani}\ \emph
  {et~al.}(2021{\natexlab{a}})\citenamefont {Lavasani}, \citenamefont
  {Alavirad},\ and\ \citenamefont {Barkeshli}}]{Lavasani2021a}%
  \BibitemOpen
  \bibfield  {author} {\bibinfo {author} {\bibfnamefont {A.}~\bibnamefont
  {Lavasani}}, \bibinfo {author} {\bibfnamefont {Y.}~\bibnamefont {Alavirad}},\
  and\ \bibinfo {author} {\bibfnamefont {M.}~\bibnamefont {Barkeshli}},\
  }\bibfield  {title} {\bibinfo {title} {Measurement-induced topological
  entanglement transitions in symmetric random quantum circuits},\ }\href
  {https://doi.org/10.1038/s41567-020-01112-z} {\bibfield  {journal} {\bibinfo
  {journal} {Nat. Phys.}\ }\textbf {\bibinfo {volume} {17}},\ \bibinfo {pages}
  {342} (\bibinfo {year} {2021}{\natexlab{a}})}\BibitemShut {NoStop}%
\bibitem [{\citenamefont {Lavasani}\ \emph
  {et~al.}(2021{\natexlab{b}})\citenamefont {Lavasani}, \citenamefont
  {Alavirad},\ and\ \citenamefont {Barkeshli}}]{Lavasani2021b}%
  \BibitemOpen
  \bibfield  {author} {\bibinfo {author} {\bibfnamefont {A.}~\bibnamefont
  {Lavasani}}, \bibinfo {author} {\bibfnamefont {Y.}~\bibnamefont {Alavirad}},\
  and\ \bibinfo {author} {\bibfnamefont {M.}~\bibnamefont {Barkeshli}},\
  }\bibfield  {title} {\bibinfo {title} {Topological order and criticality in
  $(2+1)\mathrm{D}$ monitored random quantum circuits},\ }\href
  {https://doi.org/10.1103/PhysRevLett.127.235701} {\bibfield  {journal}
  {\bibinfo  {journal} {Phys. Rev. Lett.}\ }\textbf {\bibinfo {volume} {127}},\
  \bibinfo {pages} {235701} (\bibinfo {year} {2021}{\natexlab{b}})}\BibitemShut
  {NoStop}%
\bibitem [{\citenamefont {Sang}\ and\ \citenamefont {Hsieh}(2021)}]{Sang2021a}%
  \BibitemOpen
  \bibfield  {author} {\bibinfo {author} {\bibfnamefont {S.}~\bibnamefont
  {Sang}}\ and\ \bibinfo {author} {\bibfnamefont {T.~H.}\ \bibnamefont
  {Hsieh}},\ }\bibfield  {title} {\bibinfo {title} {Measurement-protected
  quantum phases},\ }\href {https://doi.org/10.1103/PhysRevResearch.3.023200}
  {\bibfield  {journal} {\bibinfo  {journal} {Phys. Rev. Research}\ }\textbf
  {\bibinfo {volume} {3}},\ \bibinfo {pages} {023200} (\bibinfo {year}
  {2021})}\BibitemShut {NoStop}%
\bibitem [{\citenamefont {Fisher}\ \emph {et~al.}(2023)\citenamefont {Fisher},
  \citenamefont {Khemani}, \citenamefont {Nahum},\ and\ \citenamefont
  {Vijay}}]{Fisher2022}%
  \BibitemOpen
  \bibfield  {author} {\bibinfo {author} {\bibfnamefont {M.~P.}\ \bibnamefont
  {Fisher}}, \bibinfo {author} {\bibfnamefont {V.}~\bibnamefont {Khemani}},
  \bibinfo {author} {\bibfnamefont {A.}~\bibnamefont {Nahum}},\ and\ \bibinfo
  {author} {\bibfnamefont {S.}~\bibnamefont {Vijay}},\ }\bibfield  {title}
  {\bibinfo {title} {Random quantum circuits},\ }\href
  {https://doi.org/10.1146/annurev-conmatphys-031720-030658} {\bibfield
  {journal} {\bibinfo  {journal} {Annual Review of Condensed Matter Physics}\
  }\textbf {\bibinfo {volume} {14}},\ \bibinfo {pages} {335} (\bibinfo {year}
  {2023})}\BibitemShut {NoStop}%
\bibitem [{\citenamefont {Block}\ \emph {et~al.}(2022)\citenamefont {Block},
  \citenamefont {Bao}, \citenamefont {Choi}, \citenamefont {Altman},\ and\
  \citenamefont {Yao}}]{Block2022a}%
  \BibitemOpen
  \bibfield  {author} {\bibinfo {author} {\bibfnamefont {M.}~\bibnamefont
  {Block}}, \bibinfo {author} {\bibfnamefont {Y.}~\bibnamefont {Bao}}, \bibinfo
  {author} {\bibfnamefont {S.}~\bibnamefont {Choi}}, \bibinfo {author}
  {\bibfnamefont {E.}~\bibnamefont {Altman}},\ and\ \bibinfo {author}
  {\bibfnamefont {N.~Y.}\ \bibnamefont {Yao}},\ }\bibfield  {title} {\bibinfo
  {title} {Measurement-induced transition in long-range interacting quantum
  circuits},\ }\href {https://doi.org/10.1103/PhysRevLett.128.010604}
  {\bibfield  {journal} {\bibinfo  {journal} {Phys. Rev. Lett.}\ }\textbf
  {\bibinfo {volume} {128}},\ \bibinfo {pages} {010604} (\bibinfo {year}
  {2022})}\BibitemShut {NoStop}%
\bibitem [{\citenamefont {Sharma}\ \emph {et~al.}(2022)\citenamefont {Sharma},
  \citenamefont {Turkeshi}, \citenamefont {Fazio},\ and\ \citenamefont
  {Dalmonte}}]{Sharma2022}%
  \BibitemOpen
  \bibfield  {author} {\bibinfo {author} {\bibfnamefont {S.}~\bibnamefont
  {Sharma}}, \bibinfo {author} {\bibfnamefont {X.}~\bibnamefont {Turkeshi}},
  \bibinfo {author} {\bibfnamefont {R.}~\bibnamefont {Fazio}},\ and\ \bibinfo
  {author} {\bibfnamefont {M.}~\bibnamefont {Dalmonte}},\ }\bibfield  {title}
  {\bibinfo {title} {{Measurement-induced criticality in extended and
  long-range unitary circuits}},\ }\href
  {https://doi.org/10.21468/SciPostPhysCore.5.2.023} {\bibfield  {journal}
  {\bibinfo  {journal} {SciPost Phys. Core}\ }\textbf {\bibinfo {volume} {5}},\
  \bibinfo {pages} {023} (\bibinfo {year} {2022})}\BibitemShut {NoStop}%
\bibitem [{\citenamefont {Jian}\ \emph {et~al.}(2023)\citenamefont {Jian},
  \citenamefont {Shapourian}, \citenamefont {Bauer},\ and\ \citenamefont
  {Ludwig}}]{Jian2023}%
  \BibitemOpen
  \bibfield  {author} {\bibinfo {author} {\bibfnamefont {C.-M.}\ \bibnamefont
  {Jian}}, \bibinfo {author} {\bibfnamefont {H.}~\bibnamefont {Shapourian}},
  \bibinfo {author} {\bibfnamefont {B.}~\bibnamefont {Bauer}},\ and\ \bibinfo
  {author} {\bibfnamefont {A.~W.~W.}\ \bibnamefont {Ludwig}},\ }\href
  {https://doi.org/10.48550/arxiv.2302.09094} {\bibinfo {title}
  {Measurement-induced entanglement transitions in quantum circuits of
  non-interacting fermions: {Born}-rule versus forced measurements}} (\bibinfo
  {year} {2023}),\ \Eprint {https://arxiv.org/abs/2302.09094}
  {arXiv:2302.09094} \BibitemShut {NoStop}%
\bibitem [{\citenamefont {Kelly}\ \emph {et~al.}(2023)\citenamefont {Kelly},
  \citenamefont {Poschinger}, \citenamefont {Schmidt-Kaler}, \citenamefont
  {Fisher},\ and\ \citenamefont {Marino}}]{Kelly2023}%
  \BibitemOpen
  \bibfield  {author} {\bibinfo {author} {\bibfnamefont {S.~P.}\ \bibnamefont
  {Kelly}}, \bibinfo {author} {\bibfnamefont {U.}~\bibnamefont {Poschinger}},
  \bibinfo {author} {\bibfnamefont {F.}~\bibnamefont {Schmidt-Kaler}}, \bibinfo
  {author} {\bibfnamefont {M.~P.~A.}\ \bibnamefont {Fisher}},\ and\ \bibinfo
  {author} {\bibfnamefont {J.}~\bibnamefont {Marino}},\ }\href@noop {}
  {\bibinfo {title} {Coherence requirements for quantum communication from
  hybrid circuit dynamics}} (\bibinfo {year} {2023}),\ \Eprint
  {https://arxiv.org/abs/2210.11547} {arXiv:2210.11547 [quant-ph]} \BibitemShut
  {NoStop}%
\bibitem [{\citenamefont {Cao}\ \emph {et~al.}(2019)\citenamefont {Cao},
  \citenamefont {Tilloy},\ and\ \citenamefont {{De~Luca}}}]{Cao2019a}%
  \BibitemOpen
  \bibfield  {author} {\bibinfo {author} {\bibfnamefont {X.}~\bibnamefont
  {Cao}}, \bibinfo {author} {\bibfnamefont {A.}~\bibnamefont {Tilloy}},\ and\
  \bibinfo {author} {\bibfnamefont {A.}~\bibnamefont {{De~Luca}}},\ }\bibfield
  {title} {\bibinfo {title} {Entanglement in a fermion chain under continuous
  monitoring},\ }\href {https://doi.org/10.21468/SciPostPhys.7.2.024}
  {\bibfield  {journal} {\bibinfo  {journal} {SciPost Phys.}\ }\textbf
  {\bibinfo {volume} {7}},\ \bibinfo {pages} {024} (\bibinfo {year}
  {2019})}\BibitemShut {NoStop}%
\bibitem [{\citenamefont {Alberton}\ \emph {et~al.}(2021)\citenamefont
  {Alberton}, \citenamefont {Buchhold},\ and\ \citenamefont
  {Diehl}}]{Alberton2021a}%
  \BibitemOpen
  \bibfield  {author} {\bibinfo {author} {\bibfnamefont {O.}~\bibnamefont
  {Alberton}}, \bibinfo {author} {\bibfnamefont {M.}~\bibnamefont {Buchhold}},\
  and\ \bibinfo {author} {\bibfnamefont {S.}~\bibnamefont {Diehl}},\ }\bibfield
   {title} {\bibinfo {title} {Entanglement transition in a monitored
  free-fermion chain: {F}rom extended criticality to area law},\ }\href
  {https://doi.org/10.1103/PhysRevLett.126.170602} {\bibfield  {journal}
  {\bibinfo  {journal} {Phys. Rev. Lett.}\ }\textbf {\bibinfo {volume} {126}},\
  \bibinfo {pages} {170602} (\bibinfo {year} {2021})}\BibitemShut {NoStop}%
\bibitem [{\citenamefont {Chen}\ \emph {et~al.}(2020)\citenamefont {Chen},
  \citenamefont {Li}, \citenamefont {Fisher},\ and\ \citenamefont
  {Lucas}}]{Chen2020a}%
  \BibitemOpen
  \bibfield  {author} {\bibinfo {author} {\bibfnamefont {X.}~\bibnamefont
  {Chen}}, \bibinfo {author} {\bibfnamefont {Y.}~\bibnamefont {Li}}, \bibinfo
  {author} {\bibfnamefont {M.~P.~A.}\ \bibnamefont {Fisher}},\ and\ \bibinfo
  {author} {\bibfnamefont {A.}~\bibnamefont {Lucas}},\ }\bibfield  {title}
  {\bibinfo {title} {Emergent conformal symmetry in nonunitary random dynamics
  of free fermions},\ }\href {https://doi.org/10.1103/PhysRevResearch.2.033017}
  {\bibfield  {journal} {\bibinfo  {journal} {Phys. Rev. Research}\ }\textbf
  {\bibinfo {volume} {2}},\ \bibinfo {pages} {033017} (\bibinfo {year}
  {2020})}\BibitemShut {NoStop}%
\bibitem [{\citenamefont {Tang}\ \emph {et~al.}(2021)\citenamefont {Tang},
  \citenamefont {Chen},\ and\ \citenamefont {Zhu}}]{Tang2021a}%
  \BibitemOpen
  \bibfield  {author} {\bibinfo {author} {\bibfnamefont {Q.}~\bibnamefont
  {Tang}}, \bibinfo {author} {\bibfnamefont {X.}~\bibnamefont {Chen}},\ and\
  \bibinfo {author} {\bibfnamefont {W.}~\bibnamefont {Zhu}},\ }\bibfield
  {title} {\bibinfo {title} {Quantum criticality in the nonunitary dynamics of
  $(2+1)$-dimensional free fermions},\ }\href
  {https://doi.org/10.1103/PhysRevB.103.174303} {\bibfield  {journal} {\bibinfo
   {journal} {Phys. Rev. B}\ }\textbf {\bibinfo {volume} {103}},\ \bibinfo
  {pages} {174303} (\bibinfo {year} {2021})}\BibitemShut {NoStop}%
\bibitem [{\citenamefont {Agrawal}\ \emph {et~al.}(2022)\citenamefont
  {Agrawal}, \citenamefont {Zabalo}, \citenamefont {Chen}, \citenamefont
  {Wilson}, \citenamefont {Potter}, \citenamefont {Pixley}, \citenamefont
  {Gopalakrishnan},\ and\ \citenamefont {Vasseur}}]{Agrawal2022}%
  \BibitemOpen
  \bibfield  {author} {\bibinfo {author} {\bibfnamefont {U.}~\bibnamefont
  {Agrawal}}, \bibinfo {author} {\bibfnamefont {A.}~\bibnamefont {Zabalo}},
  \bibinfo {author} {\bibfnamefont {K.}~\bibnamefont {Chen}}, \bibinfo {author}
  {\bibfnamefont {J.~H.}\ \bibnamefont {Wilson}}, \bibinfo {author}
  {\bibfnamefont {A.~C.}\ \bibnamefont {Potter}}, \bibinfo {author}
  {\bibfnamefont {J.~H.}\ \bibnamefont {Pixley}}, \bibinfo {author}
  {\bibfnamefont {S.}~\bibnamefont {Gopalakrishnan}},\ and\ \bibinfo {author}
  {\bibfnamefont {R.}~\bibnamefont {Vasseur}},\ }\bibfield  {title} {\bibinfo
  {title} {Entanglement and charge-sharpening transitions in u(1) symmetric
  monitored quantum circuits},\ }\href
  {https://doi.org/10.1103/PhysRevX.12.041002} {\bibfield  {journal} {\bibinfo
  {journal} {Phys. Rev. X}\ }\textbf {\bibinfo {volume} {12}},\ \bibinfo
  {pages} {041002} (\bibinfo {year} {2022})}\BibitemShut {NoStop}%
\bibitem [{\citenamefont {Barratt}\ \emph {et~al.}(2022)\citenamefont
  {Barratt}, \citenamefont {Agrawal}, \citenamefont {Gopalakrishnan},
  \citenamefont {Huse}, \citenamefont {Vasseur},\ and\ \citenamefont
  {Potter}}]{Barratt2022}%
  \BibitemOpen
  \bibfield  {author} {\bibinfo {author} {\bibfnamefont {F.}~\bibnamefont
  {Barratt}}, \bibinfo {author} {\bibfnamefont {U.}~\bibnamefont {Agrawal}},
  \bibinfo {author} {\bibfnamefont {S.}~\bibnamefont {Gopalakrishnan}},
  \bibinfo {author} {\bibfnamefont {D.~A.}\ \bibnamefont {Huse}}, \bibinfo
  {author} {\bibfnamefont {R.}~\bibnamefont {Vasseur}},\ and\ \bibinfo {author}
  {\bibfnamefont {A.~C.}\ \bibnamefont {Potter}},\ }\bibfield  {title}
  {\bibinfo {title} {Field theory of charge sharpening in symmetric monitored
  quantum circuits},\ }\href {https://doi.org/10.1103/PhysRevLett.129.120604}
  {\bibfield  {journal} {\bibinfo  {journal} {Phys. Rev. Lett.}\ }\textbf
  {\bibinfo {volume} {129}},\ \bibinfo {pages} {120604} (\bibinfo {year}
  {2022})}\BibitemShut {NoStop}%
\bibitem [{\citenamefont {Coppola}\ \emph {et~al.}(2022)\citenamefont
  {Coppola}, \citenamefont {Tirrito}, \citenamefont {Karevski},\ and\
  \citenamefont {Collura}}]{Coppola2022}%
  \BibitemOpen
  \bibfield  {author} {\bibinfo {author} {\bibfnamefont {M.}~\bibnamefont
  {Coppola}}, \bibinfo {author} {\bibfnamefont {E.}~\bibnamefont {Tirrito}},
  \bibinfo {author} {\bibfnamefont {D.}~\bibnamefont {Karevski}},\ and\
  \bibinfo {author} {\bibfnamefont {M.}~\bibnamefont {Collura}},\ }\bibfield
  {title} {\bibinfo {title} {Growth of entanglement entropy under local
  projective measurements},\ }\href
  {https://doi.org/10.1103/PhysRevB.105.094303} {\bibfield  {journal} {\bibinfo
   {journal} {Phys. Rev. B}\ }\textbf {\bibinfo {volume} {105}},\ \bibinfo
  {pages} {094303} (\bibinfo {year} {2022})}\BibitemShut {NoStop}%
\bibitem [{\citenamefont {Ladewig}\ \emph {et~al.}(2022)\citenamefont
  {Ladewig}, \citenamefont {Diehl},\ and\ \citenamefont
  {Buchhold}}]{Ladewig2022}%
  \BibitemOpen
  \bibfield  {author} {\bibinfo {author} {\bibfnamefont {B.}~\bibnamefont
  {Ladewig}}, \bibinfo {author} {\bibfnamefont {S.}~\bibnamefont {Diehl}},\
  and\ \bibinfo {author} {\bibfnamefont {M.}~\bibnamefont {Buchhold}},\
  }\bibfield  {title} {\bibinfo {title} {Monitored open fermion dynamics:
  {E}xploring the interplay of measurement, decoherence, and free {Hamiltonian}
  evolution},\ }\href {https://doi.org/10.1103/PhysRevResearch.4.033001}
  {\bibfield  {journal} {\bibinfo  {journal} {Phys. Rev. Research}\ }\textbf
  {\bibinfo {volume} {4}},\ \bibinfo {pages} {033001} (\bibinfo {year}
  {2022})}\BibitemShut {NoStop}%
\bibitem [{\citenamefont {Carollo}\ and\ \citenamefont
  {Alba}(2022)}]{Carollo2022}%
  \BibitemOpen
  \bibfield  {author} {\bibinfo {author} {\bibfnamefont {F.}~\bibnamefont
  {Carollo}}\ and\ \bibinfo {author} {\bibfnamefont {V.}~\bibnamefont {Alba}},\
  }\bibfield  {title} {\bibinfo {title} {Entangled multiplets and spreading of
  quantum correlations in a continuously monitored tight-binding chain},\
  }\href {https://doi.org/10.1103/PhysRevB.106.L220304} {\bibfield  {journal}
  {\bibinfo  {journal} {Phys. Rev. B}\ }\textbf {\bibinfo {volume} {106}},\
  \bibinfo {pages} {L220304} (\bibinfo {year} {2022})}\BibitemShut {NoStop}%
\bibitem [{\citenamefont {Buchhold}\ \emph {et~al.}(2022)\citenamefont
  {Buchhold}, \citenamefont {M{\"u}ller},\ and\ \citenamefont
  {Diehl}}]{Buchhold2022}%
  \BibitemOpen
  \bibfield  {author} {\bibinfo {author} {\bibfnamefont {M.}~\bibnamefont
  {Buchhold}}, \bibinfo {author} {\bibfnamefont {T.}~\bibnamefont
  {M{\"u}ller}},\ and\ \bibinfo {author} {\bibfnamefont {S.}~\bibnamefont
  {Diehl}},\ }\href {https://doi.org/10.48550/arxiv.2208.10506} {\bibinfo
  {title} {Revealing measurement-induced phase transitions by pre-selection}}
  (\bibinfo {year} {2022}),\ \Eprint {https://arxiv.org/abs/2208.10506}
  {arXiv:2208.10506} \BibitemShut {NoStop}%
\bibitem [{\citenamefont {Yang}\ \emph
  {et~al.}(2023{\natexlab{a}})\citenamefont {Yang}, \citenamefont {Zuo},\ and\
  \citenamefont {Liu}}]{Yang2022}%
  \BibitemOpen
  \bibfield  {author} {\bibinfo {author} {\bibfnamefont {Q.}~\bibnamefont
  {Yang}}, \bibinfo {author} {\bibfnamefont {Y.}~\bibnamefont {Zuo}},\ and\
  \bibinfo {author} {\bibfnamefont {D.~E.}\ \bibnamefont {Liu}},\ }\bibfield
  {title} {\bibinfo {title} {Keldysh nonlinear sigma model for a free-fermion
  gas under continuous measurements},\ }\href
  {https://doi.org/10.1103/PhysRevResearch.5.033174} {\bibfield  {journal}
  {\bibinfo  {journal} {Phys. Rev. Res.}\ }\textbf {\bibinfo {volume} {5}},\
  \bibinfo {pages} {033174} (\bibinfo {year} {2023}{\natexlab{a}})}\BibitemShut
  {NoStop}%
\bibitem [{\citenamefont {Szyniszewski}\ \emph {et~al.}(2023)\citenamefont
  {Szyniszewski}, \citenamefont {Lunt},\ and\ \citenamefont
  {Pal}}]{Szyniszewski2022}%
  \BibitemOpen
  \bibfield  {author} {\bibinfo {author} {\bibfnamefont {M.}~\bibnamefont
  {Szyniszewski}}, \bibinfo {author} {\bibfnamefont {O.}~\bibnamefont {Lunt}},\
  and\ \bibinfo {author} {\bibfnamefont {A.}~\bibnamefont {Pal}},\ }\bibfield
  {title} {\bibinfo {title} {Disordered monitored free fermions},\ }\href
  {https://doi.org/10.1103/PhysRevB.108.165126} {\bibfield  {journal} {\bibinfo
   {journal} {Phys. Rev. B}\ }\textbf {\bibinfo {volume} {108}},\ \bibinfo
  {pages} {165126} (\bibinfo {year} {2023})}\BibitemShut {NoStop}%
\bibitem [{\citenamefont {Buchhold}\ \emph {et~al.}(2021)\citenamefont
  {Buchhold}, \citenamefont {Minoguchi}, \citenamefont {Altland},\ and\
  \citenamefont {Diehl}}]{Buchhold2021a}%
  \BibitemOpen
  \bibfield  {author} {\bibinfo {author} {\bibfnamefont {M.}~\bibnamefont
  {Buchhold}}, \bibinfo {author} {\bibfnamefont {Y.}~\bibnamefont {Minoguchi}},
  \bibinfo {author} {\bibfnamefont {A.}~\bibnamefont {Altland}},\ and\ \bibinfo
  {author} {\bibfnamefont {S.}~\bibnamefont {Diehl}},\ }\bibfield  {title}
  {\bibinfo {title} {Effective theory for the measurement-induced phase
  transition of {Dirac} fermions},\ }\href
  {https://doi.org/10.1103/PhysRevX.11.041004} {\bibfield  {journal} {\bibinfo
  {journal} {Phys. Rev. X}\ }\textbf {\bibinfo {volume} {11}},\ \bibinfo
  {pages} {041004} (\bibinfo {year} {2021})}\BibitemShut {NoStop}%
\bibitem [{\citenamefont {Van~Regemortel}\ \emph {et~al.}(2021)\citenamefont
  {Van~Regemortel}, \citenamefont {Cian}, \citenamefont {Seif}, \citenamefont
  {Dehghani},\ and\ \citenamefont {Hafezi}}]{VanRegemortel2021a}%
  \BibitemOpen
  \bibfield  {author} {\bibinfo {author} {\bibfnamefont {M.}~\bibnamefont
  {Van~Regemortel}}, \bibinfo {author} {\bibfnamefont {Z.-P.}\ \bibnamefont
  {Cian}}, \bibinfo {author} {\bibfnamefont {A.}~\bibnamefont {Seif}}, \bibinfo
  {author} {\bibfnamefont {H.}~\bibnamefont {Dehghani}},\ and\ \bibinfo
  {author} {\bibfnamefont {M.}~\bibnamefont {Hafezi}},\ }\bibfield  {title}
  {\bibinfo {title} {Entanglement entropy scaling transition under competing
  monitoring protocols},\ }\href
  {https://doi.org/10.1103/PhysRevLett.126.123604} {\bibfield  {journal}
  {\bibinfo  {journal} {Phys. Rev. Lett.}\ }\textbf {\bibinfo {volume} {126}},\
  \bibinfo {pages} {123604} (\bibinfo {year} {2021})}\BibitemShut {NoStop}%
\bibitem [{\citenamefont {Gal}\ \emph {et~al.}(2023)\citenamefont {Gal},
  \citenamefont {Turkeshi},\ and\ \citenamefont {Schirò}}]{Youenn2023}%
  \BibitemOpen
  \bibfield  {author} {\bibinfo {author} {\bibfnamefont {Y.~L.}\ \bibnamefont
  {Gal}}, \bibinfo {author} {\bibfnamefont {X.}~\bibnamefont {Turkeshi}},\ and\
  \bibinfo {author} {\bibfnamefont {M.}~\bibnamefont {Schirò}},\ }\bibfield
  {title} {\bibinfo {title} {{Volume-to-area law entanglement transition in a
  non-Hermitian free fermionic chain}},\ }\href
  {https://doi.org/10.21468/SciPostPhys.14.5.138} {\bibfield  {journal}
  {\bibinfo  {journal} {SciPost Phys.}\ }\textbf {\bibinfo {volume} {14}},\
  \bibinfo {pages} {138} (\bibinfo {year} {2023})}\BibitemShut {NoStop}%
\bibitem [{\citenamefont {L\'oio}\ \emph {et~al.}(2023)\citenamefont {L\'oio},
  \citenamefont {De~Luca}, \citenamefont {De~Nardis},\ and\ \citenamefont
  {Turkeshi}}]{Loio2023}%
  \BibitemOpen
  \bibfield  {author} {\bibinfo {author} {\bibfnamefont {H.}~\bibnamefont
  {L\'oio}}, \bibinfo {author} {\bibfnamefont {A.}~\bibnamefont {De~Luca}},
  \bibinfo {author} {\bibfnamefont {J.}~\bibnamefont {De~Nardis}},\ and\
  \bibinfo {author} {\bibfnamefont {X.}~\bibnamefont {Turkeshi}},\ }\bibfield
  {title} {\bibinfo {title} {Purification timescales in monitored fermions},\
  }\href {https://doi.org/10.1103/PhysRevB.108.L020306} {\bibfield  {journal}
  {\bibinfo  {journal} {Phys. Rev. B}\ }\textbf {\bibinfo {volume} {108}},\
  \bibinfo {pages} {L020306} (\bibinfo {year} {2023})}\BibitemShut {NoStop}%
\bibitem [{\citenamefont {Turkeshi}\ \emph
  {et~al.}(2022{\natexlab{a}})\citenamefont {Turkeshi}, \citenamefont
  {Piroli},\ and\ \citenamefont {Schir\'o}}]{Turkeshi2022b}%
  \BibitemOpen
  \bibfield  {author} {\bibinfo {author} {\bibfnamefont {X.}~\bibnamefont
  {Turkeshi}}, \bibinfo {author} {\bibfnamefont {L.}~\bibnamefont {Piroli}},\
  and\ \bibinfo {author} {\bibfnamefont {M.}~\bibnamefont {Schir\'o}},\
  }\bibfield  {title} {\bibinfo {title} {Enhanced entanglement negativity in
  boundary-driven monitored fermionic chains},\ }\href
  {https://doi.org/10.1103/PhysRevB.106.024304} {\bibfield  {journal} {\bibinfo
   {journal} {Phys. Rev. B}\ }\textbf {\bibinfo {volume} {106}},\ \bibinfo
  {pages} {024304} (\bibinfo {year} {2022}{\natexlab{a}})}\BibitemShut
  {NoStop}%
\bibitem [{\citenamefont {Kells}\ \emph {et~al.}(2023)\citenamefont {Kells},
  \citenamefont {Meidan},\ and\ \citenamefont {Romito}}]{Kells2023}%
  \BibitemOpen
  \bibfield  {author} {\bibinfo {author} {\bibfnamefont {G.}~\bibnamefont
  {Kells}}, \bibinfo {author} {\bibfnamefont {D.}~\bibnamefont {Meidan}},\ and\
  \bibinfo {author} {\bibfnamefont {A.}~\bibnamefont {Romito}},\ }\bibfield
  {title} {\bibinfo {title} {{Topological transitions in weakly monitored free
  fermions}},\ }\href {https://doi.org/10.21468/SciPostPhys.14.3.031}
  {\bibfield  {journal} {\bibinfo  {journal} {SciPost Phys.}\ }\textbf
  {\bibinfo {volume} {14}},\ \bibinfo {pages} {031} (\bibinfo {year}
  {2023})}\BibitemShut {NoStop}%
\bibitem [{\citenamefont {Fava}\ \emph {et~al.}(2023)\citenamefont {Fava},
  \citenamefont {Piroli}, \citenamefont {Swann}, \citenamefont {Bernard},\ and\
  \citenamefont {Nahum}}]{Fava2023}%
  \BibitemOpen
  \bibfield  {author} {\bibinfo {author} {\bibfnamefont {M.}~\bibnamefont
  {Fava}}, \bibinfo {author} {\bibfnamefont {L.}~\bibnamefont {Piroli}},
  \bibinfo {author} {\bibfnamefont {T.}~\bibnamefont {Swann}}, \bibinfo
  {author} {\bibfnamefont {D.}~\bibnamefont {Bernard}},\ and\ \bibinfo {author}
  {\bibfnamefont {A.}~\bibnamefont {Nahum}},\ }\href
  {https://doi.org/10.48550/arxiv.2302.12820} {\bibinfo {title} {Nonlinear
  sigma models for monitored dynamics of free fermions}} (\bibinfo {year}
  {2023}),\ \Eprint {https://arxiv.org/abs/2302.12820} {arXiv:2302.12820}
  \BibitemShut {NoStop}%
\bibitem [{\citenamefont {Swann}\ \emph {et~al.}(2023)\citenamefont {Swann},
  \citenamefont {Bernard},\ and\ \citenamefont {Nahum}}]{Swann2023}%
  \BibitemOpen
  \bibfield  {author} {\bibinfo {author} {\bibfnamefont {T.}~\bibnamefont
  {Swann}}, \bibinfo {author} {\bibfnamefont {D.}~\bibnamefont {Bernard}},\
  and\ \bibinfo {author} {\bibfnamefont {A.}~\bibnamefont {Nahum}},\ }\href
  {https://doi.org/10.48550/arxiv.2302.12212} {\bibinfo {title} {Spacetime
  picture for entanglement generation in noisy fermion chains}} (\bibinfo
  {year} {2023}),\ \Eprint {https://arxiv.org/abs/2302.12212}
  {arXiv:2302.12212} \BibitemShut {NoStop}%
\bibitem [{\citenamefont {Lang}\ and\ \citenamefont
  {B\"uchler}(2020)}]{Lang2020a}%
  \BibitemOpen
  \bibfield  {author} {\bibinfo {author} {\bibfnamefont {N.}~\bibnamefont
  {Lang}}\ and\ \bibinfo {author} {\bibfnamefont {H.~P.}\ \bibnamefont
  {B\"uchler}},\ }\bibfield  {title} {\bibinfo {title} {Entanglement transition
  in the projective transverse field {Ising} model},\ }\href
  {https://doi.org/10.1103/PhysRevB.102.094204} {\bibfield  {journal} {\bibinfo
   {journal} {Phys. Rev. B}\ }\textbf {\bibinfo {volume} {102}},\ \bibinfo
  {pages} {094204} (\bibinfo {year} {2020})}\BibitemShut {NoStop}%
\bibitem [{\citenamefont {Rossini}\ and\ \citenamefont
  {Vicari}(2020)}]{Rossini2020a}%
  \BibitemOpen
  \bibfield  {author} {\bibinfo {author} {\bibfnamefont {D.}~\bibnamefont
  {Rossini}}\ and\ \bibinfo {author} {\bibfnamefont {E.}~\bibnamefont
  {Vicari}},\ }\bibfield  {title} {\bibinfo {title} {Measurement-induced
  dynamics of many-body systems at quantum criticality},\ }\href
  {https://doi.org/10.1103/PhysRevB.102.035119} {\bibfield  {journal} {\bibinfo
   {journal} {Phys. Rev. B}\ }\textbf {\bibinfo {volume} {102}},\ \bibinfo
  {pages} {035119} (\bibinfo {year} {2020})}\BibitemShut {NoStop}%
\bibitem [{\citenamefont {Biella}\ and\ \citenamefont
  {Schir{\`o}}(2021)}]{Biella2021a}%
  \BibitemOpen
  \bibfield  {author} {\bibinfo {author} {\bibfnamefont {A.}~\bibnamefont
  {Biella}}\ and\ \bibinfo {author} {\bibfnamefont {M.}~\bibnamefont
  {Schir{\`o}}},\ }\bibfield  {title} {\bibinfo {title} {Many-body quantum
  {Z}eno effect and measurement-induced subradiance transition},\ }\href
  {https://doi.org/10.22331/q-2021-08-19-528} {\bibfield  {journal} {\bibinfo
  {journal} {{Quantum}}\ }\textbf {\bibinfo {volume} {5}},\ \bibinfo {pages}
  {528} (\bibinfo {year} {2021})}\BibitemShut {NoStop}%
\bibitem [{\citenamefont {Turkeshi}\ \emph {et~al.}(2021)\citenamefont
  {Turkeshi}, \citenamefont {Biella}, \citenamefont {Fazio}, \citenamefont
  {Dalmonte},\ and\ \citenamefont {Schir{\`o}}}]{Turkeshi2021}%
  \BibitemOpen
  \bibfield  {author} {\bibinfo {author} {\bibfnamefont {X.}~\bibnamefont
  {Turkeshi}}, \bibinfo {author} {\bibfnamefont {A.}~\bibnamefont {Biella}},
  \bibinfo {author} {\bibfnamefont {R.}~\bibnamefont {Fazio}}, \bibinfo
  {author} {\bibfnamefont {M.}~\bibnamefont {Dalmonte}},\ and\ \bibinfo
  {author} {\bibfnamefont {M.}~\bibnamefont {Schir{\`o}}},\ }\bibfield  {title}
  {\bibinfo {title} {Measurement-induced entanglement transitions in the
  quantum {Ising} chain: From infinite to zero clicks},\ }\href
  {https://doi.org/10.1103/PhysRevB.103.224210} {\bibfield  {journal} {\bibinfo
   {journal} {Phys. Rev. B}\ }\textbf {\bibinfo {volume} {103}},\ \bibinfo
  {pages} {224210} (\bibinfo {year} {2021})}\BibitemShut {NoStop}%
\bibitem [{\citenamefont {Tirrito}\ \emph {et~al.}(2023)\citenamefont
  {Tirrito}, \citenamefont {Santini}, \citenamefont {Fazio},\ and\
  \citenamefont {Collura}}]{Tirrito2022}%
  \BibitemOpen
  \bibfield  {author} {\bibinfo {author} {\bibfnamefont {E.}~\bibnamefont
  {Tirrito}}, \bibinfo {author} {\bibfnamefont {A.}~\bibnamefont {Santini}},
  \bibinfo {author} {\bibfnamefont {R.}~\bibnamefont {Fazio}},\ and\ \bibinfo
  {author} {\bibfnamefont {M.}~\bibnamefont {Collura}},\ }\bibfield  {title}
  {\bibinfo {title} {{Full counting statistics as probe of measurement-induced
  transitions in the quantum Ising chain}},\ }\href
  {https://doi.org/10.21468/SciPostPhys.15.3.096} {\bibfield  {journal}
  {\bibinfo  {journal} {SciPost Phys.}\ }\textbf {\bibinfo {volume} {15}},\
  \bibinfo {pages} {096} (\bibinfo {year} {2023})}\BibitemShut {NoStop}%
\bibitem [{\citenamefont {Yang}\ \emph
  {et~al.}(2023{\natexlab{b}})\citenamefont {Yang}, \citenamefont {Mao},\ and\
  \citenamefont {Jian}}]{Yang2023}%
  \BibitemOpen
  \bibfield  {author} {\bibinfo {author} {\bibfnamefont {Z.}~\bibnamefont
  {Yang}}, \bibinfo {author} {\bibfnamefont {D.}~\bibnamefont {Mao}},\ and\
  \bibinfo {author} {\bibfnamefont {C.-M.}\ \bibnamefont {Jian}},\ }\bibfield
  {title} {\bibinfo {title} {Entanglement in a one-dimensional critical state
  after measurements},\ }\href {https://doi.org/10.1103/PhysRevB.108.165120}
  {\bibfield  {journal} {\bibinfo  {journal} {Phys. Rev. B}\ }\textbf {\bibinfo
  {volume} {108}},\ \bibinfo {pages} {165120} (\bibinfo {year}
  {2023}{\natexlab{b}})}\BibitemShut {NoStop}%
\bibitem [{\citenamefont {Weinstein}\ \emph {et~al.}(2023)\citenamefont
  {Weinstein}, \citenamefont {Sajith}, \citenamefont {Altman},\ and\
  \citenamefont {Garratt}}]{Weinstein2023}%
  \BibitemOpen
  \bibfield  {author} {\bibinfo {author} {\bibfnamefont {Z.}~\bibnamefont
  {Weinstein}}, \bibinfo {author} {\bibfnamefont {R.}~\bibnamefont {Sajith}},
  \bibinfo {author} {\bibfnamefont {E.}~\bibnamefont {Altman}},\ and\ \bibinfo
  {author} {\bibfnamefont {S.~J.}\ \bibnamefont {Garratt}},\ }\bibfield
  {title} {\bibinfo {title} {Nonlocality and entanglement in measured critical
  quantum ising chains},\ }\href {https://doi.org/10.1103/PhysRevB.107.245132}
  {\bibfield  {journal} {\bibinfo  {journal} {Phys. Rev. B}\ }\textbf {\bibinfo
  {volume} {107}},\ \bibinfo {pages} {245132} (\bibinfo {year}
  {2023})}\BibitemShut {NoStop}%
\bibitem [{\citenamefont {Murciano}\ \emph {et~al.}(2023)\citenamefont
  {Murciano}, \citenamefont {Sala}, \citenamefont {Liu}, \citenamefont {Mong},\
  and\ \citenamefont {Alicea}}]{Murciano2023}%
  \BibitemOpen
  \bibfield  {author} {\bibinfo {author} {\bibfnamefont {S.}~\bibnamefont
  {Murciano}}, \bibinfo {author} {\bibfnamefont {P.}~\bibnamefont {Sala}},
  \bibinfo {author} {\bibfnamefont {Y.}~\bibnamefont {Liu}}, \bibinfo {author}
  {\bibfnamefont {R.~S.~K.}\ \bibnamefont {Mong}},\ and\ \bibinfo {author}
  {\bibfnamefont {J.}~\bibnamefont {Alicea}},\ }\href
  {https://doi.org/10.48550/arxiv.2302.04325} {\bibinfo {title}
  {Measurement-altered {Ising} quantum criticality}} (\bibinfo {year} {2023}),\
  \Eprint {https://arxiv.org/abs/2302.04325} {arXiv:2302.04325} \BibitemShut
  {NoStop}%
\bibitem [{\citenamefont {Sierant}\ \emph {et~al.}(2022)\citenamefont
  {Sierant}, \citenamefont {Chiriac{\`o}}, \citenamefont {Surace},
  \citenamefont {Sharma}, \citenamefont {Turkeshi}, \citenamefont {Dalmonte},
  \citenamefont {Fazio},\ and\ \citenamefont {Pagano}}]{Sierant2022a}%
  \BibitemOpen
  \bibfield  {author} {\bibinfo {author} {\bibfnamefont {P.}~\bibnamefont
  {Sierant}}, \bibinfo {author} {\bibfnamefont {G.}~\bibnamefont
  {Chiriac{\`o}}}, \bibinfo {author} {\bibfnamefont {F.~M.}\ \bibnamefont
  {Surace}}, \bibinfo {author} {\bibfnamefont {S.}~\bibnamefont {Sharma}},
  \bibinfo {author} {\bibfnamefont {X.}~\bibnamefont {Turkeshi}}, \bibinfo
  {author} {\bibfnamefont {M.}~\bibnamefont {Dalmonte}}, \bibinfo {author}
  {\bibfnamefont {R.}~\bibnamefont {Fazio}},\ and\ \bibinfo {author}
  {\bibfnamefont {G.}~\bibnamefont {Pagano}},\ }\bibfield  {title} {\bibinfo
  {title} {Dissipative {Floquet} dynamics: from steady state to measurement
  induced criticality in trapped-ion chains},\ }\href
  {https://doi.org/10.22331/q-2022-02-02-638} {\bibfield  {journal} {\bibinfo
  {journal} {{Quantum}}\ }\textbf {\bibinfo {volume} {6}},\ \bibinfo {pages}
  {638} (\bibinfo {year} {2022})}\BibitemShut {NoStop}%
\bibitem [{\citenamefont {Turkeshi}\ \emph
  {et~al.}(2022{\natexlab{b}})\citenamefont {Turkeshi}, \citenamefont
  {Dalmonte}, \citenamefont {Fazio},\ and\ \citenamefont
  {Schir\`o}}]{Turkeshi2022a}%
  \BibitemOpen
  \bibfield  {author} {\bibinfo {author} {\bibfnamefont {X.}~\bibnamefont
  {Turkeshi}}, \bibinfo {author} {\bibfnamefont {M.}~\bibnamefont {Dalmonte}},
  \bibinfo {author} {\bibfnamefont {R.}~\bibnamefont {Fazio}},\ and\ \bibinfo
  {author} {\bibfnamefont {M.}~\bibnamefont {Schir\`o}},\ }\bibfield  {title}
  {\bibinfo {title} {Entanglement transitions from stochastic resetting of
  non-{H}ermitian quasiparticles},\ }\href
  {https://doi.org/10.1103/PhysRevB.105.L241114} {\bibfield  {journal}
  {\bibinfo  {journal} {Phys. Rev. B}\ }\textbf {\bibinfo {volume} {105}},\
  \bibinfo {pages} {L241114} (\bibinfo {year}
  {2022}{\natexlab{b}})}\BibitemShut {NoStop}%
\bibitem [{\citenamefont {Tang}\ and\ \citenamefont {Zhu}(2020)}]{Tang2020a}%
  \BibitemOpen
  \bibfield  {author} {\bibinfo {author} {\bibfnamefont {Q.}~\bibnamefont
  {Tang}}\ and\ \bibinfo {author} {\bibfnamefont {W.}~\bibnamefont {Zhu}},\
  }\bibfield  {title} {\bibinfo {title} {Measurement-induced phase transition:
  A case study in the nonintegrable model by density-matrix renormalization
  group calculations},\ }\href
  {https://doi.org/10.1103/PhysRevResearch.2.013022} {\bibfield  {journal}
  {\bibinfo  {journal} {Phys. Rev. Research}\ }\textbf {\bibinfo {volume}
  {2}},\ \bibinfo {pages} {013022} (\bibinfo {year} {2020})}\BibitemShut
  {NoStop}%
\bibitem [{\citenamefont {Goto}\ and\ \citenamefont
  {Danshita}(2020)}]{Goto2020a}%
  \BibitemOpen
  \bibfield  {author} {\bibinfo {author} {\bibfnamefont {S.}~\bibnamefont
  {Goto}}\ and\ \bibinfo {author} {\bibfnamefont {I.}~\bibnamefont
  {Danshita}},\ }\bibfield  {title} {\bibinfo {title} {Measurement-induced
  transitions of the entanglement scaling law in ultracold gases with
  controllable dissipation},\ }\href
  {https://doi.org/10.1103/PhysRevA.102.033316} {\bibfield  {journal} {\bibinfo
   {journal} {Phys. Rev. A}\ }\textbf {\bibinfo {volume} {102}},\ \bibinfo
  {pages} {033316} (\bibinfo {year} {2020})}\BibitemShut {NoStop}%
\bibitem [{\citenamefont {Fuji}\ and\ \citenamefont
  {Ashida}(2020)}]{Fuji2020a}%
  \BibitemOpen
  \bibfield  {author} {\bibinfo {author} {\bibfnamefont {Y.}~\bibnamefont
  {Fuji}}\ and\ \bibinfo {author} {\bibfnamefont {Y.}~\bibnamefont {Ashida}},\
  }\bibfield  {title} {\bibinfo {title} {Measurement-induced quantum
  criticality under continuous monitoring},\ }\href
  {https://doi.org/10.1103/PhysRevB.102.054302} {\bibfield  {journal} {\bibinfo
   {journal} {Phys. Rev. B}\ }\textbf {\bibinfo {volume} {102}},\ \bibinfo
  {pages} {054302} (\bibinfo {year} {2020})}\BibitemShut {NoStop}%
\bibitem [{\citenamefont {Doggen}\ \emph {et~al.}(2022)\citenamefont {Doggen},
  \citenamefont {Gefen}, \citenamefont {Gornyi}, \citenamefont {Mirlin},\ and\
  \citenamefont {Polyakov}}]{Doggen2022a}%
  \BibitemOpen
  \bibfield  {author} {\bibinfo {author} {\bibfnamefont {E.~V.~H.}\
  \bibnamefont {Doggen}}, \bibinfo {author} {\bibfnamefont {Y.}~\bibnamefont
  {Gefen}}, \bibinfo {author} {\bibfnamefont {I.~V.}\ \bibnamefont {Gornyi}},
  \bibinfo {author} {\bibfnamefont {A.~D.}\ \bibnamefont {Mirlin}},\ and\
  \bibinfo {author} {\bibfnamefont {D.~G.}\ \bibnamefont {Polyakov}},\
  }\bibfield  {title} {\bibinfo {title} {Generalized quantum measurements with
  matrix product states: {E}ntanglement phase transition and clusterization},\
  }\href {https://doi.org/10.1103/PhysRevResearch.4.023146} {\bibfield
  {journal} {\bibinfo  {journal} {Phys. Rev. Research}\ }\textbf {\bibinfo
  {volume} {4}},\ \bibinfo {pages} {023146} (\bibinfo {year}
  {2022})}\BibitemShut {NoStop}%
\bibitem [{\citenamefont {Doggen}\ \emph {et~al.}(2023)\citenamefont {Doggen},
  \citenamefont {Gefen}, \citenamefont {Gornyi}, \citenamefont {Mirlin},\ and\
  \citenamefont {Polyakov}}]{Doggen2023}%
  \BibitemOpen
  \bibfield  {author} {\bibinfo {author} {\bibfnamefont {E.~V.~H.}\
  \bibnamefont {Doggen}}, \bibinfo {author} {\bibfnamefont {Y.}~\bibnamefont
  {Gefen}}, \bibinfo {author} {\bibfnamefont {I.~V.}\ \bibnamefont {Gornyi}},
  \bibinfo {author} {\bibfnamefont {A.~D.}\ \bibnamefont {Mirlin}},\ and\
  \bibinfo {author} {\bibfnamefont {D.~G.}\ \bibnamefont {Polyakov}},\
  }\bibfield  {title} {\bibinfo {title} {Evolution of many-body systems under
  ancilla quantum measurements},\ }\href
  {https://doi.org/10.1103/PhysRevB.107.214203} {\bibfield  {journal} {\bibinfo
   {journal} {Phys. Rev. B}\ }\textbf {\bibinfo {volume} {107}},\ \bibinfo
  {pages} {214203} (\bibinfo {year} {2023})}\BibitemShut {NoStop}%
\bibitem [{\citenamefont {Lunt}\ and\ \citenamefont {Pal}(2020)}]{Lunt2020a}%
  \BibitemOpen
  \bibfield  {author} {\bibinfo {author} {\bibfnamefont {O.}~\bibnamefont
  {Lunt}}\ and\ \bibinfo {author} {\bibfnamefont {A.}~\bibnamefont {Pal}},\
  }\bibfield  {title} {\bibinfo {title} {Measurement-induced entanglement
  transitions in many-body localized systems},\ }\href
  {https://doi.org/10.1103/PhysRevResearch.2.043072} {\bibfield  {journal}
  {\bibinfo  {journal} {Phys. Rev. Research}\ }\textbf {\bibinfo {volume}
  {2}},\ \bibinfo {pages} {043072} (\bibinfo {year} {2020})}\BibitemShut
  {NoStop}%
\bibitem [{\citenamefont {Yamamoto}\ and\ \citenamefont
  {Hamazaki}(2023)}]{Yamamoto2023}%
  \BibitemOpen
  \bibfield  {author} {\bibinfo {author} {\bibfnamefont {K.}~\bibnamefont
  {Yamamoto}}\ and\ \bibinfo {author} {\bibfnamefont {R.}~\bibnamefont
  {Hamazaki}},\ }\bibfield  {title} {\bibinfo {title} {Localization properties
  in disordered quantum many-body dynamics under continuous measurement},\
  }\href {https://doi.org/10.1103/PhysRevB.107.L220201} {\bibfield  {journal}
  {\bibinfo  {journal} {Phys. Rev. B}\ }\textbf {\bibinfo {volume} {107}},\
  \bibinfo {pages} {L220201} (\bibinfo {year} {2023})}\BibitemShut {NoStop}%
\bibitem [{\citenamefont {Jian}\ \emph {et~al.}(2021)\citenamefont {Jian},
  \citenamefont {Liu}, \citenamefont {Chen}, \citenamefont {Swingle},\ and\
  \citenamefont {Zhang}}]{Jian2021a}%
  \BibitemOpen
  \bibfield  {author} {\bibinfo {author} {\bibfnamefont {S.-K.}\ \bibnamefont
  {Jian}}, \bibinfo {author} {\bibfnamefont {C.}~\bibnamefont {Liu}}, \bibinfo
  {author} {\bibfnamefont {X.}~\bibnamefont {Chen}}, \bibinfo {author}
  {\bibfnamefont {B.}~\bibnamefont {Swingle}},\ and\ \bibinfo {author}
  {\bibfnamefont {P.}~\bibnamefont {Zhang}},\ }\bibfield  {title} {\bibinfo
  {title} {Measurement-induced phase transition in the monitored
  {Sachdev-Ye-Kitaev} model},\ }\href
  {https://doi.org/10.1103/PhysRevLett.127.140601} {\bibfield  {journal}
  {\bibinfo  {journal} {Phys. Rev. Lett.}\ }\textbf {\bibinfo {volume} {127}},\
  \bibinfo {pages} {140601} (\bibinfo {year} {2021})}\BibitemShut {NoStop}%
\bibitem [{\citenamefont {Altland}\ \emph {et~al.}(2022)\citenamefont
  {Altland}, \citenamefont {Buchhold}, \citenamefont {Diehl},\ and\
  \citenamefont {Micklitz}}]{Altland2022}%
  \BibitemOpen
  \bibfield  {author} {\bibinfo {author} {\bibfnamefont {A.}~\bibnamefont
  {Altland}}, \bibinfo {author} {\bibfnamefont {M.}~\bibnamefont {Buchhold}},
  \bibinfo {author} {\bibfnamefont {S.}~\bibnamefont {Diehl}},\ and\ \bibinfo
  {author} {\bibfnamefont {T.}~\bibnamefont {Micklitz}},\ }\bibfield  {title}
  {\bibinfo {title} {Dynamics of measured many-body quantum chaotic systems},\
  }\href {https://doi.org/10.1103/PhysRevResearch.4.L022066} {\bibfield
  {journal} {\bibinfo  {journal} {Phys. Rev. Research}\ }\textbf {\bibinfo
  {volume} {4}},\ \bibinfo {pages} {L022066} (\bibinfo {year}
  {2022})}\BibitemShut {NoStop}%
\bibitem [{\citenamefont {Noel}\ \emph {et~al.}(2022)\citenamefont {Noel},
  \citenamefont {Niroula}, \citenamefont {Zhu}, \citenamefont {Risinger},
  \citenamefont {Egan}, \citenamefont {Biswas}, \citenamefont {Cetina},
  \citenamefont {Gorshkov}, \citenamefont {Gullans}, \citenamefont {Huse},\
  and\ \citenamefont {Monroe}}]{Noel2022a}%
  \BibitemOpen
  \bibfield  {author} {\bibinfo {author} {\bibfnamefont {C.}~\bibnamefont
  {Noel}}, \bibinfo {author} {\bibfnamefont {P.}~\bibnamefont {Niroula}},
  \bibinfo {author} {\bibfnamefont {D.}~\bibnamefont {Zhu}}, \bibinfo {author}
  {\bibfnamefont {A.}~\bibnamefont {Risinger}}, \bibinfo {author}
  {\bibfnamefont {L.}~\bibnamefont {Egan}}, \bibinfo {author} {\bibfnamefont
  {D.}~\bibnamefont {Biswas}}, \bibinfo {author} {\bibfnamefont
  {M.}~\bibnamefont {Cetina}}, \bibinfo {author} {\bibfnamefont {A.~V.}\
  \bibnamefont {Gorshkov}}, \bibinfo {author} {\bibfnamefont {M.~J.}\
  \bibnamefont {Gullans}}, \bibinfo {author} {\bibfnamefont {D.~A.}\
  \bibnamefont {Huse}},\ and\ \bibinfo {author} {\bibfnamefont
  {C.}~\bibnamefont {Monroe}},\ }\bibfield  {title} {\bibinfo {title}
  {Measurement-induced quantum phases realized in a trapped-ion quantum
  computer},\ }\href {https://doi.org/10.1038/s41567-022-01619-7} {\bibfield
  {journal} {\bibinfo  {journal} {Nat. Phys.}\ }\textbf {\bibinfo {volume}
  {18}},\ \bibinfo {pages} {760} (\bibinfo {year} {2022})}\BibitemShut
  {NoStop}%
\bibitem [{\citenamefont {Koh}\ \emph {et~al.}(2023)\citenamefont {Koh},
  \citenamefont {Sun}, \citenamefont {Motta},\ and\ \citenamefont
  {Minnich}}]{Koh2022}%
  \BibitemOpen
  \bibfield  {author} {\bibinfo {author} {\bibfnamefont {J.~M.}\ \bibnamefont
  {Koh}}, \bibinfo {author} {\bibfnamefont {S.-N.}\ \bibnamefont {Sun}},
  \bibinfo {author} {\bibfnamefont {M.}~\bibnamefont {Motta}},\ and\ \bibinfo
  {author} {\bibfnamefont {A.~J.}\ \bibnamefont {Minnich}},\ }\bibfield
  {title} {\bibinfo {title} {Measurement-induced entanglement phase transition
  on a superconducting quantum processor with mid-circuit readout},\ }\bibfield
   {journal} {\bibinfo  {journal} {Nat. Phys.}\ }\textbf {\bibinfo {volume}
  {19}},\ \href {https://doi.org/10.1038/s41567-023-02076-6}
  {10.1038/s41567-023-02076-6} (\bibinfo {year} {2023})\BibitemShut {NoStop}%
\bibitem [{\citenamefont {Hoke~\textit{et al.,} {Google Quantum AI and
  Collaborators}}(2023)}]{Hoke2023}%
  \BibitemOpen
  \bibfield  {author} {\bibinfo {author} {\bibfnamefont {J.~C.}\ \bibnamefont
  {Hoke~\textit{et al.,} {Google Quantum AI and Collaborators}}},\ }\bibfield
  {title} {\bibinfo {title} {Measurement-induced entanglement and teleportation
  on a noisy quantum processor},\ }\href
  {https://doi.org/10.1038/s41586-023-06505-7} {\bibfield  {journal} {\bibinfo
  {journal} {Nature}\ }\textbf {\bibinfo {volume} {622}},\ \bibinfo {pages}
  {481} (\bibinfo {year} {2023})}\BibitemShut {NoStop}%
\bibitem [{\citenamefont {P\"opperl}\ \emph {et~al.}(2023)\citenamefont
  {P\"opperl}, \citenamefont {Gornyi},\ and\ \citenamefont {Gefen}}]{Paul2023}%
  \BibitemOpen
  \bibfield  {author} {\bibinfo {author} {\bibfnamefont {P.}~\bibnamefont
  {P\"opperl}}, \bibinfo {author} {\bibfnamefont {I.~V.}\ \bibnamefont
  {Gornyi}},\ and\ \bibinfo {author} {\bibfnamefont {Y.}~\bibnamefont
  {Gefen}},\ }\bibfield  {title} {\bibinfo {title} {Measurements on an
  {A}nderson chain},\ }\href {https://doi.org/10.1103/PhysRevB.107.174203}
  {\bibfield  {journal} {\bibinfo  {journal} {Phys. Rev. B}\ }\textbf {\bibinfo
  {volume} {107}},\ \bibinfo {pages} {174203} (\bibinfo {year}
  {2023})}\BibitemShut {NoStop}%
\bibitem [{\citenamefont {Klich}\ and\ \citenamefont
  {Levitov}(2009)}]{KlichLevitov}%
  \BibitemOpen
  \bibfield  {author} {\bibinfo {author} {\bibfnamefont {I.}~\bibnamefont
  {Klich}}\ and\ \bibinfo {author} {\bibfnamefont {L.}~\bibnamefont
  {Levitov}},\ }\bibfield  {title} {\bibinfo {title} {Quantum noise as an
  entanglement meter},\ }\href {https://doi.org/10.1103/PhysRevLett.102.100502}
  {\bibfield  {journal} {\bibinfo  {journal} {Phys. Rev. Lett.}\ }\textbf
  {\bibinfo {volume} {102}},\ \bibinfo {pages} {100502} (\bibinfo {year}
  {2009})}\BibitemShut {NoStop}%
\bibitem [{\citenamefont {Song}\ \emph {et~al.}(2011)\citenamefont {Song},
  \citenamefont {Flindt}, \citenamefont {Rachel}, \citenamefont {Klich},\ and\
  \citenamefont {Le~Hur}}]{Song2011}%
  \BibitemOpen
  \bibfield  {author} {\bibinfo {author} {\bibfnamefont {H.~F.}\ \bibnamefont
  {Song}}, \bibinfo {author} {\bibfnamefont {C.}~\bibnamefont {Flindt}},
  \bibinfo {author} {\bibfnamefont {S.}~\bibnamefont {Rachel}}, \bibinfo
  {author} {\bibfnamefont {I.}~\bibnamefont {Klich}},\ and\ \bibinfo {author}
  {\bibfnamefont {K.}~\bibnamefont {Le~Hur}},\ }\bibfield  {title} {\bibinfo
  {title} {Entanglement entropy from charge statistics: {E}xact relations for
  noninteracting many-body systems},\ }\href
  {https://doi.org/10.1103/PhysRevB.83.161408} {\bibfield  {journal} {\bibinfo
  {journal} {Phys. Rev. B}\ }\textbf {\bibinfo {volume} {83}},\ \bibinfo
  {pages} {161408(R)} (\bibinfo {year} {2011})}\BibitemShut {NoStop}%
\bibitem [{\citenamefont {Song}\ \emph {et~al.}(2012)\citenamefont {Song},
  \citenamefont {Rachel}, \citenamefont {Flindt}, \citenamefont {Klich},
  \citenamefont {Laflorencie},\ and\ \citenamefont {Le~Hur}}]{Song2012}%
  \BibitemOpen
  \bibfield  {author} {\bibinfo {author} {\bibfnamefont {H.~F.}\ \bibnamefont
  {Song}}, \bibinfo {author} {\bibfnamefont {S.}~\bibnamefont {Rachel}},
  \bibinfo {author} {\bibfnamefont {C.}~\bibnamefont {Flindt}}, \bibinfo
  {author} {\bibfnamefont {I.}~\bibnamefont {Klich}}, \bibinfo {author}
  {\bibfnamefont {N.}~\bibnamefont {Laflorencie}},\ and\ \bibinfo {author}
  {\bibfnamefont {K.}~\bibnamefont {Le~Hur}},\ }\bibfield  {title} {\bibinfo
  {title} {Bipartite fluctuations as a probe of many-body entanglement},\
  }\href {https://doi.org/10.1103/PhysRevB.85.035409} {\bibfield  {journal}
  {\bibinfo  {journal} {Phys. Rev. B}\ }\textbf {\bibinfo {volume} {85}},\
  \bibinfo {pages} {035409} (\bibinfo {year} {2012})}\BibitemShut {NoStop}%
\bibitem [{\citenamefont {Thomas}\ and\ \citenamefont
  {Flindt}(2015)}]{Thomas2015}%
  \BibitemOpen
  \bibfield  {author} {\bibinfo {author} {\bibfnamefont {K.~H.}\ \bibnamefont
  {Thomas}}\ and\ \bibinfo {author} {\bibfnamefont {C.}~\bibnamefont
  {Flindt}},\ }\bibfield  {title} {\bibinfo {title} {Entanglement entropy in
  dynamic quantum-coherent conductors},\ }\href
  {https://doi.org/10.1103/PhysRevB.91.125406} {\bibfield  {journal} {\bibinfo
  {journal} {Phys. Rev. B}\ }\textbf {\bibinfo {volume} {91}},\ \bibinfo
  {pages} {125406} (\bibinfo {year} {2015})}\BibitemShut {NoStop}%
\bibitem [{\citenamefont {Burmistrov}\ \emph {et~al.}(2017)\citenamefont
  {Burmistrov}, \citenamefont {Tikhonov}, \citenamefont {Gornyi},\ and\
  \citenamefont {Mirlin}}]{Burmistrov2017}%
  \BibitemOpen
  \bibfield  {author} {\bibinfo {author} {\bibfnamefont {I.}~\bibnamefont
  {Burmistrov}}, \bibinfo {author} {\bibfnamefont {K.}~\bibnamefont
  {Tikhonov}}, \bibinfo {author} {\bibfnamefont {I.}~\bibnamefont {Gornyi}},\
  and\ \bibinfo {author} {\bibfnamefont {A.}~\bibnamefont {Mirlin}},\
  }\bibfield  {title} {\bibinfo {title} {Entanglement entropy and particle
  number cumulants of disordered fermions},\ }\href
  {https://doi.org/https://doi.org/10.1016/j.aop.2017.05.011} {\bibfield
  {journal} {\bibinfo  {journal} {Annals of Physics (NY)}\ }\textbf {\bibinfo
  {volume} {383}},\ \bibinfo {pages} {140} (\bibinfo {year}
  {2017})}\BibitemShut {NoStop}%
\bibitem [{\citenamefont {Jin}\ \emph {et~al.}(2022)\citenamefont {Jin},
  \citenamefont {Ferreira}, \citenamefont {Filippone},\ and\ \citenamefont
  {Giamarchi}}]{Giamarchi2022}%
  \BibitemOpen
  \bibfield  {author} {\bibinfo {author} {\bibfnamefont {T.}~\bibnamefont
  {Jin}}, \bibinfo {author} {\bibfnamefont {J.~a.~S.}\ \bibnamefont
  {Ferreira}}, \bibinfo {author} {\bibfnamefont {M.}~\bibnamefont
  {Filippone}},\ and\ \bibinfo {author} {\bibfnamefont {T.}~\bibnamefont
  {Giamarchi}},\ }\bibfield  {title} {\bibinfo {title} {Exact description of
  quantum stochastic models as quantum resistors},\ }\href
  {https://doi.org/10.1103/PhysRevResearch.4.013109} {\bibfield  {journal}
  {\bibinfo  {journal} {Phys. Rev. Res.}\ }\textbf {\bibinfo {volume} {4}},\
  \bibinfo {pages} {013109} (\bibinfo {year} {2022})}\BibitemShut {NoStop}%
\bibitem [{\citenamefont {Bernard}\ \emph {et~al.}(2018)\citenamefont
  {Bernard}, \citenamefont {Jin},\ and\ \citenamefont
  {Shpielberg}}]{Bernard2018}%
  \BibitemOpen
  \bibfield  {author} {\bibinfo {author} {\bibfnamefont {D.}~\bibnamefont
  {Bernard}}, \bibinfo {author} {\bibfnamefont {T.}~\bibnamefont {Jin}},\ and\
  \bibinfo {author} {\bibfnamefont {O.}~\bibnamefont {Shpielberg}},\ }\bibfield
   {title} {\bibinfo {title} {Transport in quantum chains under strong
  monitoring},\ }\href {https://doi.org/10.1209/0295-5075/121/60006} {\bibfield
   {journal} {\bibinfo  {journal} {Europhysics Letters}\ }\textbf {\bibinfo
  {volume} {121}},\ \bibinfo {pages} {60006} (\bibinfo {year}
  {2018})}\BibitemShut {NoStop}%
\bibitem [{\citenamefont {Bernard}\ \emph {et~al.}(2022)\citenamefont
  {Bernard}, \citenamefont {Essler}, \citenamefont {Hruza},\ and\ \citenamefont
  {Medenjak}}]{Bernard2022}%
  \BibitemOpen
  \bibfield  {author} {\bibinfo {author} {\bibfnamefont {D.}~\bibnamefont
  {Bernard}}, \bibinfo {author} {\bibfnamefont {F.~H.~L.}\ \bibnamefont
  {Essler}}, \bibinfo {author} {\bibfnamefont {L.}~\bibnamefont {Hruza}},\ and\
  \bibinfo {author} {\bibfnamefont {M.}~\bibnamefont {Medenjak}},\ }\bibfield
  {title} {\bibinfo {title} {{Dynamics of fluctuations in quantum simple
  exclusion processes}},\ }\href
  {https://doi.org/10.21468/SciPostPhys.12.1.042} {\bibfield  {journal}
  {\bibinfo  {journal} {SciPost Phys.}\ }\textbf {\bibinfo {volume} {12}},\
  \bibinfo {pages} {042} (\bibinfo {year} {2022})}\BibitemShut {NoStop}%
\bibitem [{\citenamefont {Nosov}\ \emph {et~al.}(2023)\citenamefont {Nosov},
  \citenamefont {Shapiro}, \citenamefont {Goldstein},\ and\ \citenamefont
  {Burmistrov}}]{nosov2023}%
  \BibitemOpen
  \bibfield  {author} {\bibinfo {author} {\bibfnamefont {P.~A.}\ \bibnamefont
  {Nosov}}, \bibinfo {author} {\bibfnamefont {D.~S.}\ \bibnamefont {Shapiro}},
  \bibinfo {author} {\bibfnamefont {M.}~\bibnamefont {Goldstein}},\ and\
  \bibinfo {author} {\bibfnamefont {I.~S.}\ \bibnamefont {Burmistrov}},\
  }\bibfield  {title} {\bibinfo {title} {Reaction-diffusive dynamics of
  number-conserving dissipative quantum state preparation},\ }\href
  {https://doi.org/10.1103/PhysRevB.107.174312} {\bibfield  {journal} {\bibinfo
   {journal} {Phys. Rev. B}\ }\textbf {\bibinfo {volume} {107}},\ \bibinfo
  {pages} {174312} (\bibinfo {year} {2023})}\BibitemShut {NoStop}%
\bibitem [{\citenamefont {Esposito}\ and\ \citenamefont
  {Gaspard}(2005)}]{Esposito2005}%
  \BibitemOpen
  \bibfield  {author} {\bibinfo {author} {\bibfnamefont {M.}~\bibnamefont
  {Esposito}}\ and\ \bibinfo {author} {\bibfnamefont {P.}~\bibnamefont
  {Gaspard}},\ }\bibfield  {title} {\bibinfo {title} {Exactly solvable model of
  quantum diffusion},\ }\href
  {https://doi.org/https://doi.org/10.1007/s10955-005-7577-x} {\bibfield
  {journal} {\bibinfo  {journal} {Journal of Statistical Physics}\ }\textbf
  {\bibinfo {volume} {121}},\ \bibinfo {pages} {463} (\bibinfo {year}
  {2005})}\BibitemShut {NoStop}%
\bibitem [{\citenamefont {Jian}\ \emph {et~al.}(2022)\citenamefont {Jian},
  \citenamefont {Bauer}, \citenamefont {Keselman},\ and\ \citenamefont
  {Ludwig}}]{Jian2022}%
  \BibitemOpen
  \bibfield  {author} {\bibinfo {author} {\bibfnamefont {C.-M.}\ \bibnamefont
  {Jian}}, \bibinfo {author} {\bibfnamefont {B.}~\bibnamefont {Bauer}},
  \bibinfo {author} {\bibfnamefont {A.}~\bibnamefont {Keselman}},\ and\
  \bibinfo {author} {\bibfnamefont {A.~W.~W.}\ \bibnamefont {Ludwig}},\
  }\href@noop {} {\bibinfo {title} {Criticality and entanglement in non-unitary
  quantum circuits and tensor networks of non-interacting fermions}} (\bibinfo
  {year} {2022}),\ \Eprint {https://arxiv.org/abs/2012.04666}
  {arXiv:2012.04666} \BibitemShut {NoStop}%
\bibitem [{\citenamefont {Kamenev}\ and\ \citenamefont
  {Levchenko}(2009)}]{KamenevLevchenko}%
  \BibitemOpen
  \bibfield  {author} {\bibinfo {author} {\bibfnamefont {A.}~\bibnamefont
  {Kamenev}}\ and\ \bibinfo {author} {\bibfnamefont {A.}~\bibnamefont
  {Levchenko}},\ }\bibfield  {title} {\bibinfo {title} {Keldysh technique and
  non-linear $\sigma$-model: basic principles and applications},\ }\href
  {https://doi.org/10.1080/00018730902850504} {\bibfield  {journal} {\bibinfo
  {journal} {Advances in Physics}\ }\textbf {\bibinfo {volume} {58}},\ \bibinfo
  {pages} {197} (\bibinfo {year} {2009})}\BibitemShut {NoStop}%
\bibitem [{\citenamefont {Ostrovsky}\ \emph {et~al.}(2014)\citenamefont
  {Ostrovsky}, \citenamefont {Protopopov}, \citenamefont {K\"onig},
  \citenamefont {Gornyi}, \citenamefont {Mirlin},\ and\ \citenamefont
  {Skvortsov}}]{Ostrovsky2014}%
  \BibitemOpen
  \bibfield  {author} {\bibinfo {author} {\bibfnamefont {P.~M.}\ \bibnamefont
  {Ostrovsky}}, \bibinfo {author} {\bibfnamefont {I.~V.}\ \bibnamefont
  {Protopopov}}, \bibinfo {author} {\bibfnamefont {E.~J.}\ \bibnamefont
  {K\"onig}}, \bibinfo {author} {\bibfnamefont {I.~V.}\ \bibnamefont {Gornyi}},
  \bibinfo {author} {\bibfnamefont {A.~D.}\ \bibnamefont {Mirlin}},\ and\
  \bibinfo {author} {\bibfnamefont {M.~A.}\ \bibnamefont {Skvortsov}},\
  }\bibfield  {title} {\bibinfo {title} {Density of states in a two-dimensional
  chiral metal with vacancies},\ }\href
  {https://doi.org/10.1103/PhysRevLett.113.186803} {\bibfield  {journal}
  {\bibinfo  {journal} {Phys. Rev. Lett.}\ }\textbf {\bibinfo {volume} {113}},\
  \bibinfo {pages} {186803} (\bibinfo {year} {2014})}\BibitemShut {NoStop}%
\bibitem [{\citenamefont {Hikami}(1981)}]{HikamiRG}%
  \BibitemOpen
  \bibfield  {author} {\bibinfo {author} {\bibfnamefont {S.}~\bibnamefont
  {Hikami}},\ }\bibfield  {title} {\bibinfo {title} {Three-loop
  $\beta$-functions of non-linear $\sigma$ models on symmetric spaces},\ }\href
  {https://doi.org/https://doi.org/10.1016/0370-2693(81)90989-8} {\bibfield
  {journal} {\bibinfo  {journal} {Physics Letters B}\ }\textbf {\bibinfo
  {volume} {98}},\ \bibinfo {pages} {208} (\bibinfo {year} {1981})}\BibitemShut
  {NoStop}%
\bibitem [{\citenamefont {Wegner}(1989)}]{WegnerRG}%
  \BibitemOpen
  \bibfield  {author} {\bibinfo {author} {\bibfnamefont {F.}~\bibnamefont
  {Wegner}},\ }\bibfield  {title} {\bibinfo {title} {Four-loop-order
  $\beta$-function of nonlinear $\sigma$-models in symmetric spaces},\ }\href
  {https://doi.org/https://doi.org/10.1016/0550-3213(89)90063-1} {\bibfield
  {journal} {\bibinfo  {journal} {Nuclear Physics B}\ }\textbf {\bibinfo
  {volume} {316}},\ \bibinfo {pages} {663} (\bibinfo {year}
  {1989})}\BibitemShut {NoStop}%
\bibitem [{\citenamefont {Evers}\ and\ \citenamefont {Mirlin}(2008)}]{evers08}%
  \BibitemOpen
  \bibfield  {author} {\bibinfo {author} {\bibfnamefont {F.}~\bibnamefont
  {Evers}}\ and\ \bibinfo {author} {\bibfnamefont {A.~D.}\ \bibnamefont
  {Mirlin}},\ }\bibfield  {title} {\bibinfo {title} {Anderson transitions},\
  }\href {https://doi.org/10.1103/RevModPhys.80.1355} {\bibfield  {journal}
  {\bibinfo  {journal} {Rev. Mod. Phys.}\ }\textbf {\bibinfo {volume} {80}},\
  \bibinfo {pages} {1355} (\bibinfo {year} {2008})}\BibitemShut {NoStop}%
\bibitem [{\citenamefont {Gade}\ and\ \citenamefont
  {Wegner}(1991)}]{gade1991the}%
  \BibitemOpen
  \bibfield  {author} {\bibinfo {author} {\bibfnamefont {R.}~\bibnamefont
  {Gade}}\ and\ \bibinfo {author} {\bibfnamefont {F.}~\bibnamefont {Wegner}},\
  }\bibfield  {title} {\bibinfo {title} {{The n = 0 replica limit of U(n) and
  U(n)/SO(n) models}},\ }\href
  {https://doi.org/https://doi.org/10.1016/0550-3213(91)90401-I} {\bibfield
  {journal} {\bibinfo  {journal} {Nuclear Physics B}\ }\textbf {\bibinfo
  {volume} {360}},\ \bibinfo {pages} {213} (\bibinfo {year}
  {1991})}\BibitemShut {NoStop}%
\bibitem [{\citenamefont {Finkel'stein}(1984)}]{Finkelstein1984}%
  \BibitemOpen
  \bibfield  {author} {\bibinfo {author} {\bibfnamefont {A.~M.}\ \bibnamefont
  {Finkel'stein}},\ }\bibfield  {title} {\bibinfo {title} {Weak localization
  and {C}oulomb interaction in disordered systems},\ }\href@noop {} {\bibfield
  {journal} {\bibinfo  {journal} {Zeitschrift f{\"u}r Physik B Condensed
  Matter}\ }\textbf {\bibinfo {volume} {56}},\ \bibinfo {pages} {189} (\bibinfo
  {year} {1984})}\BibitemShut {NoStop}%
\bibitem [{\citenamefont {Belitz}\ and\ \citenamefont
  {Kirkpatrick}(1994)}]{BelitzKirkpatrick1994}%
  \BibitemOpen
  \bibfield  {author} {\bibinfo {author} {\bibfnamefont {D.}~\bibnamefont
  {Belitz}}\ and\ \bibinfo {author} {\bibfnamefont {T.~R.}\ \bibnamefont
  {Kirkpatrick}},\ }\bibfield  {title} {\bibinfo {title} {The {Anderson-Mott}
  transition},\ }\href {https://doi.org/10.1103/RevModPhys.66.261} {\bibfield
  {journal} {\bibinfo  {journal} {Rev. Mod. Phys.}\ }\textbf {\bibinfo {volume}
  {66}},\ \bibinfo {pages} {261} (\bibinfo {year} {1994})}\BibitemShut
  {NoStop}%
\bibitem [{\citenamefont {Schwiete}\ and\ \citenamefont
  {Finkel'stein}(2014)}]{SchwieteFinkelstein2014}%
  \BibitemOpen
  \bibfield  {author} {\bibinfo {author} {\bibfnamefont {G.}~\bibnamefont
  {Schwiete}}\ and\ \bibinfo {author} {\bibfnamefont {A.~M.}\ \bibnamefont
  {Finkel'stein}},\ }\bibfield  {title} {\bibinfo {title} {Keldysh approach to
  the renormalization group analysis of the disordered electron liquid},\
  }\href {https://doi.org/10.1103/PhysRevB.89.075437} {\bibfield  {journal}
  {\bibinfo  {journal} {Phys. Rev. B}\ }\textbf {\bibinfo {volume} {89}},\
  \bibinfo {pages} {075437} (\bibinfo {year} {2014})}\BibitemShut {NoStop}%
\bibitem [{\citenamefont {Finkelstein}\ and\ \citenamefont
  {Schwiete}(2023)}]{Finkelstein2023}%
  \BibitemOpen
  \bibfield  {author} {\bibinfo {author} {\bibfnamefont {A.}~\bibnamefont
  {Finkelstein}}\ and\ \bibinfo {author} {\bibfnamefont {G.}~\bibnamefont
  {Schwiete}},\ }\bibfield  {title} {\bibinfo {title} {Scale-dependent theory
  of the disordered electron liquid},\ }\href
  {https://doi.org/https://doi.org/10.1016/j.aop.2023.169260} {\bibfield
  {journal} {\bibinfo  {journal} {Annals of Physics}\ }\textbf {\bibinfo
  {volume} {456}},\ \bibinfo {pages} {169260} (\bibinfo {year}
  {2023})}\BibitemShut {NoStop}%
\bibitem [{\citenamefont {Calabrese}\ and\ \citenamefont
  {Cardy}(2004)}]{Calabrese2004}%
  \BibitemOpen
  \bibfield  {author} {\bibinfo {author} {\bibfnamefont {P.}~\bibnamefont
  {Calabrese}}\ and\ \bibinfo {author} {\bibfnamefont {J.}~\bibnamefont
  {Cardy}},\ }\bibfield  {title} {\bibinfo {title} {Entanglement entropy and
  quantum field theory},\ }\href
  {https://doi.org/10.1088/1742-5468/2004/06/p06002} {\bibfield  {journal}
  {\bibinfo  {journal} {J. Stat. Mech.: Theory Exp.}\ }\textbf {\bibinfo
  {volume} {2004}}\bibinfo  {number} { (06)},\ \bibinfo {pages}
  {P06002}}\BibitemShut {NoStop}%
\bibitem [{\citenamefont {Cardy}\ \emph {et~al.}(2008)\citenamefont {Cardy},
  \citenamefont {Castro-Alvaredo},\ and\ \citenamefont {Doyon}}]{Cardy2008}%
  \BibitemOpen
\bibfield  {number} {  }\bibfield  {author} {\bibinfo {author} {\bibfnamefont
  {J.~L.}\ \bibnamefont {Cardy}}, \bibinfo {author} {\bibfnamefont {O.~A.}\
  \bibnamefont {Castro-Alvaredo}},\ and\ \bibinfo {author} {\bibfnamefont
  {B.}~\bibnamefont {Doyon}},\ }\bibfield  {title} {\bibinfo {title} {Form
  factors of branch-point twist fields in quantum integrable models and
  entanglement entropy},\ }\href {https://doi.org/10.1007/s10955-007-9422-x}
  {\bibfield  {journal} {\bibinfo  {journal} {Journal of Statistical Physics}\
  }\textbf {\bibinfo {volume} {130}},\ \bibinfo {pages} {129} (\bibinfo {year}
  {2008})}\BibitemShut {NoStop}%
\bibitem [{\citenamefont {Calabrese}\ and\ \citenamefont
  {Cardy}(2009)}]{Calabrese2009}%
  \BibitemOpen
  \bibfield  {author} {\bibinfo {author} {\bibfnamefont {P.}~\bibnamefont
  {Calabrese}}\ and\ \bibinfo {author} {\bibfnamefont {J.}~\bibnamefont
  {Cardy}},\ }\bibfield  {title} {\bibinfo {title} {Entanglement entropy and
  conformal field theory},\ }\href
  {https://doi.org/10.1088/1751-8113/42/50/504005} {\bibfield  {journal}
  {\bibinfo  {journal} {J. Phys. A}\ }\textbf {\bibinfo {volume} {42}},\
  \bibinfo {pages} {504005} (\bibinfo {year} {2009})}\BibitemShut {NoStop}%
\bibitem [{\citenamefont {Efetov}(1999)}]{EfetovBook}%
  \BibitemOpen
  \bibfield  {author} {\bibinfo {author} {\bibfnamefont {K.}~\bibnamefont
  {Efetov}},\ }\href@noop {} {\emph {\bibinfo {title} {Supersymmetry in
  Disorder and Chaos}}}\ (\bibinfo  {publisher} {Cambridge University Press},\
  \bibinfo {year} {1999})\BibitemShut {NoStop}%
\end{thebibliography}%

\end{document}